\documentclass{aa}  

\usepackage{graphicx}
\usepackage{txfonts}
\usepackage{natbib}
\usepackage{xcolor}
\bibpunct{(}{)}{;}{a}{}{,}
\usepackage{xcolor}

\begin{document} 

\def\hi {H\,{\sc i}}
\def\hii {H\,{\sc ii}}
\def\water {H$_2$O}
\def\meth { CH$_{3}$OH}
\def\dg{$^{\circ}$}
\def\kms{km\,s$^{-1}$}
\def\ms{m\,s$^{-1}$}
\def\jyb{Jy\,beam$^{-1}$}
\def\mjyb{mJy\,beam$^{-1}$}
\def\solmass {\hbox{M$_{\odot}$}}
\def\solum {\hbox{L$_{\odot}$}} 
\def\d {$^{\circ}$}
\def\n {$n_{\rm{H_{2}}}$}
\def\kmsg{km\,s$^{-1}$\,G$^{-1}$}
\def\tbo {$T_{\rm{b}}\Delta\Omega$}
\def\tb {$T_{\rm{b}}$}
\def\om{$\Delta\Omega$}
\def\dvi {$\Delta V_{\rm{i}}$}
\def\dvz {$\Delta V_{\rm{Z}}$}
\def\code {FRTM code}
\def\NW {Nedoluha \& Watson}
\title{EVN observations of 6.7~GHz methanol maser polarization in massive star-forming regions}
\subtitle{V. Completion of the flux-limited sample. }

\author{G.\ Surcis  \inst{1}
  \and 
  W.H.T. \ Vlemmings \inst{2}
 \and
  H.J.~van Langevelde \inst{3,4}
  \and
  B. \ Hutawarakorn Kramer \inst{5,6}
  \and
  A. Bartkiewicz \inst{7}
  }

\institute{INAF - Osservatorio Astronomico di Cagliari, Via della Scienza 5, I-09047, Selargius, Italy\\
 \email{gabriele.surcis@inaf.it}
 \and
 Department of Space, Earth and Environment, Chalmers University of Technology, Onsala Space Observatory, SE-439 92 Onsala, Sweden
 \and
 Joint Institute for VLBI ERIC, Oude Hoogeveensedijk 4, 7991 PD Dwingeloo, The Netherlands
 \and
 Sterrewacht Leiden, Leiden University, Postbus 9513, 2300 RA Leiden, The Netherlands
 \and
 Max-Planck-Institut f\"{u}r Radioastronomie, Auf dem H\"{u}gel 69, 53121 Bonn, Germany
 \and
 National Astronomical Research Institute of Thailand, 260 Moo 4, T. Donkaew, A. Maerim, Chiang Mai, 50180, Thailand
 \and
 Institute of Astronomy, Faculty of Physics, Astronomy and Informatics, Nicolaus Copernicus University, Grudziadzka 5, 87-100 Torun, Poland
   }

\date{Received 01 September 2021; accepted 04 November 2021}
\abstract
{Although the role of magnetic fields in launching molecular outflows in massive young stellar objects has 
been convincingly demonstrated by theoretical arguments, observationally, the alignment of the magnetic field
lines with the molecular outflows is still under debate.}
{We aim to complete the measurements of the direction of the magnetic fields at milliarcsecond resolution
around a sample of massive star-forming regions to determine whether the magnetic
field and outflows are aligned.} 
{In 2012, we started a large very long baseline interferometry campaign with the European VLBI Network to
measure the magnetic field orientation and strength toward a sample of 31 massive star-forming regions 
(called the flux-limited sample) by analyzing the polarized emission of 6.7~GHz \meth ~masers. In the previous 
papers of the series, we have presented 80\% of the sample. Here, we report the linearly and
circularly polarized emission of 6.7~GHz \meth ~masers toward the last five massive star-forming regions
of the flux-limited sample. The sources are G30.70-0.07, G30.76-0.05, G31.28+0.06, G32.03+0.06, and
G69.52-0.97.}
{We detected a total of 209 \meth ~maser cloudlets, 15\% of which show linearly polarized emission
(0.07\%--16.7\%), and 2\% of which show circularly polarized emission (0.2\%--4.2\%). As reported in 
previous papers, in the last five sources of the flux-limited sample, we also measured well-ordered 
linear polarization vectors. Zeeman splitting was measured toward G30.70-0.07, G32.03+0.06, and
G69.52-0.97.}
{The statistical analysis of the entire flux-limited sample shows that the observations are 
consistent with a bimodal distribution in the difference between the 3D magnetic field direction and 
the outflow axis, with half the magnetic field directions being perpendicular and the other half being
parallel to the outflow. In addition, we determined that typical values 
of the linear and circular polarization fractions for 6.7~GHz \meth ~masers are $P_{\rm{l}}=1.0\%-2.5\%$ and
$P_{\rm{V}}=0.5\%-0.75\%$, respectively. From the circularly polarized spectra of the \meth ~maser features, 
we found that a typical Zeeman splitting is in the range between 0.5~\ms ~and 2.0~\ms. This would correspond 
to $9~\rm{mG}<|B_{\rm{||}}|<40~\rm{mG}$ if $F=3\rightarrow4$ is the most favored of the eight hyperfine
transitions that might contribute to the maser emission.}
\keywords{Stars: formation - masers - polarization - magnetic fields}
\titlerunning{EVN obs of 6.7~GHz methanol maser polarization in MSFRs V.}
\authorrunning{Surcis et al.}

\maketitle
\section{Introduction}
\label{intro}
Molecular outflows are a common and essential component in the formation process of low- and
high-mass stars. In the past 40 years, astronomers have mapped outflows in the whole
mass range of young stellar objects \citep[YSOs; e.g.,][]{fra14,bal16, ang18, ray21}. 
Magnetohydrodynamical (MHD) simulations have shown that the magnetic field plays a crucial role in the
launching of molecular outflows \citep[e.g.,][]{pud19}, more significantly so in the case of massive YSOs
\citep[e.g.,][]{mat18}. Here, for instance, the presence of a magnetic field leads to the formation 
of early outflows. These reduce the radiation pressure, which allows the protostar mass to grow further
\citep{ban07,ros20}. In addition, the intensity of the magnetic field may influence the collimation of the
outflows in massive YSOs. The outflows are well collimated for weak fields and poorly collimated for strong
fields \citep{hen11,sei12}. In case of strong magnetic fields, the structure of the outflows is determined
by the large-scale geometry of the magnetic field lines \citep{mat17}. Recently, \cite{mac20} have 
found a strong dependence of the evolution of outflows in massive YSOs on the initial magnetic field
strength of the prestellar cloud for different accretion rates. In
their 3D MHD simulations, they grouped the results into three categories: successful outflows,
failed outflows, and delayed outflows. In the successful outflows, the outflows appear only when the 
prestellar cloud is strongly magnetized ($\mu\footnote{$\mu=(M/\Phi)/(M/\Phi)_{\rm{crit}}$, where
$(\rm{M/\Phi})$ is the mass-to-magnetic flux ratio and $(M/\Phi)_{\rm{crit}}\approx0.12/\sqrt{G}$ is the
critical value of this ratio, where $G$ is the gravitational constant \citep{tom88}. The critical value
indicates the maximum mass supported by the magnetic field \citep{tom88}. The stronger the magnetic field,
the lower $\mu$.}=2,3$), and after an evolution time of $\sim10^4$~yr, they reach a distance from the
central protostar of about $10^4$~au. When the magnetic field is weak ($\mu\geq5$), we have failed and delayed 
outflows; even though small outflows of about 100-1000~au are
observed in both cases, only in delayed outflows they can overcome the ram pressure and can ultimately
grow. In a massive YSO, a large molecular outflow is therefore formed only if the initial 
magnetic field strength is $B_{\rm{0}}\gtrsim B_{\rm{0,cr}}=10^{-4}(M_{\rm{cl}}/100$\solmass$)$~G, where 
$M_{\rm{cl}}$ is the cloud mass \citep{mac20}.\\
\indent Even though consensus has now been reached on the theoretical importance of magnetic fields in launching 
the outflows both in low- and high-mass protostellar objects, there 
are still some open issues from an observational
point of view. One of this regards the alignment of the magnetic field lines with the
molecular outflows \citep[e.g.,][]{hul19}. \cite{cha13} studied the orientation of the magnetic field in 
a sample of seven low-mass YSOs and found a good alignment on large scales ($>2 \times 10^3$~au) 
between the magnetic field and the outflow axis. In contrast, by analyzing the results from several 
authors, \cite{hul19} concluded that in low-mass protostars the magnetic field has no preferential 
orientation with respect to the outflow axis on scales of $\sim$1000~au. However, only a few sources have 
well-aligned outflows and magnetic fields (\citealt{hul19} and references therein). \cite{gal18} observed
a sample of 12 low-mass protostars (on scales of 600-5000~au) with the Submillimiter Array (SMA) and
found a bimodal distribution: the envelope-scale magnetic field is either aligned or perpendicular to the
molecular outflow axis. In particular, the magnetic field is aligned in single YSOs with small disks, while
there is a preferentially 90\d ~misalignment in wide multiple YSOs systems with higher rotational energies
and/or large disks. Based on 21 high-mass YSOs (on scales $>1000$~au), \cite{zhaq14} found a slight
preference of about 0\d ~and 90\d ~in angles between the magnetic field and the outflow axis. However, because
the size of the sample was small, they concluded that the data are consistent with a random distribution. 
\citet[hereafter Papers~I-IV]{sur12,sur13,sur15,sur19} determined the magnetic field orientation (on scales 
$<100$~au) around a number of high-mass YSOs that are part of the flux-limited sample (Paper~III) by
analyzing the polarized emission of 6.7~GHz \meth ~masers. From an incomplete statistical analysis (based on 
 80\% of the total sample), they suggested the possibility that the magnetic field and
molecular outflows might be aligned. \\
\indent Here, in the fifth and last paper of the series, we present the results of the 
last five sources in the flux-limited sample, which we briefly describe in Sect.~\ref{SEVNG}. In 
Sect.~\ref{obssect}
we describe the observations and analysis of the data. The results of the individual five sources are 
presented in Sect.~\ref{res} and are discussed in Sect.~\ref{discussion}. Here, we also statistically
analyze the flux-limited sample for the orientation of the magnetic field and for the
polarimetric properties of the observed 6.7~GHz \meth ~maser.
\section{Massive star-forming regions}
\label{SEVNG}
The flux-limited sample was selected from the catalog of 6.7~GHz \meth ~masers compiled by \cite{pes05}.
We considered massive star-forming regions (SFRs) with declination $>-9$\d ~and with a total \meth
~maser single-dish flux
density greater than 50~Jy. In addition, we selected the massive SFRs that in 2013 (when the proposal 
was submitted) showed a single-dish total flux density $>$20~Jy \citep{vle11}. This selection
yielded 31 massive SFRs. All the sources but one were observed with the European VLBI
Network\footnote{The European VLBI Network is a joint facility of European, Chinese, South African and
other radio astronomy institutes funded by their national research councils.} (EVN) by 2015, and 26 of
them were analyzed and the results published in a series of papers (\citealt{vle10, sur09, sur111,
sur142}; Papers~I-IV). The last five sources are described below
in Sects.~\ref{G307_intro}-\ref{G69_intro}.
\subsection{\object{G30.70-0.07 (W43-MM2)}}
\label{G307_intro}
The source G30.70-0.07, also known as W43-MM2 ($V_{\rm{lsr}}^{\rm{^{SiO(2-1)}}}=+91.0$~\kms; \citealt{ngu13}), is a
massive dense core in the molecular cloud W43-Main that is part
of the massive SFR W43 \citep{mot03} at a parallax distance of $5.49^{+0.39}_{-0.34}$~kpc
\citep{zhab14}. \cite{bal10} observed the infrared emission of the massive SFR W43 and
found a Z-shaped ridge of warm dust centered on W43-Main, where a giant \hii ~region is powered by a
cluster of OB and Wolf-Rayet stars \citep[e.g.,][]{blu99}. G30.70-0.07 is then located in the southern
segment of the Z-shaped ridge, which is oriented southeast-northwest, and has an estimated mass of
$3.5-5.3\times 10^3$~\solmass ~\citep{bal10}. Most recently, W43-MM2 has been studied by \cite{cor19} in
more detail with the Atacama Large Millimeter/submillimeter Array (ALMA) telescope ($\sim0''\!\!.5$).
\cite{cor19} detected ten different cores (named from A to J) whose total mass is about 600~\solmass 
~($~\sim 70\%$ of which is provided by source A). Three different masers have been detected
toward G30.70-0.07. These are 6.7~GHz \meth, 1.6~GHz OH, and
22~GHz \water ~masers \citep{cas95,fuj14,bre15,arg00,szy05,xi15}. In particular, the 6.7~GHz \meth ~maser
features are associated with source~G \citep[$M=13.7$~\solmass;][]{cor19} of W43-MM2 \citep{fuj14,cor19}.
Source G is located $\sim9''$ south of source~A. 
No molecular outflows have been detected toward G30.70-0.07 so far \citep[e.g.,][]{ngu13, car13, cor19}.\\
\indent \cite{cor19} determined from dust polarized emission that the magnetic fields at subarcsecond
resolution around sources A-D and F of W43-MM2 are connected and likely spirally dragged into the 
gravitational potential of
source~A (see their Fig.~17). The magnetic field of source~G is not connected with that around source~A, its vectors are oriented east-west at the millimeter-peak position, and toward the south, they rotate 
northeast-southwest. The magnetic field strength on the plane of the sky has also been estimated by using
the Davis-Chandrasekhar-Fermi method. It is on the order of mG (sources A-F and J; \citealt{cor19}).
\begin {table*}[t]
\caption []{Observational details.} 
\begin{center}
\scriptsize
\begin{tabular}{ l c c c c c c c c c c c}
\hline
\hline
Source               & Observation      & Calibrator   & Polarization & Beam size        & Position & rms     &$\sigma_{\rm{s.-n.}}$\tablefootmark{b}  & \multicolumn{4}{c}{Estimated absolute position using FRMAP} \\ 
name\tablefootmark{a}& date             &              &  angle       &                  & Angle    &         &         &    $\alpha_{2000}$           & $\delta_{2000}$            & $\Delta\alpha$\tablefootmark{c} & $\Delta\delta$\tablefootmark{c}     \\ 
                     &                  &              &  (\d)        &(mas~$\times$~mas)& (\d)     & ($\frac{\rm{mJy}}{\rm{beam}}$) &   ($\frac{\rm{mJy}}{\rm{beam}}$) &($\rm{^{h}:~^{m}:~^{s}}$) & ($\rm{^{\circ}:\,':\,''}$) &     (mas)             & (mas) \\ 
\hline
G30.70-0.07         & 16 October 2014    & J2202+4216\tablefootmark{d} & $+1\pm1$   & $8.5\times4.2$   & -40.93    &  2      & 25      & +18:47:36.900             &  -02:01:05.025           & 0.4 & 13.1 \\
G30.76-0.05         & 17 October 2014    & J2202+4216\tablefootmark{e} & $+1\pm1$   & $8.5\times3.9$   & -36.28    & 5       & 29      & +18:47:39.732             &  -01:57:21.975           & 0.3  & 10.2\\
G31.28+0.06         & 20 October 2015    & J2202+4216\tablefootmark{f} & $-15\pm3$   & $5.3\times4.4$   & -66.03   &  3      & 60      & +18:48:12.390             &  -01:26:22.629             & 0.8 & 37.4\\
G32.03+0.06         & 21 October 2015    & J2202+4216\tablefootmark{g} & $-16\pm5$   & $5.1\times4.4$   & -76.73   & 3       & 40      & +18:49:36.580             & -00:45:46.891              & 0.2 & 17.5 \\
G69.52-0.97         & 22 October 2015    & J2202+4216\tablefootmark{h} & $-27\pm4$   & $5.9\times3.1$   & -67.04   & 2       &  28     & +20:10:09.0699             & +31:31:34.399              & 0.7 & 1.8 \\
\hline
\end{tabular}
\end{center}
\tablefoot{
\tablefoottext{a}{Source name provided in Galactic coordinates.}
\tablefoottext{b}{Self-noise in the maser emission channels \citep[e.g.,][]{sau12}. When more than one maser feature shows circularly polarized emission, we present the self-noise of the weakest feature. 
When no circularly polarized emission is detected, we consider the self-noise of the brightest maser feature.}
\tablefoottext{c}{Formal errors of the fringe-rate mapping.}
\tablefoottext{d}{Calibrated using results from G30.76-0.05.}
\tablefoottext{e}{Calibrated using 3C286 ($I=1.26$~\jyb, $P_{\rm{l}}=6.5\%$).}
\tablefoottext{f}{Calibrated using 3C286 ($I=0.60$~\jyb, $P_{\rm{l}}=7.7\%$).}
\tablefoottext{g}{Calibrated using 3C286 ($I=0.73$~\jyb, $P_{\rm{l}}=9.0\%$).}
\tablefoottext{h}{Calibrated using 3C286 ($I=0.65$~\jyb, $P_{\rm{l}}=9.2\%$).}
}
\label{Obs}
\end{table*}
\subsection{\object{G30.76-0.05 (W43-MM11)}}
\label{G308_intro}
The source G30.76-0.05 is a massive dense core in the molecular cloud W43-Main, called W43-MM11 \citep{mot03}.
It is located about $3''$ north of W43-MM2 at the east edge of the giant \hii ~region 
\citep[e.g.,][]{blu99, mot03}. We assume a parallax distance equal to that of G30.70-0.07, that is,
$5.49^{+0.39}_{-0.34}$~kpc \citep{zhab14}, and a systemic velocity of
$V_{\rm{lsr}}^{\rm{^{SiO(2-1)}}}=+94.5$~\kms \citep{ngu13}. Compared with the Z-shaped ridge of warm
dust reported by \cite{bal10}, G30.76-0.05 is located at the center, where the ridge might be
responsible for confining the eastern lobe of the \hii ~region \citep{bal10}. No molecular outflow has 
been detected so far \citep[e.g.,][]{ngu13}. The 6.7~GHz \meth ~maser features that are associated with
G30.76-0.05 \citep{cas95,wal97,wal98,szy02,bre15} are grouped into two clusters separated by 34~mas
\citep{fuj14}, and their velocities agree with those of the 1.6 and 1.7 GHz OH maser features
\citep{szy04}. 12~GHz \meth ~maser and 22~GHz \water ~maser emissions have also been detected
\citep{szy05,xi15,bre16}.
\subsection{\object{G31.28+0.06}}
\label{G31_intro}
The source G31.28+0.06 is an ultracompact \hii ~region (UC\hii, e.g., \citealt{kur94,tho06}) at a parallax 
distance of $4.27^{+0.85}_{-0.61}$~kpc from the Sun \citep{zhab14} and with a systemic velocity
$V_{\rm{lsr}}^{\rm{^{^{13}CO}}}=+108.3$~\kms ~\citep{yan18}. G31.28+0.06 is associated with the
cloud~A of the infrared dark cloud (IRDC) G31.23+0.05 \citep{liu17}. \cite{liu17} estimated the
age of the UC\hii ~region to be 0.03--0.09~Myr, suggesting the presence of a YSO at the center.
\cite{yan18} detected a $^{13}$CO outflow with blueshifted
(+101.2~\kms$<\rm{V_{blue}^{^{13}\rm{CO}}} <$+105.7~\kms) and redshifted
(+110.7~\kms$<\rm{V_{red}^{^{13}\rm{CO}}} <$+113.7~\kms) lobes oriented on the plane of the sky
with position angles $\rm{PA_{outflow, blue}^{\rm{^{13}CO}}}=-78$\d ~and
$\rm{PA_{outflow,red}^{\rm{^{13}CO}}}=+154$\d, respectively. Toward the UC\hii ~region, several
maser emissions have been detected; these are \water, OH, and \meth ~maser emissions
\citep[e.g.,][]{min00,cas13,zhab14,fuj14,bre15,xi15,bre16,kim19}. The velocities of the 6.7~GHz
\meth ~masers agree with the velocities of both the blue- and redshifted lobes of the
outflow, but their distribution is complex \citep{fuj14}. The 6.7~GHz \meth ~masers are located
about 7~arcsec north and 1.6~arcsec west of the OH masers \citep[e.g.,][]{cas13,fuj14}, which
are 0.5~arcsec north from the UC\hii ~region \citep{tho06}.\\
\indent With the Nançay Radio Telescope (NRT), \cite{szy09} were able to measure  a linear
polarization fraction of the 1.665~GHz OH maser of $P_{\rm{l}}=62.4\%$ and a linear polarization
angle of $\chi=+35$\d. In addition, \cite{vle08} measured a Zeeman
splitting of the 6.7~GHz \meth ~maser \dvz$=(2.06\pm0.36)$~\ms\ with the Effelsberg
telescope.
\subsection{\object{G32.03+0.06 (MM1)}}
\label{G32_intro}
The source G32.03+0.06 is a massive protostellar core \citep[MM1;][]{rat06} located at a parallax distance of
$5.18^{+0.22}_{-0.21}$~kpc \citep{sat14} in the IRDC~G31.97+0.07 \citep[e.g;][]{zho19}, and its 
systemic velocity is $V_{\rm{lsr}}^{\rm{^{^{18}CO}}}=+96.3$~\kms ~\citep{are18}. The mass and
luminosity of MM1 are $M=2500$~\solmass ~and $L=1.33\times 10^4$~\solum
\citep[ID=16;][]{zho19}. By analyzing the blue profile of the detected CO emissions, \cite{zho19}
reported an infall motion of the gas rather than an outflow. The absence of an outflow was also
reported by \cite{yan18}. Both 12.2~GHz and 6.7~GHz \meth ~maser emissions have been detected
toward G32.03+0.06 \citep{gay94,wal95,bla04,bre16}. \cite{fuj14} observed  the 6.7~GHz \meth ~masers with the East-Asian VLBI Network (EAVN) and grouped them into two clusters: one red- and one
blueshifted cluster. The blueshifted group is located at about 500~mas ($\sim2600$~au) southwest of
the redshifted group. The total velocity range of the \meth ~masers observed by \cite{fuj14} is
+92~\kms$~<V_{\rm{lsr}}<$~+102~\kms. 

\subsection{\object{G69.52-0.97}}
\label{G69_intro}
The source G69.52-0.97 is better known by the name Onsala~1 (ON~1); it is a massive SFR at a parallax
distance of $2.57^{+0.34}_{-0.27}$~kpc \citep{ryg10}. This region contains several YSOs, among
which lies an UC\hii ~region \citep[e.g.,][]{hab74} with an assumed systemic velocity of
$V_{\rm{lsr}}^{\rm{CS}}=+11.6$~\kms ~\citep{bro96}. \cite{kum04} identified two molecular
outflows that might be ejected from the YSO at the center of the UC\hii ~region. One outflow is traced by
$\rm{H^{13}CO^+}$ and SiO (J=2-1) emissions and is oriented on the plane of the sky with a
$\rm{PA_{outflow}^{H^{13}CO^+}}=+44^{\circ}$, and the other is traced by CO (J=2-1), oriented roughly 
west-east ($\rm{PA_{outflow}^{CO}}=-69^{\circ}$). The $\rm{H^{13}CO^+}$/SiO outflow shows that the
blue- (+7~\kms$<\rm{V_{blue}^{\rm{H^{13}CO^+,SiO}}}<$+12~\kms) and redshifted 
(+12~\kms$<\rm{V_{red}^{\rm{H^{13}CO^+,SiO}}}<$+17~\kms) lobes are located northeast and southwest 
of the UC\hii ~region, respectively. In the CO outflow, the blue-
(0~\kms$<\rm{V_{blue}^{\rm{CO}}} <$+12~\kms) and redshifted lobes 
(+12~\kms$<\rm{V_{red}^{\rm{CO}}} <$+24~\kms) are located east and west. Although
the driving sources of each of the outflows are unclear, they could be part of possible massive
binary protostars  \citep{kum03,kum04}. Three maser species were detected and associated
with the UC\hii ~region: \meth, \water, and OH masers 
\citep[e.g.,][]{des98,fis05,fis06,nag08,ryg10,sug11,kim19}. \meth ~masers at 6.7~GHz are grouped
into three clusters (called I-III by \citealt{sug11}). Cluster I is located at the north edge of the
UC\hii ~region, while clusters II and III lie at the south edge \citep{sug11}. They coincide in
position with the 1.6~GHz and 6.0~GHz OH masers \citep{fis05,gre07,fis071,fis076,fis10}. The study
of the proper motion of the 6.7~GHz \meth ~masers and of the 1.6~GHz OH masers indicates that the
gas around the UC\hii ~region is expanding outward \citep{sug11,fis071}. The \water ~masers are
instead located about 1~arcsec ($\sim$2500~au) east with respect to the \meth ~and the OH masers,
where the shock front of the blueshifted lobe of the CO-outflow hits the surrounding gas
\citep{nag08}. \\
\indent The polarized emission of these OH masers was studied several times in the past
\cite[e.g.,][]{fis05,gre07,fis071,fis06,fis10}. \cite{fis05} detected linearly polarized emission
toward 24 OH maser spots at 1.665~GHz with the VLBA ($\langle P_{\rm{l}}\rangle=36\%$). From
these detections, they measured a mean polarization angle of $+59$\d. Under the assumption that the magnetic
field is perpendicular to the linear polarization vectors, the magnetic field is orientated on the
plane of the sky with an angle $\Phi_{\rm{B}}^{\rm{1.6~GHz~OH}}=-31$\d. From the 6.0~GHz OH
masers, the linearly polarized emission was also detected, which led \cite{gre07} and \cite{fis10} to
measure $\langle P_{\rm{l}}\rangle=13.6\%$ and $19\%$, respectively. \cite{gre07} measured a mean linear 
polarization angle of
$-58$\d ~($\Phi_{\rm{B}}^{\rm{6.0~GHz~OH}}=+32$\d) with
the Multi-Element Radio Linked Interferometer Network (MERLIN), while \cite{fis10} measured 
$\langle \chi \rangle=-60$\d ~($\Phi_{\rm{B}}^{\rm{6.0~GHz~OH}}=+30$\d) with a global VLBI
network. In addition, from
the right circular and left circular polarization spectra of the 1.6 and 6.0~GHz OH masers, it was
possible to measure the magnetic field strength. From 11 and 10 Zeeman pairs of 1.7~GHz OH
masers, a magnetic field on the plane of the sky $B_{||}=-3.3$~mG \citep{fis05} and $-4.5$~mG
\citep{fis071} was measured, respectively; while from 7, 6, and 11 Zeeman pairs of
6.0~GHz OH masers, $B_{||}=-5.6$~mG \citep{fis06}, $-3.5$~mG \citep{gre07}, and $-4.8$~mG
\citep{fis10} was estimated, respectively. \cite{gre07} reported a linearly polarized emission from two 
6.7~GHz \meth ~maser features with the
MERLIN (beam size 
43~mas$\times$~43~mas). In
particular, for their features C ($V_{\rm{lsr}}^{\rm{C}}=+14.62$~\kms, $I^{\rm{C}}=53.56$~\jyb,
$\Delta v\rm{_{L}}^{\rm{C}}=0.27$~\kms, $P_{\rm{l}}^{\rm{C}}=0.2\%$) and D 
($V_{\rm{lsr}}^{\rm{D}}=+14.62$~\kms, $I^{\rm{D}}=20.03$~\jyb, $\Delta v\rm{_{L}}^{\rm{D}}=0.27$~\kms,
$P_{\rm{l}}^{\rm{D}}=1.3\%$), they measured a linear polarization angle of $\chi^{\rm{C}}=+20.6$\d 
~and $\chi^{\rm{D}}=-76.7$\d, respectively. \cite{gre07} also reported the very first
Zeeman-splitting measurement for 6.7~GHz \meth ~maser emission using the 
cross-correlation method \citep{mod05}; for maser feature D, they measured 
$P_{\rm{V}}^{\rm{D}}=0.6\%$ and \dvz$^{D}=0.9\pm0.3$~\ms. 
\section{Observations and analysis}
\label{obssect}
The last five massive SFRs in the flux-limited sample were observed in full polarization spectral 
mode at 6.7~GHz with seven and ten of the EVN antennas in October 2014 (Ef, Wb, Jb, On, Nt, Tr, 
and Ys) and in October 2015 (Ef, Wb, Jb, Mc, On, Nt, Tr, Ys, Sr, and Ir). We 
observed a bandwidth of 2~MHz in order to cover a velocity range of $\sim$100~\kms, with a total 
observing time of 35~h (7~h per source). The data were correlated with the EVN software correlator
\citep[SFXC;][]{kei15} at the Joint Institute for VLBI ERIC (JIVE, the Netherlands) using 2048 
channels (spectral resolution of $\sim$1~kHz corresponding to $\sim$0.05~\kms) and producing all 
four polarization combinations (RR, LL, RL, and LR). We present the observational details in 
Table~\ref{Obs}. In particular, Cols.~3 and 4 list the polarization calibrators with their 
polarization angles. We list the restoring beam size and the corresponding position 
angle, the thermal noise, and the self-noise\footnote{The self-noise is high when the power 
contributed by the astronomical maser is a significant portion of the total received power 
\citep{sau12}.}, which is produced by the maser emission in the maser emission channels, in 
Col.~5 to Col.~8 (see Paper~III for details). \\
\indent The data were calibrated and imaged using the Astronomical Image Processing Software 
package (AIPS), following the procedure used previously in Papers~I-IV. We used the source 
J2202+4216 to calibrate the bandpass, the delay, the phase, and the D terms. We performed the 
calibration of the polarization angles on the calibrator 3C286 (known also as J1331+3030); more 
details are given in the caption of Table~\ref{Obs}. Afterward, we performed the fringe-fitting 
and the self-calibration on the brightest maser feature of each SFR (the reference maser feature 
in Tables~\ref{G307_tab}-\ref{G69_tab}). We imaged the four \textit{I}, \textit{Q}, \textit{U}, 
and \textit{V} Stokes cubes using the AIPS task IMAGR, then we combined the \textit{Q} and
\textit{U} cubes to produce the polarized intensity ($POLI=\sqrt{Q^{2}+U^{2}}$) and polarization 
angle ($POLA=1/2\times~\rm{atan}(U/Q)$) cubes. The formal error on $POLA$ due to the thermal noise
is given by $\sigma_{POLA}=0.5 ~(\sigma_{P}/POLI) \times (180^{\circ}/\pi)$  \citep{war74}, where 
$\sigma_{P}$ is the rms error of POLI. Because the observations were not performed in
phase-reference mode, the absolute positions of the masers could be estimated, with uncertainties 
on the order of a few milliarcseconds, only through fringe-rate mapping (for details, see Papers~I-IV). 
We used 
the AIPS task FRMAP and the estimated absolute position of the reference maser. The FRMAP 
uncertainties are listed in Col.~9 to Col.~12 of Table~\ref{Obs} (see Paper III for details).\\
\indent To analyze the polarimetric data, we followed the procedure reported in Paper~IV, to which
we refer for details. After identifying the \meth ~maser features and determining which
of them were linearly polarized by measuring the mean linear polarization fraction ($P_{\rm{l}}$) and
the mean linear polarization angle ($\chi$) across the spectrum of each feature, we modeled their 
total intensity and linearly polarized spectra with the modified \code ~for \meth ~masers (see
Paper~IV). In particular, as in Paper~IV, we followed \cite{lan18} in assuming the hyperfine
transition with the largest Einstein coefficient for stimulated emission, that is, $F=3\rightarrow4$,
as more favored among the eight hyperfine transitions that might contribute to the maser emission.
We note here that the emerging brightness temperature (\tbo), the intrinsic thermal linewidth
(\dvi),~and the angle between the magnetic field and the maser propagation direction ($\theta$) 
are the outputs of the \code. We consider that if $\theta>\theta_{\rm{crit}}=55$\d, 
where $\theta_{\rm{crit}}$ is the Van Vleck angle, the magnetic field is perpendicular to the
linear polarization vectors; otherwise, it is parallel \citep{gol73}. We also measured the Zeeman
splitting ($\Delta V_{\rm{Z}}$) of the maser features that show circularly polarized emission. 
The Zeeman splitting was measured including the best estimates of \tbo ~and \dvi ~in the FRTM 
code to produce the models used for fitting the total intensity and
circularly polarized spectra of the corresponding \meth ~maser feature. We 
estimated the magnetic field along the line of sight ($B_{||}=\Delta V_{\rm{Z}}/\alpha_{\rm{Z}}$) 
by considering the Zeeman coefficient of the hyperfine transition $F=3\rightarrow4$, that is,
$\alpha_{\rm{Z}}=-0.051$~\kmsg~($\alpha_{\rm{Z}}=-1.135$~Hz~mG$^{-1}$, \citealt{lan18}). Because
$\alpha_{\rm{Z}}$ for $F=3\rightarrow4$ is the largest of the eight hyperfine transitions, 
our estimate of $B_{||}$ is a lower limit.
\section{Results}
\label{res}
In Sects.~\ref{G307_sec}--\ref{G69_sec} the 6.7 GHz \meth ~maser distribution and the polarization
results for each of the five massive SFRs observed with the EVN are reported separately. The 
properties of all the maser features can be found in Tables~\ref{G307_tab}--\ref{G69_tab}.
\begin{figure}[t!]
\centering
\includegraphics[width = 9 cm]{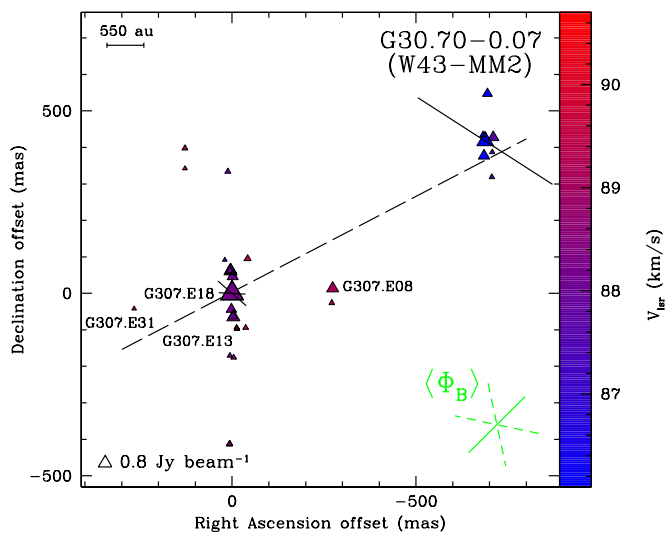}
\caption{View of the \meth ~maser features detected around G30.70-0.07 (W43-MM2).
The reference position is the estimated absolute position from Table~\ref{Obs}.
Triangles identify \meth ~maser features whose side length is scaled logarithmically according 
to their peak flux densities (Table~\ref{G307_tab}). 
Maser local standard of rest radial velocities are indicated by color (the assumed velocity of the
region is $V_{\rm{lsr}}^{\rm{^{SiO(2-1)}}}=+91.0$~\kms; \citealt{ngu13}). The 0.8~\jyb ~symbol is
plotted for comparison.
The linear polarization vectors, scaled logarithmically according to the polarization 
fraction $P_{\rm{l}}$ ($P\rm{_l}=0.07-1.2\%$; see Table~\ref{G307_tab}), are overplotted. In the
bottom right corner, the corresponding error-weighted orientation of the magnetic field
($\langle\Phi_{\rm{B}}\rangle$, see Sect.\ref{Borient}) is also shown. The two dashed segments
indicate the uncertainty. The dashed line is the best least-squares linear fit of the \meth ~maser
features ($\rm{PA_{CH_{3}OH}}=-67^{\circ}\pm7$\d).} 
\label{G307_cp}
\end{figure}
\subsection{\object{G30.70-0.07 (W43-MM2)}}
\label{G307_sec}
\begin{figure*}[th!]
\centering
\includegraphics[width = 6 cm]{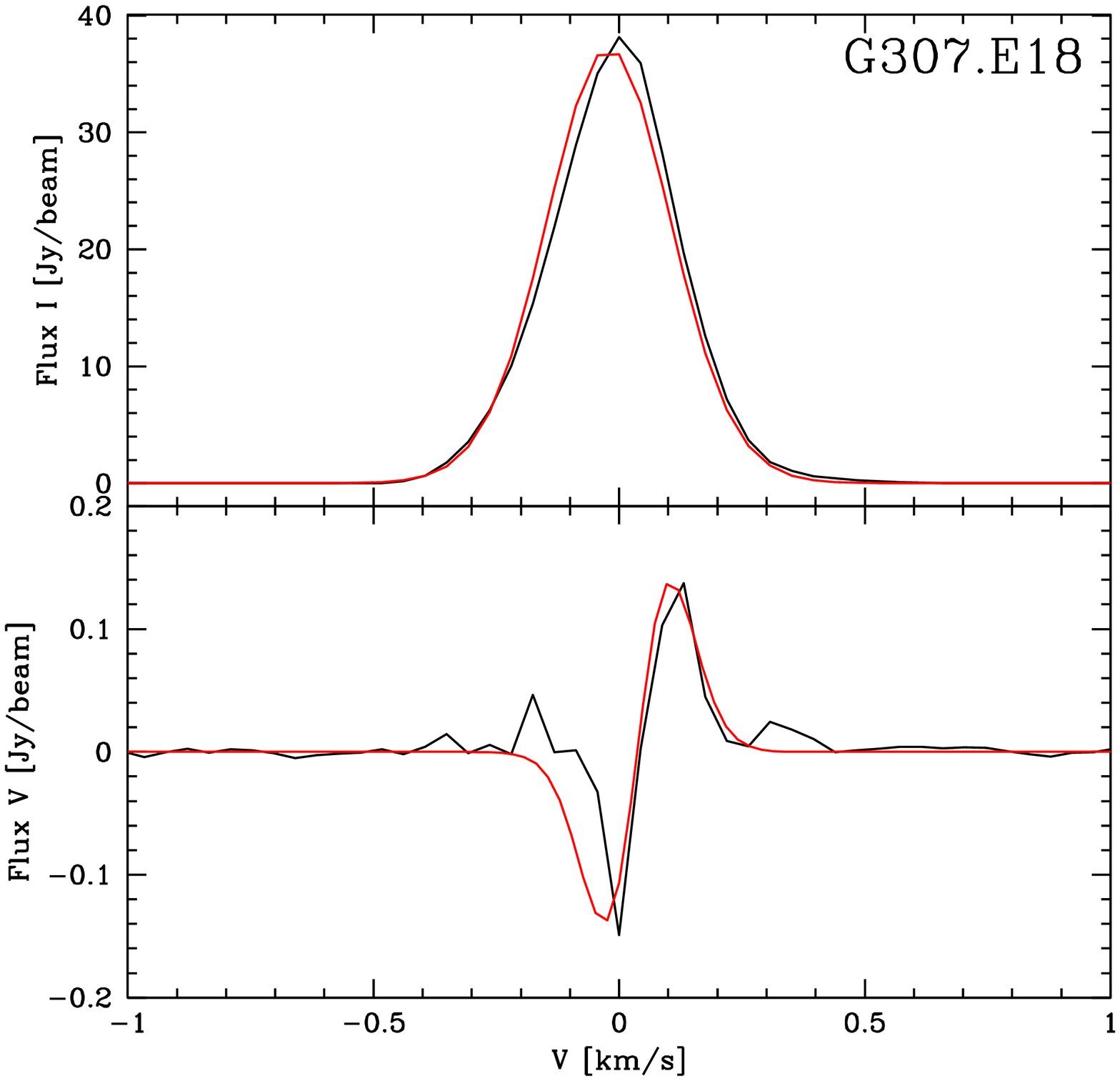}
\includegraphics[width = 6 cm]{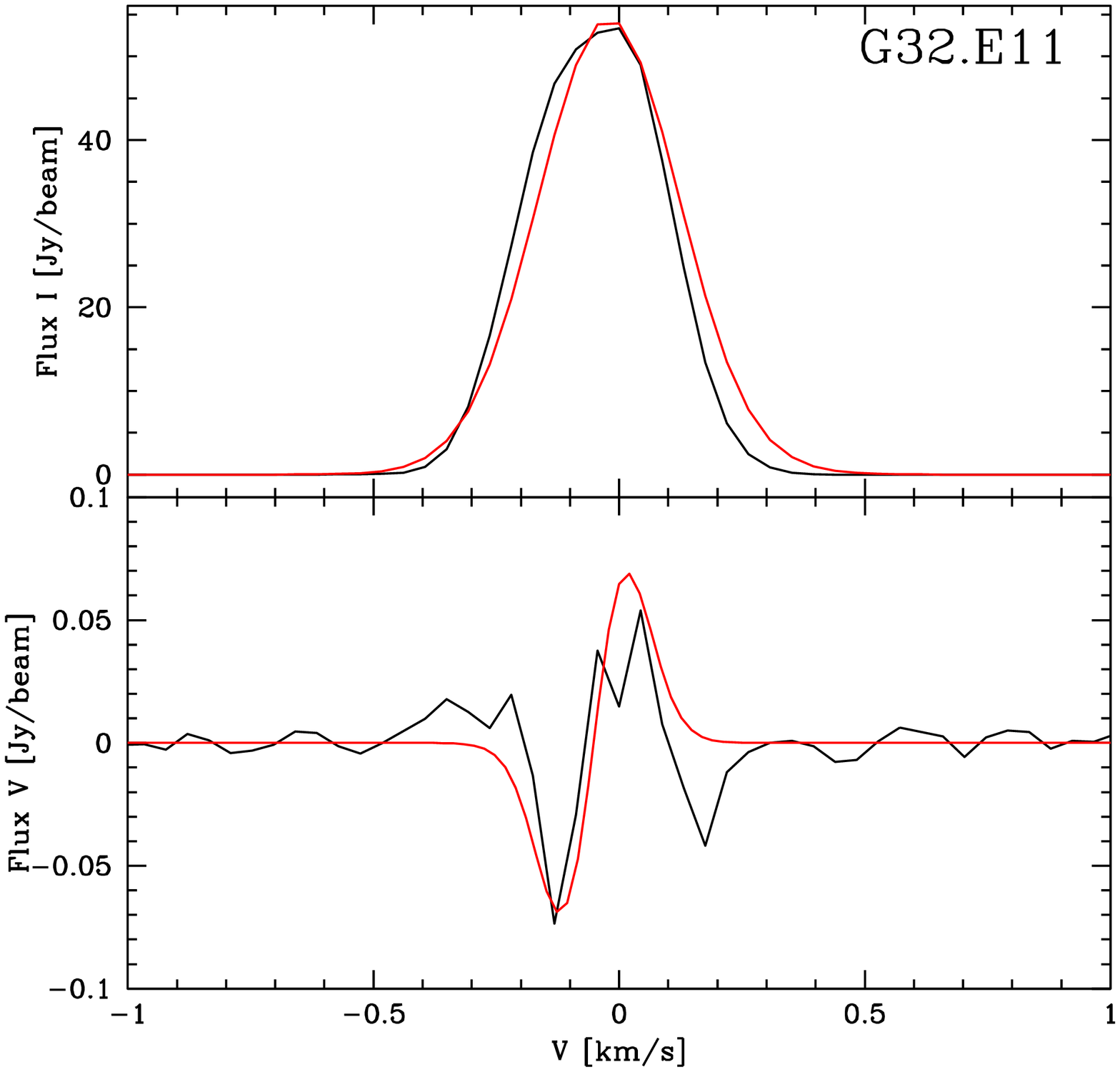}
\includegraphics[width = 6 cm]{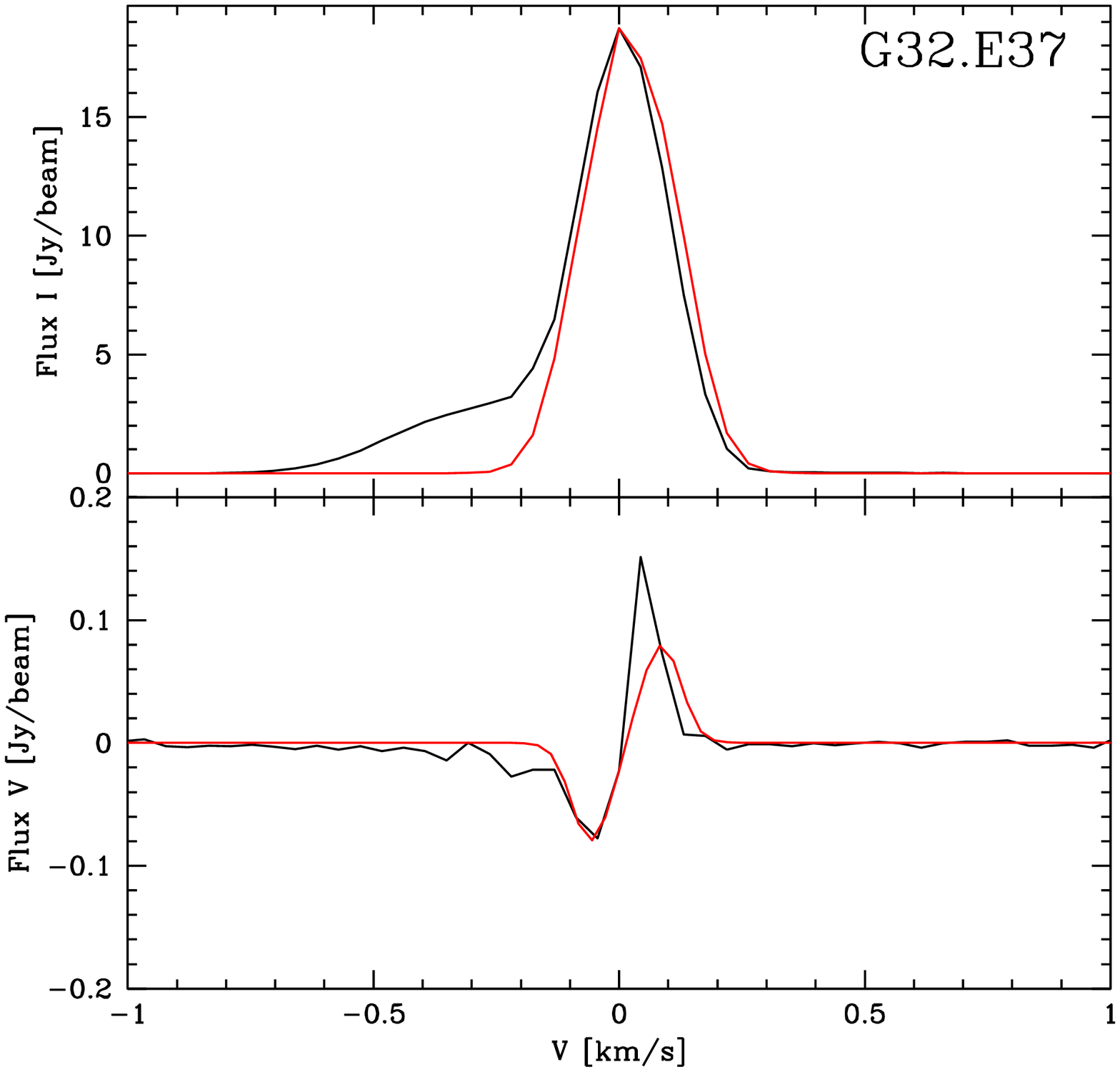}
\includegraphics[width = 6 cm]{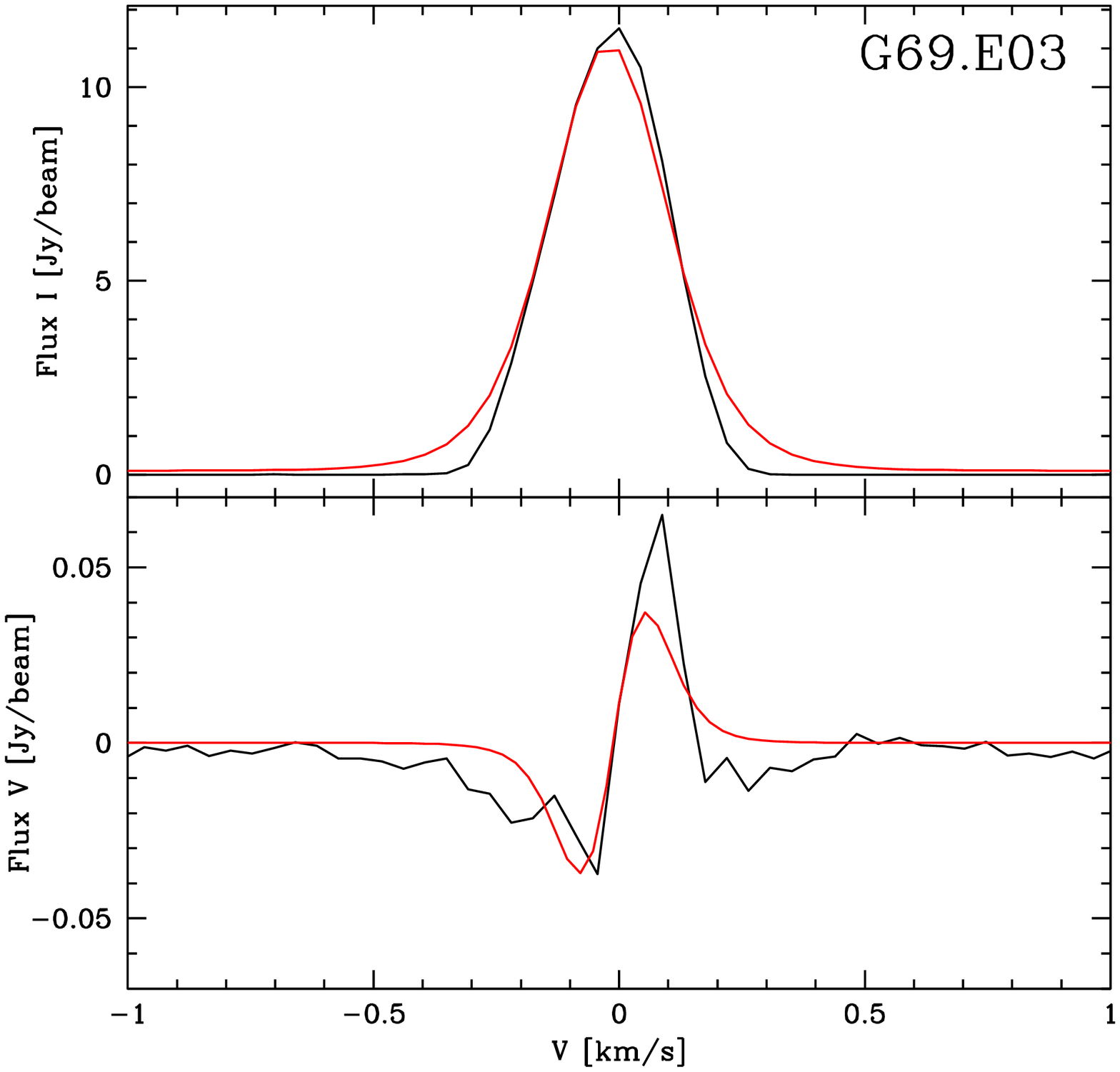}
\includegraphics[width = 6 cm]{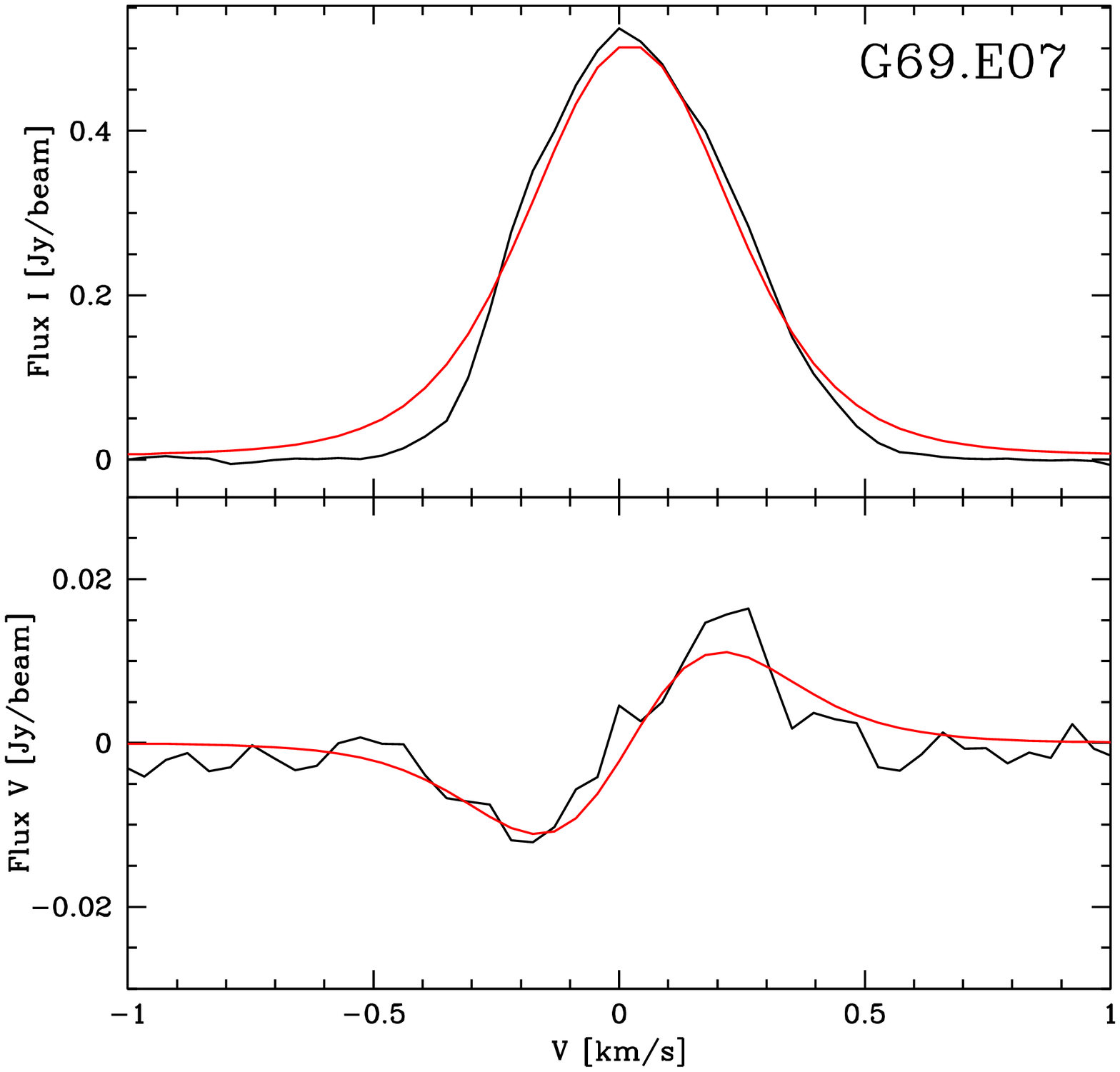}
\caption{Total intensity (\textit{I, upper panel}) and circularly polarized intensity (\textit{V, lower panel}) 
spectra for the \meth ~maser features called G307.E18, G32.E11, G32.E37, G69.E03, and G69.E07 (see 
Tables~\ref{G307_tab}, \ref{G32_tab}, and \ref{G69_tab}). The thick red lines are the best-fit models of 
\textit{I} and \textit{V} emissions obtained using the adapted FRTM code (see Sect.~\ref{obssect}). The maser 
features were centered on zero velocity.}
\label{Vfit}
\end{figure*}

We detected a total of 31 \meth ~maser features (named G307.E01--G307.E31 in Table~\ref{G307_tab})
toward G30.70-0.07 (see Fig.~\ref{G307_cp}). These maser features, which are grouped into two 
regions separated by about 800~mas southeast-northwest ($\sim$4400~au), are associated with the
southern tail of source~G of W43-MM2 (see Sect.~\ref{G307_intro}). A best least-squares linear fit 
of the maser features provided $\rm{PA}_{CH_3OH}^{G30.70}=-67^{\circ}\pm7$\d. Their velocities are
in the range +86~\kms$~<V_{\rm{lsr}}<$~+91~\kms, with the most blueshifted features located in 
the northwest region. This agrees with what \cite{fuj14} reported, even though we were able to 
detect two maser features with velocities $>+90$~\kms ~ at mas resolution (i.e., G307.E13 and
G307.E31) that were undetected by \cite{fuj14}. We note that the peak flux density ($I<70$~\mjyb) of these maser features is lower than the rms reached by \cite{fuj14}, which can be
deduced to be on the order of 100~mJy. For the same reason, we were able to detect more maser 
features than \cite{fuj14}; for instance, a group of three maser features at
$0~\rm{mas}<\rm{RA~offset}<130~\rm{mas}$ and $330~\rm{mas}<\rm{DEC~offset}<400~\rm{mas}$ (i.e.,
G307.E27, G307.E29, and G307.E30). We detected one maser feature that potentially might have been 
detected by \cite{fuj14}, but that was not detected by them. This is G307.E08 ($I=0.730$~\mjyb).\\
\indent We detected linearly polarized emission toward three maser features (G307.E05, G307.E18, 
and G307.E19), among which G307.E18 also shows circularly polarized emission ($P_{\rm{V}}=0.7$\%, 
see Fig~\ref{Vfit}). G307.E05 (northwest region) has a low linear polarization fraction
($P_{\rm{l}}=1.2$\%), while the other two show the lowest $P_{\rm{l}}$ ever measured toward a 
6.7~GHz \meth ~maser feature (i.e., $P_{\rm{l}}^{\rm{G307.E18}}=0.09$\% and
$P_{\rm{l}}^{\rm{G307.E19}}=0.07$\%, see Table~\ref{G307_tab}). Consequently, the \code ~(see
Sect.\ref{obssect}) provided the lowest values of \tbo ~ever estimated for them. These are
\tbo$~=4 \times 10^7$~K~sr and \tbo$~=5 \times 10^7$~K~sr for G307.18 and G307.E19,
respectively. The values of \dvi ~are instead typical (see Col.9 of Table~\ref{G307_tab}). Because
of the low values of $P_{\rm{l}}$ and \tbo ~of G307.E18 and G307.E19, their estimated $\theta$ 
angles are small. However, only for the feature G307.E18 is the magnetic field parallel to the
linear polarization vector. For this feature, we have that 
$|\theta^{\rm{+}}-55$\d$|<|\theta^{\rm{-}}-55$\d$|$,
where $\theta^{\rm{\pm}}=\theta\pm\varepsilon^{\rm{\pm}}$ and with $\varepsilon^{\rm{\pm}}$ the
errors associated with $\theta$, which indicates that the magnetic field is indeed parallel, as 
explained in Paper~III. For the other two maser features, the magnetic field is perpendicular
($|\theta^{\rm{+}}-55$\d$|>|\theta^{\rm{-}}-55$\d$|$).
\subsection{\object{G30.76-0.05 (W43-MM11)}}
\label{G308_sec}
\begin{figure*}[t!]
\centering
\includegraphics[width = 9 cm]{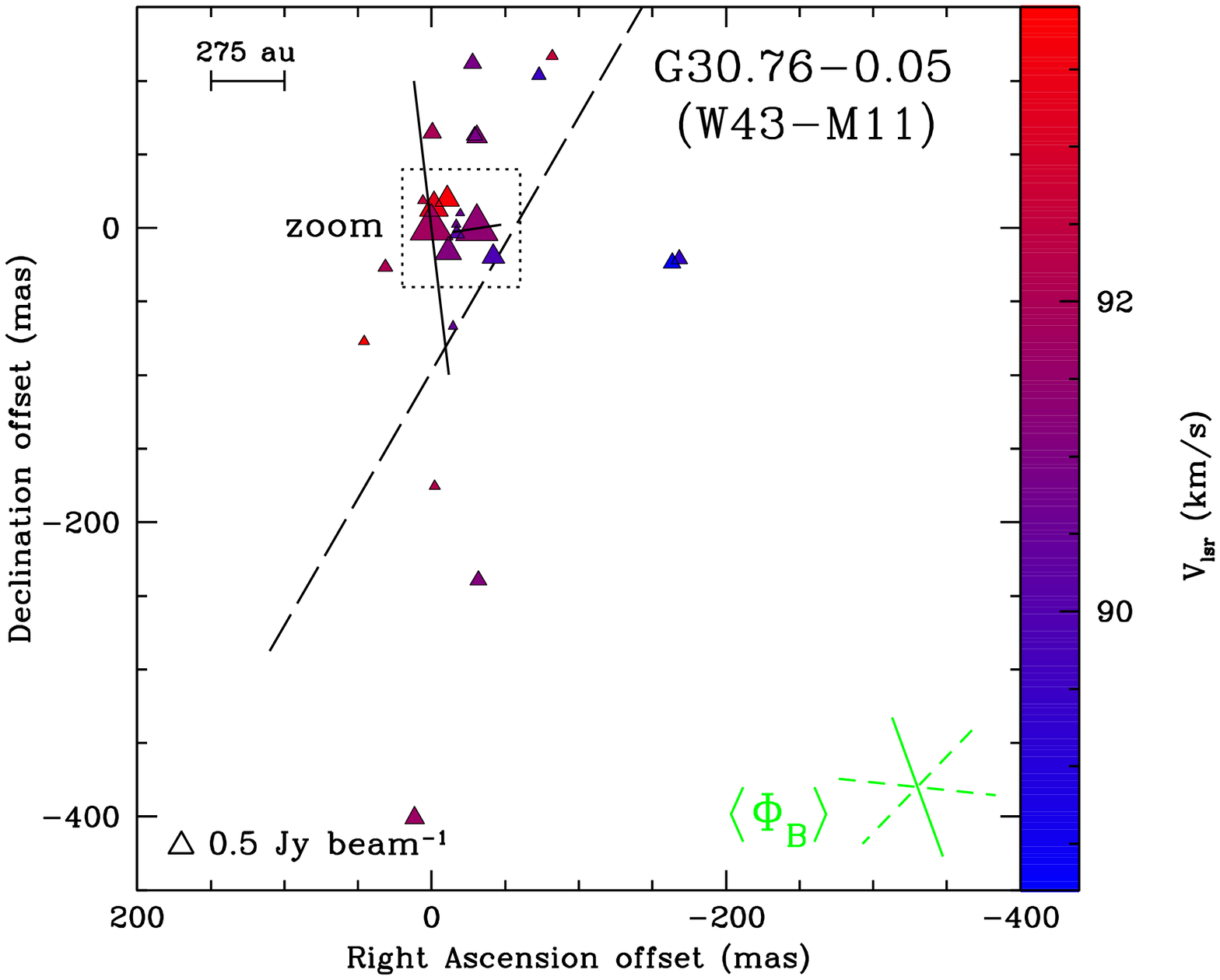}
\includegraphics[width = 9 cm]{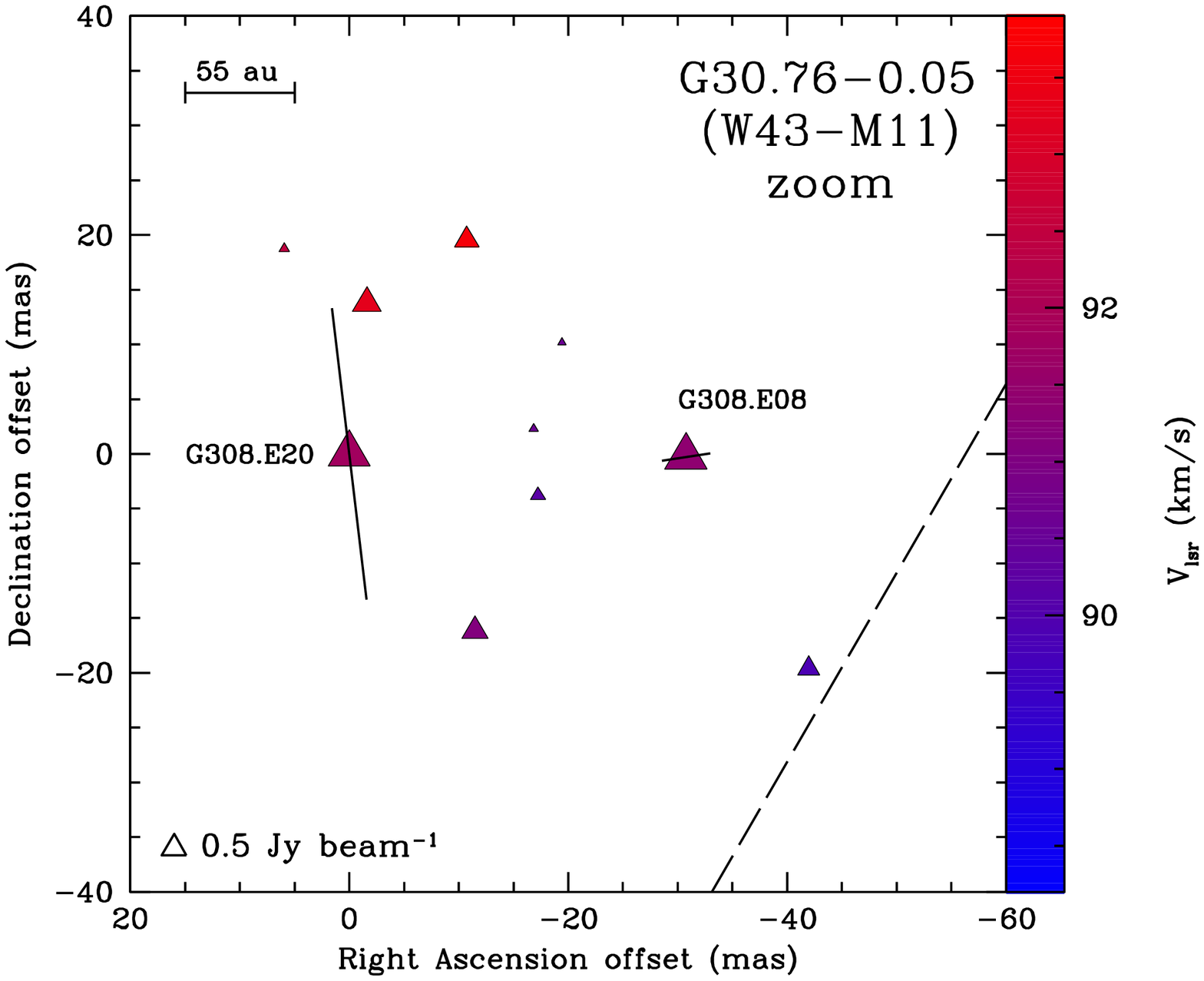}
\caption{\textbf{\textit{}} View of the \meth ~maser features detected around G30.76-0.05 (Table~\ref{G308_tab}) (\textit{left panel}). The symbols
are the same as in Fig.~\ref{G307_cp}. The polarization fraction is in the range $P\rm{_l}=0.4-1.9\%$
(Table~\ref{G308_tab}). The assumed velocity of the YSO is $V_{\rm{lsr}}^{\rm{SiO(2-1)}}=+94.5$~\kms 
~\citep{ngu13}. The dashed line is the best least-squares linear fit of the \meth ~maser features 
($\rm{PA_{CH_{3}OH}}=-30^{\circ}\pm48$\d). \textit{Right panel}: Zoom-in view.}  
\label{G308_cp}
\end{figure*}

We list in Table~\ref{G308_tab} the 24 \meth ~maser features (named G308.E01--G308.E24) that we 
detected toward G30.76-0.05 (see Fig.\ref{G308_cp}). In addition to the strong maser features reported 
by \cite{fuj14} in the velocity range +90~\kms$~<V_{\rm{lsr}}<$~+93~\kms, we were able to recover 
on the milliarcsecond scale not only the weak maser features in this velocity range, but also the weak 
maser features at lower and higher velocities (i.e., +88~\kms$~<V_{\rm{lsr}}<$~+94~\kms). These 
maser features are located farther from the two clusters reported by \cite{fuj14}: north 
($>+50$~mas with respect to the reference position in Fig.~\ref{G308_cp}), south ($<-60$~mas),
and west ($<-100$~mas).\\
\indent The two brightest maser features, G308.E08 and G308.E20, show linearly polarized emission
with $P_{\rm{l}}=0.4$\% and $1.9\%$, respectively. For both maser features, the \code ~provided 
$\theta$ angles for which the magnetic field can be estimated to be parallel to the linear
polarization vector. No circularly polarized emission was detected at 3$\sigma_{\rm{s.-n.}}$
($P_{\rm{V}}<0.6$\%).
\begin{figure}[th!]
\centering
\includegraphics[width = 9 cm]{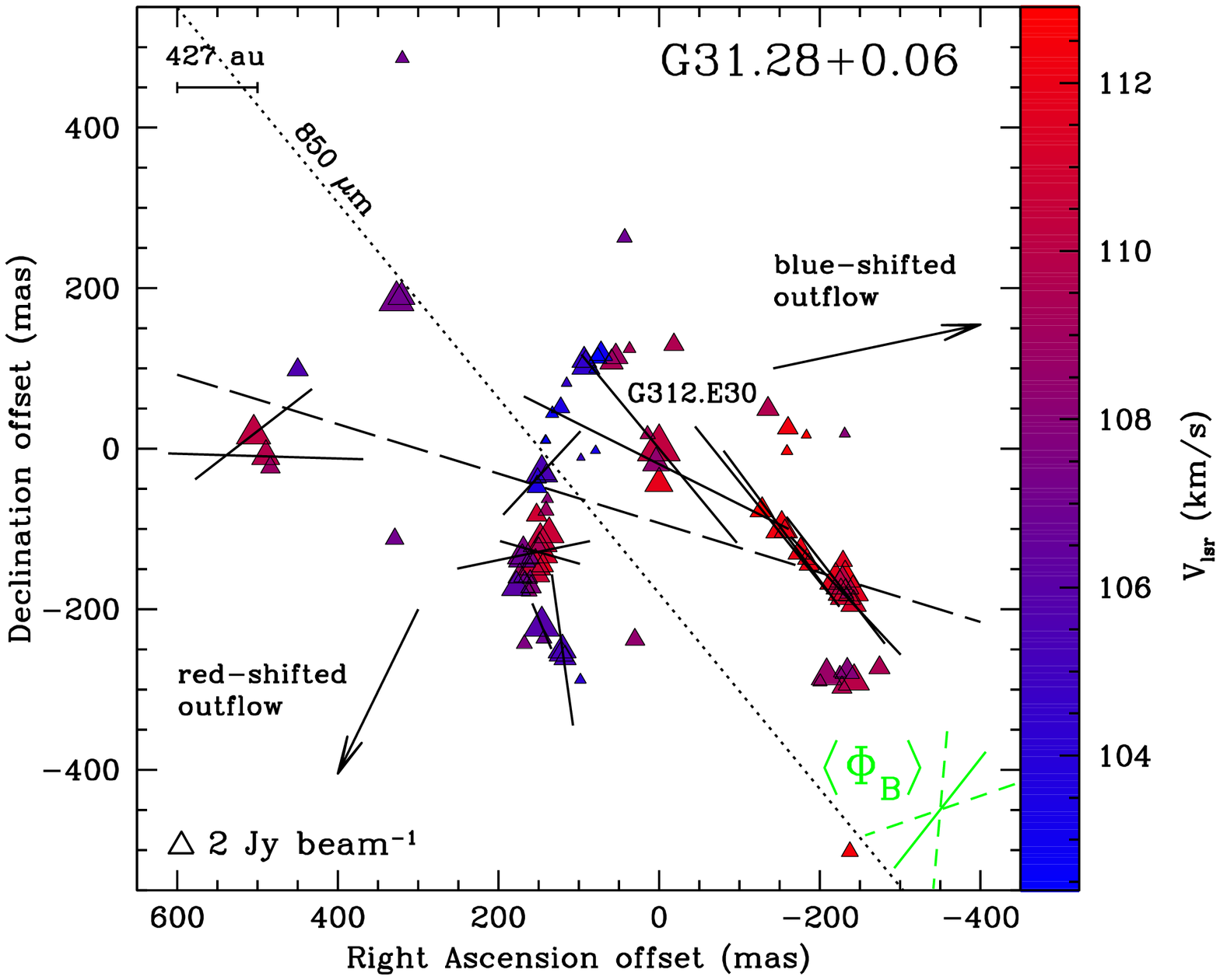}
\caption{View of the \meth ~maser features detected around G31.28+0.06 (Table~\ref{G312_tab}). The
symbols are the same as in Fig.~\ref{G307_cp}. The polarization fraction is in the range 
$P\rm{_l}=0.5-16.7\%$ (Table~\ref{G312_tab}). The assumed velocity of the YSO is 
$V_{\rm{lsr}}^{\rm{^{13}CO}}=+108.3$~\kms ~\citep{yan18}. The dashed line is the best 
least-squares linear fit of the \meth ~maser features ($\rm{PA_{CH_{3}OH}}=+73^{\circ}\pm4$\d). 
The two arrows indicate the direction and not the actual position of the red- and blueshifted 
lobe of the outflow ($\rm{PA_{outflow, blue}^{\rm{^{13}CO}}}=-78$\d ~and 
$\rm{PA_{outflow,red}^{\rm{^{13}CO}}}=+154$\d, \citealt{yan18}). The dotted line indicates
the orientation of the 850~$\rm{\mu m}$ emission detected by \cite{tho06}.}  
\label{G312_cp}
\end{figure}
\subsection{\object{G31.28+0.06}}
\label{G31_sec}
Toward G31.28+0.06, we detected 88 \meth ~maser features (named G312.E01--G312.E88 in
Table~\ref{G312_tab}) in the velocity range +102~\kms$~<V_{\rm{lsr}}<$~+113~\kms. The maser 
features show a complex distribution located at 7.5~arcsec north and 1.6~arcsec west of the 
UC\hii ~region peak and from the launching point of the $^{13}$CO outflow \citep{tho06,yan18}. 
However, the linear fit of the maser features is almost perpendicular to the redshifted lobe 
of the $^{13}$CO outflow and is quite consistent with the weak elongated emission observed at
850~$\rm{\mu m}$ and at 450~$\rm{\mu m}$ (\citealt{tho06}; see Fig.~\ref{G312_cp}). In particular, 
the western stream of redshifted maser features shows the best alignment with the
850~$\rm{\mu m}$ emission. The velocity range of
the maser features agrees well with the velocity ranges of both the red- and blueshifted 
lobes of the $^{13}$CO outflow \citep{yan18}. Although a velocity gradient is not seen, we note 
that most of the redshifted maser features ($V_{\rm{lsr}}>$~+108~\kms) are located toward 
the west of the distribution and the majority of the blueshifted ones ($V_{\rm{lsr}}<$~+108~\kms) 
toward the east of the distribution. The brightest maser feature (G312.E30, 
$V_{\rm{lsr}}=$~+110.35~\kms, $I=32.441$~\jyb; see Table~\ref{G312_tab}) is located roughly at the 
center, that is, at (0,0) position.\\
\indent About 15\% of the \meth ~maser features show linearly polarized emission. The maser
feature G312.E32 ($V_{\rm{lsr}}=$~+108.37~\kms, $I=1.485$~\jyb) has the largest 
linear polarization fraction ever measured toward a 6.7~GHz \meth ~maser, that is, 
$P_{\rm{l}}=16.7$\%
\citep[Papers~I, II, III, IV;][this work]{sur09,sur111,sur141,vle10,san15,dal17, bre19}. 
This high $P_{\rm{l}}$ might indicate that the maser feature is saturated. \cite{dal20} 
indeed indicated that when \tbo$~>10^{10}$~K~sr, the 6.7~GHz \meth ~maser starts to saturate and reaches
its maximum $P_{\rm{l}}$ when \tbo$~=10^{12}-10^{13}$~K~sr. However, the \code
~was able to properly fit all the polarized maser features. The estimated $\theta$ angles indicate
that the magnetic field is perpendicular to the linear polarization vector for all the maser 
features but G312.E61, for which $|\theta^{\rm{+}}-55$\d$|<|\theta^{\rm{-}}-55$\d$|$. No circular 
polarization was detected at 3$\sigma_{\rm{s.-n.}}$ ($P_{\rm{V}}<0.6$\%).
\begin{figure*}[ht!]
\centering
\includegraphics[width = 9 cm]{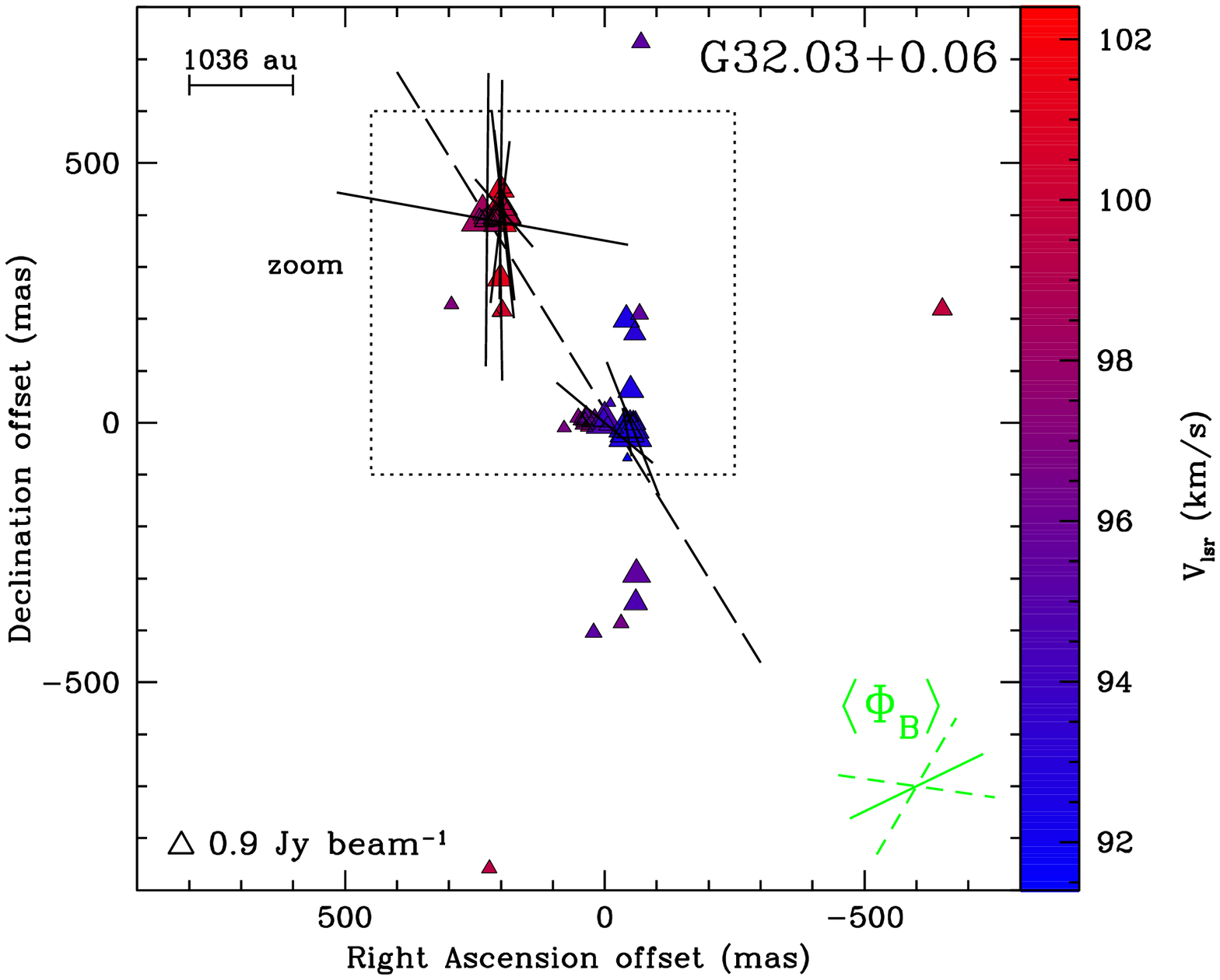}
\includegraphics[width = 9 cm]{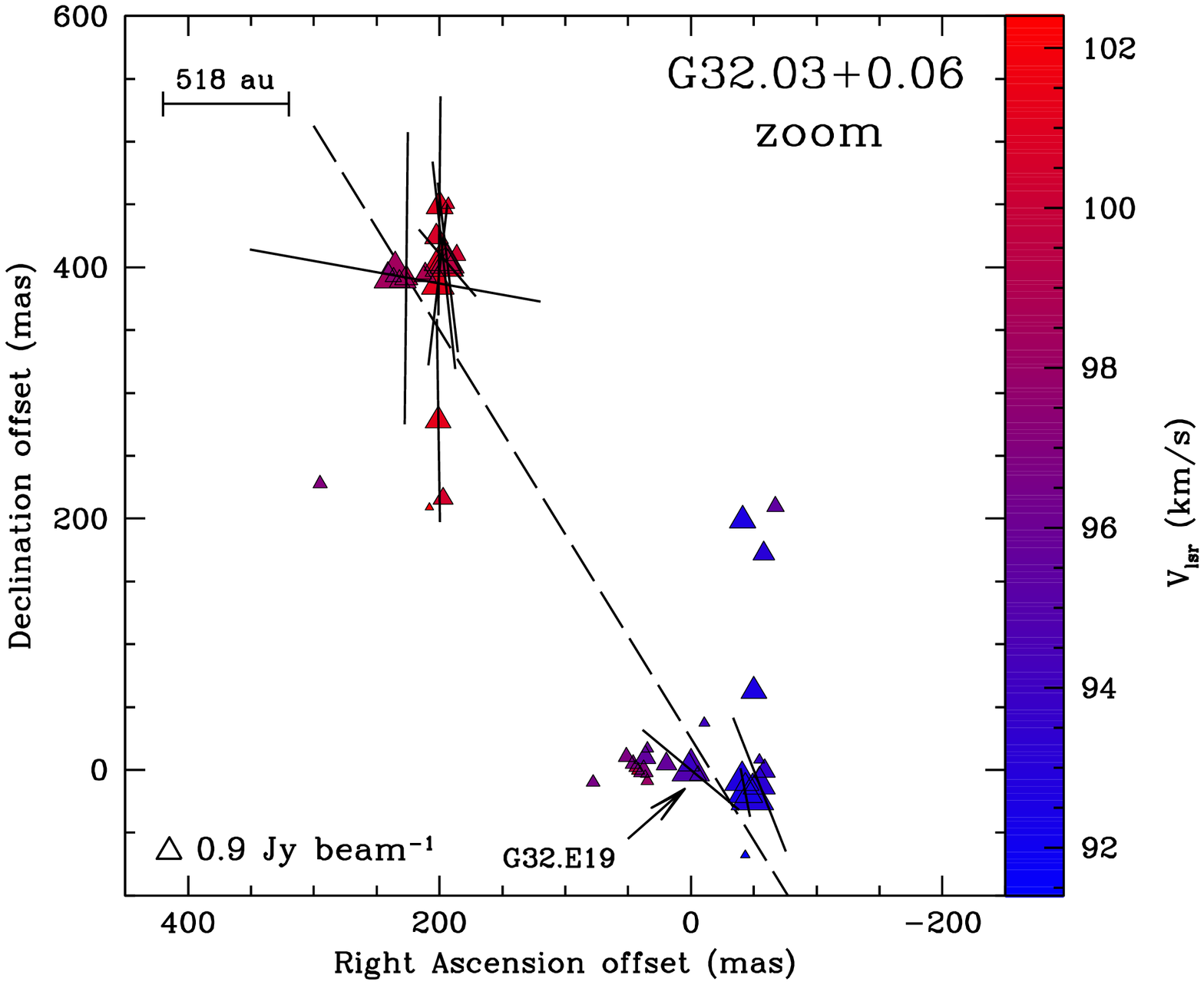}
\caption{\textit{}View of the \meth ~maser features detected around G32.03+0.06
(Table~\ref{G32_tab}) (\textit{left panel}). The symbols are the same as in Fig.~\ref{G307_cp}. The polarization fraction is in the range
$P\rm{_l}=0.4-10.0\%$ (Table~\ref{G32_tab}). The assumed velocity of the YSO is
$V_{\rm{lsr}}^{\rm{^{18}CO}}=+96.3$~\kms ~\citep{are18}. The dashed line is the best least-squares linear fit of
the \meth ~maser features ($\rm{PA_{CH_{3}OH}}=+32^{\circ}\pm7$\d). 
\textit{Right panel}: Zoom-in view.
}  
\label{G32_cp}
\end{figure*}
\subsection{\object{G32.03+0.06 (MM1)}}
\label{G32_sec}
With the EVN, we detected 53 \meth ~maser features (named G32.E01--G32.E53 in Table~\ref{G32_tab})
in the velocity range +92~\kms$~<V_{\rm{lsr}}<$~+102~\kms. They can be grouped into two main red-
and blueshifted maser clusters located at the extremes of a line oriented southwest-northeast
($\rm{PA_{CH_{3}OH}}=+32^{\circ}\pm7$\d), and they are separated from each other by about 500~mas ($\sim2600$~au;
see Fig.~\ref{G32_cp}). Although this maser distribution perfectly agrees with that reported by 
\cite{fuj14}, we also detected a few isolated maser features that were not detected by \cite{fuj14}.
These are two redshifted maser features located $\sim850$~mas south
($\sim4400$~au; G32.E47) and $\sim650$~mas west ($\sim3400$~au; G32.E01) with respect to
the reference maser feature in Fig.~\ref{G32_cp} (i.e., G32.E19), and one blueshifted maser
feature located  $\sim730$~mas north ($\sim3800$~au; G32.E02). In addition, we detected 
two groups of blueshifted maser features that were undetected by \cite{fuj14} that are aligned north-south 
with G32.E02.
One group is composed of three maser features (i.e, G32.E03, G32.E07, and G32.14) with
+92~\kms$<V_{\rm{lsr}}<$~+96~\kms ~, and they are located  $\sim$200 north of G32.E19. The other group is
composed of four maser features (i.e., G32.E04, G32.E05, G32.E16, and G32.E22) ranging between
$V_{\rm{lsr}}=$~+94.67~\kms ~and $V_{\rm{lsr}}=$~+96.52~\kms. They follow an arc structure at about
$350$~mas south of G32.E19. None of the features in the two groups shows any velocity gradient. We
note that of all the maser features  that were not detected by \cite{fuj14}, only G32.E04 and G32.E07
would have been detectable at 5$\sigma$ by them. Their nondetection suggests that these two 
maser features might have arisen only recently.\\
\indent Eleven \meth ~maser features show linearly polarized emission, most of which with a linear
polarization fraction in the range $P\rm{_l}=0.4-3.8\%$. The maser features G32.E48
($V_{\rm{lsr}}=$~+99.37~\kms, $I=0.803$\jyb) and G32.E50 ($V_{\rm{lsr}}=$~+98.48~\kms, 
$I=52.457$\jyb) have a high $P\rm{_l}$ of 9.6\% and 10.0\%, respectively. The \code ~was able to 
properly fit all the polarized maser features and, as expected, the \tbo ~of G32.E48 and G32.E50 
are very high (\tbo$~\approx1.6\cdot10^{10}$~K~sr; see Col.~10 of Table~\ref{G32_tab}). For all 
the polarized maser features but G32.E42 ($\theta=67^{\circ+7^{\circ}}_{-44^{\circ}}$), the 
estimated $\theta$ angles are such that $|\theta^{\rm{+}}-55$\d$|>|\theta^{\rm{-}}-55$\d$|$, that is, 
the magnetic field is perpendicular to the linear polarization vectors. We also measured the circular 
polarization fraction for the blueshifted maser feature G32.E11 ($V_{\rm{lsr}}=$~+92.83~\kms, 
$I=53.392$~\jyb, $P_{\rm{V}}=0.2$\%) and for the redshifted maser feature G32.E37 
($V_{\rm{lsr}}=$~+101.21~\kms, $I=18.740$~\jyb, $P_{\rm{V}}=0.8$\%). While for G32.E37 we were able 
to use the calculated \dvi ~and \tbo ~values provided by 
the \code, this was not possible for G32.E11 because this maser feature did not show linearly
polarized emission. To model the V spectra of G32.E11, we therefore considered the values
\dvi$~=1.5$~\kms ~and \tbo$~=2.0\cdot10^8~\rm{K sr}$ ~that best fit the total intensity emission
(see Fig.~\ref{Vfit}). We note that the nondetection of linearly polarized emission toward G32.E11
might be due to a $\theta$ angle equal either to the van Vleck angle of 55\d or to 0\d, rather than
to the brightness of the maser feature.
\begin{figure}[ht!]
\centering
\includegraphics[width = 9 cm]{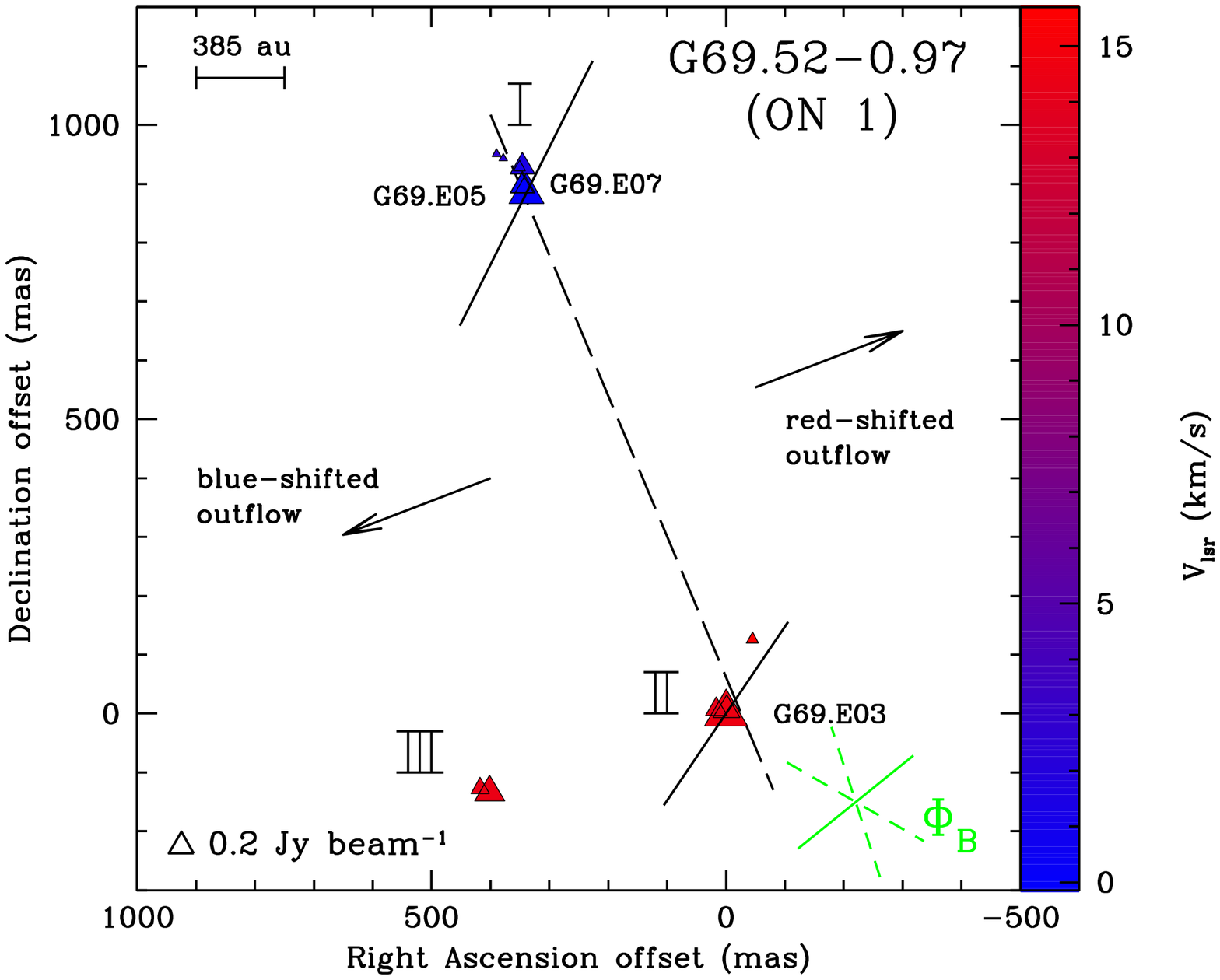}
\caption{View of the \meth ~maser features detected around G69.52-0.97
(Table~\ref{G32_tab}). Same symbols as in Fig.~\ref{G307_cp}. The polarization fraction is in the 
range $P\rm{_l}=1.3-1.5\%$ (Table~\ref{G69_tab}). The assumed velocity of the YSO is
$V_{\rm{lsr}}^{\rm{CS}}=+11.6$~\kms ~\citep{bro96}. The dashed line is the best least-squares 
linear fit of the \meth ~maser features ($\rm{PA_{CH_{3}OH}}=+23^{\circ}\pm7$\d).The two arrows 
indicate the direction and not the actual position of the red- and blueshifted lobes of the 
outflow ($\rm{PA_{outflow}^{\rm{CO}}}=-69$\d, \citealt{kum04}).
}  
\label{G69_cp}
\end{figure}
\subsection{\object{G69.52-0.97}}
\label{G69_sec}
In Figure~\ref{G69_cp} we show the 13 \meth ~maser features detected toward G69.52-0.97 that are 
listed as G69.E01--G69.E13 in Table~\ref{G69_tab}. Following \cite{sug11}, the maser 
features can be divided into three clusters. Cluster~I is blueshifted ($-0.01$~\kms$~\leq 
V_{\rm{lsr}}<$~+3~\kms) and is located in the north part of the UC\hii ~region; clusters~II and 
III are redshifted (+14~\kms$~<V_{\rm{lsr}}<$~+16~\kms) and are located in the south. 
Cluster~III lies about 400~mas ($\sim$1000~au) east of cluster~II. The velocities of clusters~II and 
III agree with those of the redshifted lobe both of the $\rm{H^{13}CO^+}$/SiO outflow and of the 
CO outflow. We can connect clusters~I and II with an imaginary line with 
$\rm{PA_{CH_{3}OH}}=+23^{\circ}\pm7$\d, which is perpendicular with the direction of the CO outflow
($\rm{PA_{CH_{3}OH}-PA_{outflow}^{\rm{CO}}}=92$\d).\\
\indent The maser features G69.E03 (cluster~II; $V_{\rm{lsr}}=$~+14.65~\kms, $I=11.524$~\jyb,
$\Delta v\rm{_{L}}=0.27$~\kms) and G69.E05 (cluster~I; $V_{\rm{lsr}}=$~-0.01~\kms, $I=3.238$~\jyb,
$\Delta v\rm{_{L}}=0.48$~\kms) show linearly polarized emission with $P\rm{_l}=1.3\%$ and $1.5\%$,
respectively.  The \code ~properly fit both maser features and provided for 
both \tbo~$\approx1.6\cdot10^{9}$~K~sr, and $\theta$ angles well below 90\d (see Cols. 11 and 15 
of Table~\ref{G69_tab}). For G69.E03 we have $\theta=69^{\circ+18^{\circ}}_{-34^{\circ}}$ and 
for G69.E05 $\theta=62^{\circ^{+14^{\circ}}}_{-45^{\circ}}$. Therefore, the magnetic field is 
perpendicular to the linear polarization vector for G69.E03 and parallel for G69.E05 (see 
Sect.\ref{obssect}). The circularly polarized emission was measured for maser features G69.E03 
($P_{\rm{V}}=0.6$\%) and G69.E07 (cluster~I; $V_{\rm{lsr}}=$~+0.03~\kms, $I=0.525$~\jyb,
$P_{\rm{V}}=4.2$\%).
Because G69.E07 does not show linearly polarized emission, we assumed 
\dvi$~=1.0$~\kms ~and \tbo$~=6.3\cdot10^9~\rm{K~sr}$ ~(see Fig.~\ref{Vfit}) to model its V
spectra. We note that most of 
the characteristics of maser feature G69.E03 (cluster, $V_{\rm{lsr}}$, $\Delta v\rm{_{L}}$, 
$P\rm{_l}$, $P\rm{_V}$) coincide perfectly with the \meth ~maser feature D detected by 
\cite{gre07}.
\section{Discussion}
\label{discussion}
\subsection{Magnetic field orientations}
\label{Borient}
The linear polarization vectors of the \meth ~masers might suffer a rotation known as Faraday 
rotation if the linearly polarized maser emission passes through a medium with a magnetic field. 
There are two main Faraday rotations that can affect the linear polarization vectors: 
internal rotation ($\Phi_{\rm{i}}$) and foreground Faraday rotation ($\Phi_{\rm{f}}$). In the 
previous papers in the series (Papers I-IV), we have analyzed these rotations and concluded that
they do not affect our study significantly. Therefore, we do not take them into account for the
\meth ~masers in our discussion. However, for completeness, we list estimates of $\Phi_{\rm{f}}$ 
for each source in Col.~2 of Table~\ref{Comp_ang}. The estimated orientation of the magnetic 
field\footnote{$\langle\Phi_{\rm{B}}\rangle$ is 
the mean error-weighted orientation of the magnetic field measured considering all the magnetic 
field vectors measured in a source. The weights are $1/e_{\rm{i}}$, where $e_{\rm{i}}$ is the 
error of the $i$th measured vector. The error on $\langle\Phi_{\rm{B}}\rangle$ is the standard 
deviation. The position angles of the magnetic field vectors are measured with respect to the north
clockwise (negative) and counterclockwise (positive), as the PA of the outflows.} in the five 
massive SFRs under investigation here are discussed separately below.\\

\noindent\textit{\textbf{G30.70-0.07.}} 
The error-weighted orientation of the magnetic field on the plane of the sky is 
$\langle\Phi_{\rm{B}}^{\rm{MM2}}\rangle=-45^{\circ}\pm57$\d, which aligns well with the imaginary 
line that connects the two groups of \meth ~maser features 
($\rm{PA_{CH_{3}OH}}=-67^{\circ}\pm7$\d). However, we are not able to compare it with a molecular 
outflow due to the lack of detections \citep[e.g.,][]{ngu13,cor19}. We can compare the magnetic 
field at milliarcsecond angular resolution ($\sim0.001''$) with that at subarcsecond resolution 
($\sim0.5''$) measured from polarized dust emission by \cite{cor19}. When we consider the most southern
magnetic field vector shown on the tail of source G (see left panel of Fig.2 of 
\citealt{cor19}), the magnetic field at the two angular resolutions shows the same orientation, 
that is, $\sim-45$\d.\\

\noindent\textit{\textbf{G30.76-0.05.}} 
The inferred magnetic field is oriented with an angle on the plane of the sky of
$\langle\Phi_{\rm{B}}^{\rm{MM11}}\rangle=+20^{\circ}\pm64$\d, which makes an angle of
+50\d$\pm80$\d ~with the linear distribution of the 6.7 GHz \meth ~masers 
($\rm{PA}_{\rm{CH_{3}OH}}=-30^{\circ}\pm48$\d, dashed line in Fig.~\ref{G308_cp}). The orientation
of the magnetic field on the plane of the sky seems to agree with the edge of the ionization front
caused by the ultraviolet radiation of the \hii ~region (compare 
$\langle\Phi_{\rm{B}}^{\rm{MM11}}\rangle$ indicated by the thick dashed white ellipse in Fig.~3 
of \citealt{mot03}).\\

\noindent\textit{\textbf{G31.28+0.06.}} 
Taking into account that for G312.E61 the magnetic field is parallel to the linear polarization 
vector, we have $\langle\Phi_{\rm{B}}\rangle=-38^{\circ}\pm33$\d. Therefore, the inferred magnetic
field is well aligned with the redshifted lobe of the $^{13}$CO outflow \citep{yan18} and almost
perpendicular to the linear fit of the maser features (see Fig.~\ref{G312_cp} and
Table~\ref{Comp_ang}). In addition, the magnetic field inferred from the 6.7~GHz \meth ~masers
is also almost aligned with the magnetic field inferred from the 1.665~GHz OH maser, assuming that
this is perpendicular to the linear polarization vector of the OH maser (i.e.,
$\Phi_{\rm{B}}^{\rm{OH}}=-55$\d; \citealt{szy09}).\\
\begin{table*}[th!]
\caption []{Comparison between position angle of magnetic field, \meth ~maser distribution, outflows, and linear polarization angles.} 
\begin{center}
\scriptsize
\begin{tabular}{ l c c c c c c c c c c}
\hline
\hline
\,\,\,\,\,(1) &(2)           & (3)                  & (4)                           & (5)                          & (6)                       & (7)        & (8)                                                   & (9)                                           &(10)                                               &(11)\\
Source & $\Phi_{\rm{f}}$\tablefootmark{a}& $\langle\chi\rangle$\tablefootmark{b} & $\langle\Phi_{\rm{B}}\rangle$\tablefootmark{b} & $\rm{PA}_{\rm{outflow}}$     & $\rm{PA}_{\rm{CH_{3}OH}}$ & $\rho$\tablefootmark{c}& $|\rm{PA}_{\rm{outflow}}-\langle\Phi_{\rm{B}}\rangle|$& $|\rm{PA}_{\rm{CH_{3}OH}}-\langle\chi\rangle|$& $|\rm{PA}_{\rm{CH_{3}OH}}-\rm{PA}_{\rm{outflow}}|$&Ref. \\ 
       & (\d)                &  (\d)                & (\d)                          & (\d)                         & (\d)                      &            & (\d)                                                  &(\d)                                           & (\d)                                              & \\ 
\hline
G30.70-0.07 (W43-MM2) & 12   & $+61\pm17$ & $-45\pm57\tablefootmark{d}$ &  $-$  & $-63\pm7$ & -0.65  & $-$ & $56\pm18\tablefootmark{e}$ & $-$ & - \\
G30.76-0.05 (W43-MM11) & 12  & $+20\pm64$ & $+20\pm64\tablefootmark{d}$ & $-$ & $-30\pm48$ & -0.41 & $-$ & $50\pm80$ & $-$ & - \\
G31.28+0.06      & 10   & $+53\pm35$ & $-38\pm33\tablefootmark{d}$ & $-26\pm15\tablefootmark{f,g}$ & $+73\pm4$ & +0.39  & $12\pm36$ & $20\pm35$ & $81\pm16$\tablefootmark{e} & (1)\\
G32.03+0.06      & 12  & $-86\pm79$ & $-64\pm34\tablefootmark{d}$ & $-$ & $+32\pm7$ & +0.91  & $-$ & $62\pm79\tablefootmark{e}$ & $-$ & - \\
G69.52-0.97      & 6  & $-29\pm5$ & $-51\pm69\tablefootmark{d}$ & $-69\pm15\tablefootmark{h,g}$ & $+23\pm7\tablefootmark{i}$ & +0.99  & $18\pm71$ & $52\pm9$ & $88\pm17\tablefootmark{e}$ & (2)\\
\hline
\multicolumn{11}{c}{From Paper~IV\tablefootmark{j}}\\
\hline
G23.44-0.18 (MM2) & 13& $+23\pm40$& $-67\pm40$                    & $-40\pm15$            & $+30\pm26$                 &  +0.69     & $27\pm43$                                         & $13\pm48$                              & $70\pm30$                         & (3)\\
                 & &           &                               &                                        & $-24\pm91$                 &  -0.37     &                                               & $47\pm99$                              & $16\pm92$                         & \\
G25.83-0.18       & 11& $+23\pm7$& $-67\pm7$                    & $+10\pm15$            & $+51\pm7$                 &  +0.69     & $77\pm17$                                         & $28\pm10$                              & $41\pm17$                         & (3)\\
G25.71-0.04       & 23& $-51\pm77$& $-80\pm43$ & $-90\pm15$            & $-71\pm9$                 &  -0.41     & $10\pm46$                                         & $20\pm78$                              & $18\pm18$                         & (3)\\
G28.31-0.39       & 24& $+40\pm44$& $-50\pm44$     & $-52\pm15$            & $+85\pm22$                 &  +0.07     & $2\pm47$                                         & $45\pm49$                              & $84\pm27$                         & (3)\\
G28.83-0.25       & 10& $-45\pm34$& $+58\pm59$ & $-40\pm15$            & $-41\pm10$                 &  -0.83     & $82\pm61$                                         & $4\pm35$                           & $1\pm18$                         & (3)\\
G29.96-0.02       & 12& $+62\pm17$& $-29\pm17$ & $-38\pm15$            & $+80\pm3$                 &  0.51     & $9\pm23$                                         & $18\pm17$                           & $62\pm16$                         & (3)\\
G43.80-0.13       & 14& $-81\pm5$& $+9\pm5$ & $+38\pm15$            & $-48\pm5$                 &     -0.94    & $29\pm16$  & $33\pm7$                           & $86\pm16$                         & (3)\\
IRAS\,20126+4104 & $4$       & $-70\pm16$           & $+20\pm16$                    & $-65\pm5$   & $+87\pm4$                 &  $+0.12$   & $85\pm17$                                             & $23\pm17$                    & $28\pm6$                         & (3)\\
G24.78+0.08-A2 &   $17$      & $-53\pm2$            & $+37\pm2$    & $-40\pm15$  & $-26\pm19$                &  $-0.77$   & $77\pm15$                                             & $79\pm19$                                     & $66\pm24$                                         & (3)\\
G25.65+1.05    &   $7$       & $-80\pm8$            & $-23\pm51$   & $-15\pm15$  & $-49\pm7$ &$-0.87$     & $8\pm53$                                              & $31\pm11$                                     & $64\pm17$                                         & (3)\\
G29.86-0.04    &   $17$      & $+46\pm41$           & $+82\pm56$   & $+6\pm15$ & $+8\pm7$ &$+0.73$     & $76\pm58$                                             & $38\pm42$                                     & $14\pm17$                                         & (3)\\
G35.03+0.35    &   $8$       & $-64\pm5$            & $+26\pm5$    & $+27\pm15$&$-26\pm19$                 &$-0.77$     & $1\pm16$                                              & $38\pm20$                                     & $53\pm24$                                         & (3)\\
G37.43+1.51    &   $4$       & $+90\pm3$            & $+90\pm3$   & $-4\pm15$   &$-64\pm5$ &$-0.87$     & $86\pm15$                           & $26\pm6$                  & $60\pm16$                                         & (3)\\
G174.20-0.08   &   $4$       & $-$                  & $-$                           & $-40\pm15$   &$-63\pm16$                 &$-0.45$     & $-$                                                   & $-$                                           & $23\pm22$                                         & (3)\\
G213.70-12.6-IRS3&$2$        & $+20\pm5$            & $-70\pm5$    & $+53\pm15$  &$+63\pm2$                  &$+0.95$     & $57\pm16$                            & $43\pm5$                                      & $10\pm15$                                         & (3)\\
Cepheus~A      &   $2$       & $-57\pm28$           & $+30\pm19$                    & $+40\pm4$                    & $-79\pm9$                 &  $-0.34$   & $10\pm19$                                             & $22\pm29$                                     & $61\pm10$                                         & (3)\\
W75N-group~A & $3$           & $-13\pm9$            & $+77\pm9$                     & $+66\pm15$                   & $+43\pm10$                &  $+0.96$   & $11\pm18$                                             & $56\pm14$                                     & $23\pm18$                                         & (3)\\
NGC7538-IRS1 &  $6$          &$-30\pm69$            & $+67\pm70$                    & $-40\pm10$                   & $+84\pm7$                 &  $+0.15$   & $73\pm71$                                               & $66\pm69$                                     & $56\pm12$                                         & (3)\\
W3(OH)-group II &  $4$       &$+21\pm45$            & $-47\pm44$                    & $-$                          & $-59\pm6$                 &  $-0.84$   & $-$                                                   & $80\pm45$                                     & $-$                                               & (3)\\
W51-e2    &  $12$            &$+33\pm16$            & $-60\pm21$                    & $-50\pm20$                   & $+57\pm8$                 &  $+0.70$   & $10\pm29$                                             & $24\pm18$                                     & $73\pm22$                                         & (3)\\
IRAS18556+0138 &  $5$        &$-2\pm11$             & $+88\pm11$                    & $+58\pm23$                   & $-40\pm2$                 &  $-0.99$   & $30\pm26$                                             & $42\pm11$                                     & $82\pm23$                                         & (3)\\
W48       & $7$              & $+23\pm7$            & $-67\pm7$                     & $-$                          & $+55\pm10$                &  $+0.70$   & $-$                                                   & $78\pm12$                                     & $-$                                               & (3) \\
IRAS06058+2138-NIRS1 &  $4$  &$+49\pm47$            & $-49\pm52$                    & $-50\pm15$                   & $+78\pm7$                 &  $+0.64$   & $1\pm54$                                              & $29\pm48$                                     & $52\pm17$                                         & (3)\\
IRAS22272+6358A &  $2$       &$-80\pm15$            & $+9\pm15$                     & $-40\pm15$                   & $-35\pm11$                &  $-0.87$   & $49\pm21$                                             & $45\pm19$                                     & $5\pm19$                                          & (3) \\
S255-IR   &  $4$             &$+36\pm12$            & $-54\pm12$                    & $+75\pm15$                   & $-63\pm49$                &  $-0.11$   &$51\pm19$                                              & $81\pm51$                                     & $42\pm51$                                         & (3) \\
S231      &  $4$             &$+28\pm49$            & $-62\pm49$                    & $-47\pm5$                    & $+28\pm8$                 &  $+0.97$   & $15\pm49$                                             & $0\pm50$                                      & $75\pm9$                                          & (3)\\
G291.27-0.70 &  $7$          &$-32\pm5$             & $+52\pm5$                     & $-$                          & $-77\pm14$                 &  $-$       & $-$                                                   & $45\pm15$                                     & $-$                                               & (3)\\
G305.21+0.21 &  $9$          &$-51\pm14$            & $28\pm14$                     & $-$                          & $+48\pm23$                &  $-$       &$-$                                                    & $81\pm27$                                     & $-$                                               & (3)\\
G309.92+0.47 &  $12$         &$+2\pm56$             & $-75\pm56$                    & $-$                          & $+35\pm5$                 &  $-$       &$-$                                                    & $33\pm56$                                     & $-$                                               & (3)\\
G316.64-0.08 &  $3$          &$-67\pm36$            & $+21\pm36$                    & $-$                          & $+34\pm29$                &  $-$       & $-$                                                   & $79\pm46$                                     & $-$                                               & (3)\\
G335.79+0.17 &  $8$          &$+44\pm28$            & $-41\pm28$                    & $-$                          & $-69\pm25$                &  $-$       & $-$                                                   & $67\pm38$                                     & $-$                                               & (3) \\
G339.88-1.26 &  $7$          &$+77\pm24$            & $-12\pm24$                    & $-$                          & $-60\pm17$                &  $-$       & $-$                                                   & $43\pm29$                                     & $-$                                               & (3) \\
G345.01+1.79 &  $5$          &$+5\pm39$             & $-86\pm39$                    & $-$                          & $+74\pm4$                 &  $-$       &$-$                                                    & $69\pm39$                                     & $-$                                               & (3) \\
NGC6334F (central) &  $5$    &$+77\pm20$            & $-13\pm20$                    & $+30\pm15$  & $-41\pm16$                &  $-$       &$43\pm25$                                              & $62\pm26$                                            & $71\pm41$                                         & (3)\\
NGC6334F (NW)&  $5$          &$-71\pm20$            & $+19\pm20$                    & $+30\pm15$  & $-80\pm38$                &  $-$       &$11\pm25$                                              & $9\pm43$                                      & $70\pm41$                       & (3)\\
\hline
\hline
\end{tabular}
\end{center}
\tablefoot{
\tablefoottext{a}{Foreground Faraday rotation. It was estimated by using Eq.~3 of Paper~I.
\tablefoottext{b}{Because of the high uncertainties of the estimated $\Phi_{\rm{f}}$, the angles are not corrected for $\Phi_{\rm{f}}$.}
\tablefoottext{c}{Pearson product-moment correlation coefficient $-1\leq\rho\leq+1$; $\rho=+1$ ( $\rho=-1$) is total positive (negative) correlation, $\rho=0$ is no correlation.}}
\tablefoottext{d}{Before averaging, we use the criterion described in Sect.~\ref{obssect} to estimate the orientation of the magnetic field w.r.t the 
linear polarization vectors.}
\tablefoottext{e}{The differences between the angles are evaluated taking into account that $\rm{PA}\equiv\rm{PA}\pm180$\d, $\langle\chi\rangle\equiv\langle\chi\rangle\pm180$\d, and $\langle\Phi_{\rm{B}}\rangle\equiv\langle\Phi_{\rm{B}}\rangle\pm180$\d.}
\tablefoottext{f}{Redshifted lobe of the outflow.}
\tablefoottext{g}{We consider an arbitrary conservative error of 15\d.}
\tablefoottext{h}{We consider the CO outflow.}
\tablefoottext{i}{We consider clusters I and II.}
\tablefoottext{j}{Here we omit all the notes that are already indicated in Table~4 of Paper~IV.}
}
\tablebib{
(1) \citet{yan18}; (2) \citet{kum04}; (3) Paper~IV and references therein.}
\label{Comp_ang}
\end{table*}

\noindent\textit{\textbf{G32.03+0.06.}} 
When the different orientation of the magnetic field with respect to the linear 
polarization vector of G32.E42 is taken into account, the inferred orientation of the magnetic field on the plane of the
sky is $\langle\Phi_{\rm{B}}\rangle=-64^{\circ}\pm34$\d. Therefore, the magnetic field is 
perpendicular to the line that connects the two main maser clusters, which has an inclination 
angle on the plane of the sky of $\rm{PA}_{\rm{CH_{3}OH}}=+32^{\circ}\pm7$\d.\\

\noindent\textit{\textbf{G69.52-0.97.}}
According to the most likely orientation of the magnetic field with respect to the linear
polarization vector of G69.E05 (parallel; see Sect.~\ref{G69_sec}), the inferred magnetic field
on the plane of the sky is $\langle\Phi_{\rm{B}}\rangle=-51^{\circ}\pm69$\d. In this case, the
magnetic field is aligned with the CO outflow and
almost perpendicular to the connecting imaginary line between clusters~I and II. If we compare the
orientation of the magnetic field with that of the $\rm{H^{13}CO^+}$/SiO outflow, we note that 
$|\langle\Phi_{\rm{B}}\rangle - \rm{PA_{outflow}^{H^{13}CO^+}}|=85^{\circ}$\footnote{The
differences between the angles are evaluated taking into account that 
$\rm{PA}\equiv\rm{PA}\pm180$\d ~and
$\langle\Phi_{\rm{B}}\rangle\equiv\langle\Phi_{\rm{B}}\rangle\pm180$\d.}. Therefore, the 6.7~GHz 
\meth ~maser features might be associated with the CO outflow rather than with the
$\rm{H^{13}CO^+}$/SiO outflow. A comparison with
the orientation of the magnetic field estimated from the linearly polarized emission of OH masers 
( $\Phi_{\rm{B}}^{\rm{1.6~GHz~OH}}=-31$\d ~and $\Phi_{\rm{B}}^{\rm{6.0~GHz~OH}}=+30$\d; 
\citealt{fis05,fis10}) suggests that $\langle\Phi_{\rm{B}}\rangle$ agrees with 
$\Phi_{\rm{B}}^{\rm{1.6~GHz~OH}}$. However, the foreground Faraday rotation for the 1.6~GHz 
OH maser is $\Phi_{\rm{f}}^{\rm{1.6~GHz}}=90$\d, much higher than that for the 6.0~GHz OH maser 
($\Phi_{\rm{f}}^{\rm{6.0~GHz}}=7$\d) and for the 6.7~GHz \meth ~maser 
($\Phi_{\rm{f}}^{\rm{6.7~GHz}}=6$\d).
To calculate $\Phi_{\rm{f}}$ , we used Eq.~6 of \cite{sur111} and the parameters reported there.
Therefore, the magnetic field orientations inferred from the two OH maser emissions agree well with 
each other, but not with that inferred from the \meth ~maser. 
It is important to note that the two OH maser transitions may trace the gas of two distinct zones
of the source, as already observed in other massive YSO \citep[e.g.,][]{eto12}, and therefore
the difference in the magnetic field orientation might be due to this rather than to the
different $\Phi_{\rm{f}}$.\\
\indent Although this is less probable, the orientation of the magnetic field estimated from the
linearly polarized emission of G69.E05 might be perpendicular to the linear polarization vector. In
this case, the inferred magnetic field of the region is 
$\langle\Phi_{\rm{B}}^{\perp}\rangle=+61^{\circ}\pm4$\d, which implies a good alignment of the 
magnetic field with the $\rm{H^{13}CO^+}$/SiO outflow rather than with the CO 
outflow. In this case, we have $|\langle\Phi_{\rm{B}}^{\perp}\rangle - \rm{PA_{outflow}^{H^{13}CO^+}}|=17^{\circ}$
and $|\langle\Phi_{\rm{B}}^{\perp}\rangle - \rm{PA_{outflow}^{CO}}|=50^{\circ5}$.
In addition, $\langle\Phi_{\rm{B}}^{\perp}\rangle$ is more consistent with the magnetic field inferred
from the OH maser emissions when we assume that they probe the magnetic field from the same zone of the
source and that the $\Phi_{\rm{f}}$ is taken into account for these emissions. \\
\indent Unfortunately, it is not possible to compare our results
with the previous linear polarization measurements of \meth ~maser made by \cite{gre07} because they
do not provide any estimate of the magnetic field orientation due to the unknown $\theta$ angles.
For the remaining discussion, we therefore consider the angle 
$\langle\Phi_{\rm{B}}\rangle=-51^{\circ}\pm69$\d ~as the inferred orientation of the magnetic field 
of the region.
\subsection{Magnetic field strength}
\label{Bstrength}
As of Paper~IV, we are able to estimate a lower limit of $B_{||}$ from the measurements of the 
Zeeman splitting of the 6.7~GHz \meth ~maser. This is possible thanks to the work of \cite{lan18}.
In the present work, we assume $\alpha_{\rm{Z}}=-0.051$~\kmsg ~\citep{lan18} as explained in 
Sect.~\ref{obssect}. In addition, by knowing the inclination of the magnetic field with respect to
the line of sight (i.e., the $\theta$ angle), we also estimated a lower limit for the 3D magnetic 
field strength ($B=\frac{B_{||}} {cos~(\theta+\varepsilon^{-}_{\rm{\theta}})}$; considering 
$\theta^{\varepsilon^{+}_{\rm{\theta}}}_{\varepsilon^{-}_{\rm{\theta}}}$). We were able to measure
the magnetic field strength toward G30.70-0.07, G32.03+0.06, and G69.52-0.97, for which Zeeman 
splitting was detected.\\

\noindent\textit{\textbf{G30.70-0.07.}} 
Using the \code ~we modeled the circularly polarized emission of G307.E18 (see Fig.~\ref{Vfit}).
We measured a Zeeman splitting of \dvz$~=+1.5\pm0.2$~\ms ~, for which we estimated a magnetic field 
strength along the line of sight of $B_{||}>30$~mG. For G307.E18, we find that
$\theta={45^{\circ}}^{+30^{\circ}}_{-28^{\circ}}$ , which implies a 3D magnetic field $B>31$~mG and 
a magnetic field on the plane of the sky of $B_{\rm{pos}}>9$~mG. \cite{cru99} empirically
determined a relation between the magnetic field strength and the density; this is $|B|\propto
n_{\rm{H_2}}^{0.47}$, where $B$ is expressed in mG and $n_{\rm{H_2}}$ in $\rm{cm^{-3}}$. The
number density of 6.7~GHz \meth ~maser is in the range
$10^7~\rm{cm^{-3}}<n_{\rm{H_2}}^{\rm{CH_3OH}}<10^9~\rm{cm^{-3}}$ \citep{cra05}. If we consider 
that the maser emission of G307.E18 is predominantly due to the hyperfine transition 
$F=3\rightarrow4$ and we assume for the dust around W43-MM2 a density of
$n_{\rm{H_2}}=10^6~\rm{cm^{-3}}$ \citep{cor19}, then we can predict for source G
$0.3~\rm{mG}<$$B_{\rm{pos}}^{\rm{dust,G}}$$<3~\rm{mG}$. As explained in \cite{cru19}, the relation
in \cite{cru99} has been modified by a more complete and statistically improved analysis by
\cite{cru10}. The new relation is $|B|\propto n_{\rm{H_2}}^{0.65}$, which in our case provides 
a range for source G $0.1~\rm{mG}<$$B_{\rm{pos}}^{\rm{dust,G}}$$<2~\rm{mG}$. \cite{cor19} did not 
provide a direct measurement of $B_{\rm{pos}}$ for source G, but they estimated a magnetic field
$B_{\rm{pos}}^{\rm{dust,A-D,F,J}}>0.5~\rm{mG}$ for the other sources identified in W43-MM2 . 
Speculatively, we might consider the magnetic field strength consistent at the two different 
scales ($0''\!\!.001$ and $0''\!\!.5$) if the preferred hyperfine transition responsible for the
6.7~GHz \meth ~maser emission were $F=3\rightarrow4$.\\

\noindent\textit{\textbf{G32.03+0.06.}} 
We measured Zeeman splitting of \dvz~$=+0.6\pm0.1$~\ms ~and \dvz~$=+1.4\pm0.2$~\ms ~by modeling 
with the \code ~the circularly polarized emission of G32.E11 (blueshifted) and G32.E37 (redshifted),
respectively (see Fig.~\ref{Vfit}). From these measurements, we estimated $B_{||}>12$~mG and 
$B_{||}>27$~mG for the blue- and redshifted clusters, respectively. When we consider that for G32.E37
$\langle\theta\rangle={90^{\circ}}^{+53^{\circ}}_{-53^{\circ}}$ ~, the 3D magnetic field is $B>34$~mG. 
We cannot estimate $B$ for G32.E11 because we do not have a direct measurement of the $\theta$ angle 
for this maser feature (see Sect.~\ref{res}).\\

\noindent\textit{\textbf{G69.52-0.97.}} 
We were able to measure the Zeeman splitting toward two \meth ~maser features. For the brightest 
and redshifted maser feature G69.E03 (cluster~II), we have \dvz~$=+1.2\pm0.2$~\ms. For the
blueshifted maser feature G69.E07 (cluster~I), we measured \dvz~$=+14.5\pm4.4$~\ms ~(see
Fig.~\ref{Vfit}). The measured \dvz ~of G69.E03 agrees perfectly with that measured from the 
6.7~GHz \meth ~maser feature D by \cite{gre07}, that is, \dvz$^{D}=0.9$~\ms$\pm0.3$~\ms, while the 
\dvz ~of G69.E07 is the highest ever measured by us toward a 6.7~GHz \meth ~maser feature. The
estimated magnetic field along the line of sight is $B_{||}>24$~mG for G69.E03 and  
$B_{||}>285$~mG for G69.E07, which is extraordinary strong. It is difficult to reconcile the absence
of circularly polarized emission from the strongest maser feature in cluster~I (G69.E05), which is
only few milliarcseconds away from G69.E07 and shows a peak flux density six times higher than 
G69.E07. Nevertheless, this might suggest that either these two maser features are physically 
well separated 
although their vicinity or the circular polarization of G69.E07 is due to a
non-Zeeman effect such as the rotation of the magnetic field along the maser path that converts linear 
polarization into circular polarization \citep{wie98}. This might also explain why we did not detect
linearly polarized emission from G69.E07. As a consequence, in our discussion, we consider only
the magnetic field estimated from G69.E03, for which the 3D magnetic field is $B>29$~mG 
($\langle\theta\rangle_{\rm{G69.E03}}={69^{\circ}}^{+18^{\circ}}_{-34^{\circ}}$). We can consider 
the relation $|B|\propto n_{\rm{H_2}}^{0.65}$ \citep{cru10} to determine whether the estimate of 
the magnetic field from the Zeeman splitting of \meth ~maser emission is consistent with
those made from the Zeeman splitting of the OH maser emissions. When we take into account the estimated 
$B_{||}$ from \meth ~maser emission, assuming that the hyperfine transition $F=3\rightarrow4$ is 
predominant (i.e., $B_{||}^{\rm{CH_3OH}}=24$~mG), the measured $B_{||}$ from the OH maser 
emissions ($B_{||}^{\rm{OH}}=5$~mG), and the range of the number density for the 6.7~GHz \meth 
~maser ($10^7~\rm{cm^{-3}}<n_{\rm{H_2}}^{\rm{CH_3OH}}<10^9~\rm{cm^{-3}}$,
\citealt{cra05}), we can estimate the number density of OH maser in the region and
compare it with the theory. We estimate that 
$10^6~\rm{cm^{-3}}<n_{\rm{H_2}}^{\rm{OH}}<10^8~\rm{cm^{-3}}$. This is consistent with the
theoretical range of $n_{\rm{H_2}}^{\rm{OH}}$ for both 1.6 and 6.0~GHz OH maser emissions
\citep{cra02}. The consistency also holds for a magnetic field $B_{||}^{\rm{CH_3OH}}<130$~mG.
\begin {table}[th!]
\caption []{Results of the Kolmogorov-Smirnov test.} 
\begin{center}
\scriptsize
\begin{tabular}{ l c c c c }
\hline
\hline
\,\,\,\,\,\,\,\,\,\,\,\,\,\,\,(1)                      &(2)   & (3)  & (4)       & (5)          \\ 
\,\,\,\,\,\,\,\,\,\,Angle                              & $N$\tablefootmark{a}  & $D$\tablefootmark{b}  & $\lambda$\tablefootmark{c} & $Q_{\rm{K-S}}(\lambda)$\tablefootmark{d}\\
\hline
\\
$|\rm{PA}_{\rm{outflow}}-\langle\Phi_{\rm{B}}\rangle|$ & 27   & 0.28 & 1.50      & 0.02 \\
$|\rm{PA}_{\rm{CH_{3}OH}}-\rm{PA_{\rm{outflow}}}|$     & 29   & 0.20 & 1.01      & 0.18 \\
$|\rm{PA}_{\rm{CH_{3}OH}}-\langle\chi\rangle|$         & 40   & 0.10 & 0.66      & 0.77 \\
\\
\hline
\hline
\end{tabular}
\end{center}
\tablefoot{
\tablefoottext{a}{$N$ is the number of elements considered in the K-S test. }
\tablefoottext{b}{$D$ is the maximum value of the absolute difference between the data set, composed of $N$ elements, and the random
distribution.}
\tablefoottext{c}{$\lambda$ is a parameter given by $\lambda=(\sqrt{N}+0.12+0.11/\sqrt{N})\times D$.}
\tablefoottext{d}{$Q_{\rm{K-S}}(\lambda)=2\sum_{j=1}^{N} (-1)^{j-1}~ e^{-2j^2\lambda^2}$ is the significance level of the K-S test.}
}
\label{KS}
\end{table}
\begin{figure}[t!]
\centering
\includegraphics[width = 9 cm]{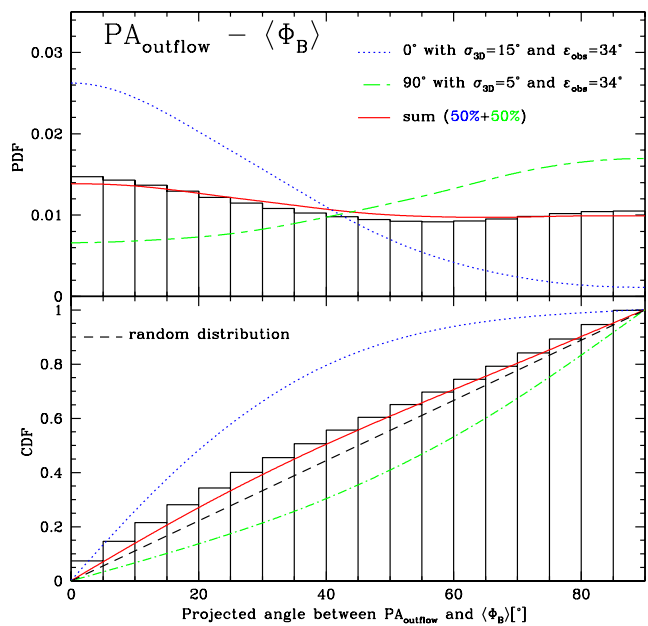}
\caption{Probability distribution function (top panel) and the cumulative
distribution function (bottom panel) of the projected angle between the magnetic field and the outflow
axes ($|\rm{PA}_{\rm{outflow}}-\langle\Phi_{\rm{B}}\rangle|$). The dotted blue line is the result of a
Monte Carlo simulation of the projection on the plane of the sky of two random 3D parallel vectors with a 
Gaussian uncertainty of 15\d ~and with a projected Gaussian error of 34\d ~(called 0 deg distribution). 
The dot-dashed green line is the result of a Monte Carlo simulation of the projection on the plane of the
sky of two random 3D perpendicular vectors with a Gaussian uncertainty of 5\d ~and with a projected error
of 34\d ~(called 90 deg distribution). The red line is the best combination of the previous two simulations
to fit the observed data. Here, both the 0 deg distribution and the 90 deg distribution contribute to the
50\%. The dashed black line is the CDF for a completely random orientation of outflows and magnetic fields,
i.e., all angular differences are equally likely. The results of the K-S test are listed in
Table~\ref{KS}.
}  
\label{CDF_B}
\end{figure}
\begin{figure*}[th!]
\centering
\includegraphics[width = 8 cm]{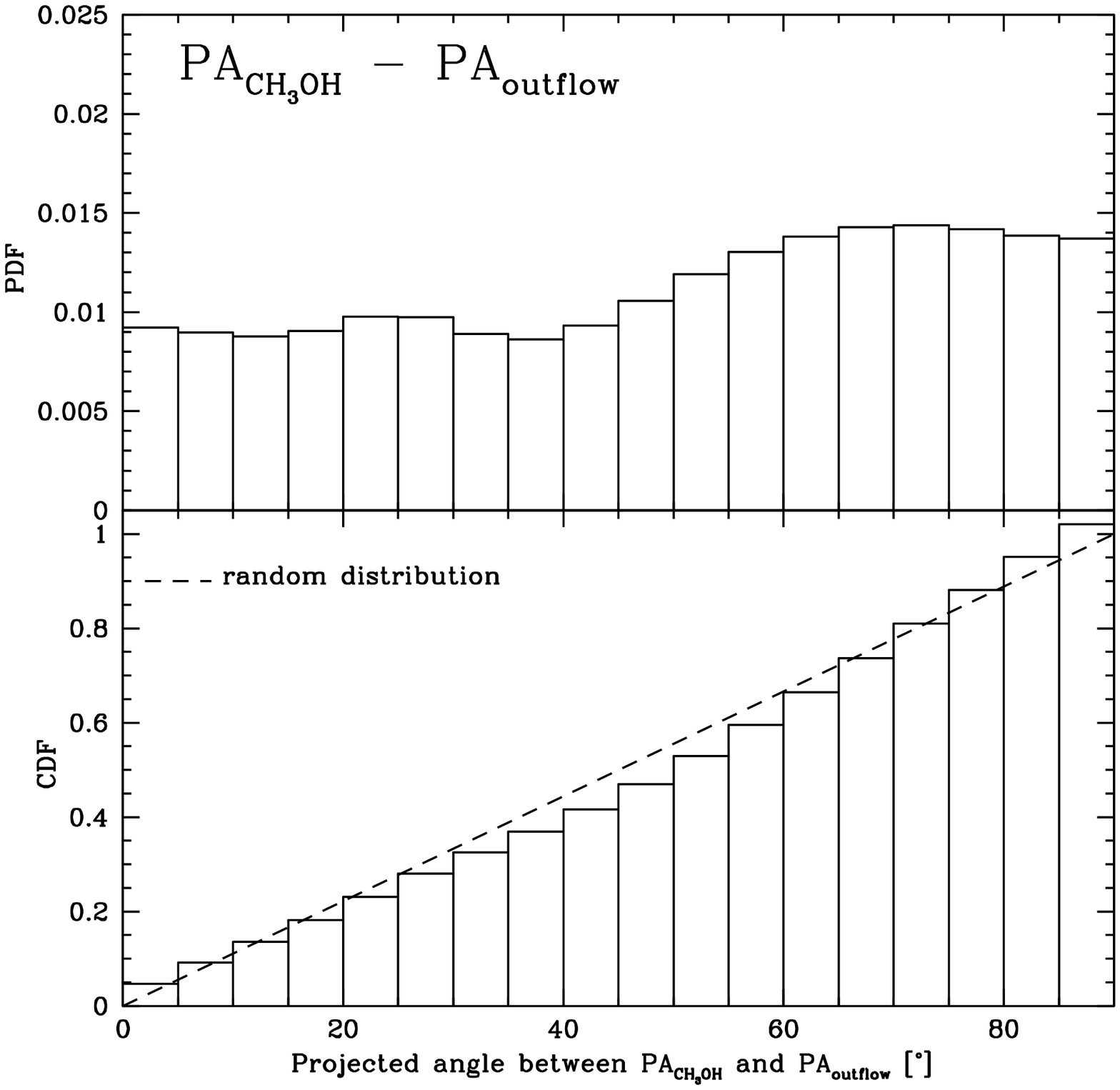}
\includegraphics[width = 8 cm]{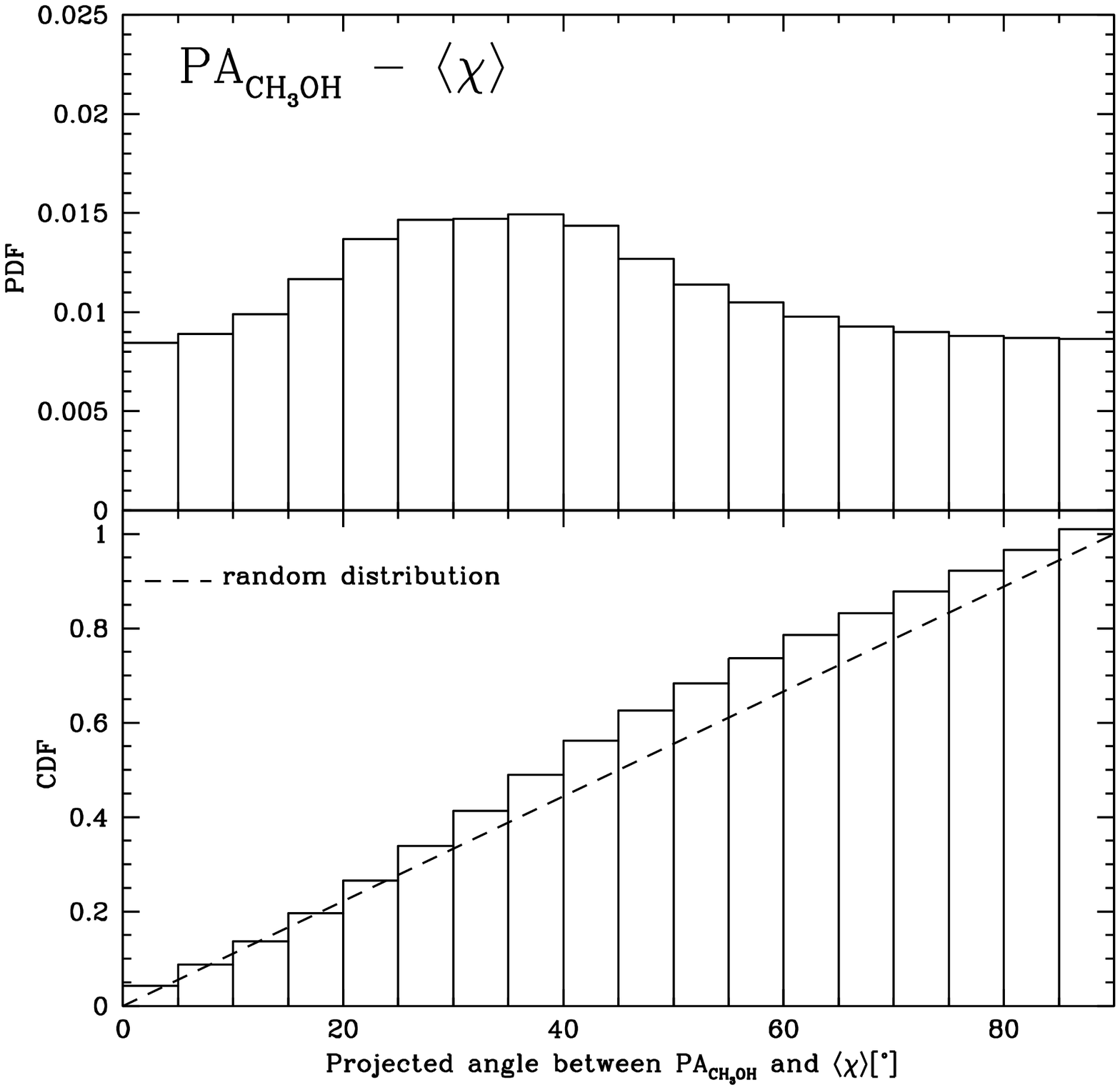}
\caption{Probability density function and CDF of the projected angle between
the PA of the \meth ~maser distribution and the outflow axis 
($|\rm{PA}_{\rm{CH_{3}OH}}-\rm{PA_{\rm{outflow}}}|$) (\textit{left}). \textit{Right}: Probability 
distribution function (top panel) and CDF (bottom panel) of the
projected angle between the PA of the \meth ~maser distribution and the linear polarization angles 
($|\rm{PA}_{\rm{CH_{3}OH}}-\langle\chi\rangle|$). In both panels the dashed line is the 
CDF for random distribution of the corresponding angles, i.e., all angular differences are equally 
likely. The results of the K-S test are listed in Table~\ref{KS}.}
\label{cdf}
\end{figure*}
\begin{figure*}[th!]
\centering
\includegraphics[width = 6 cm]{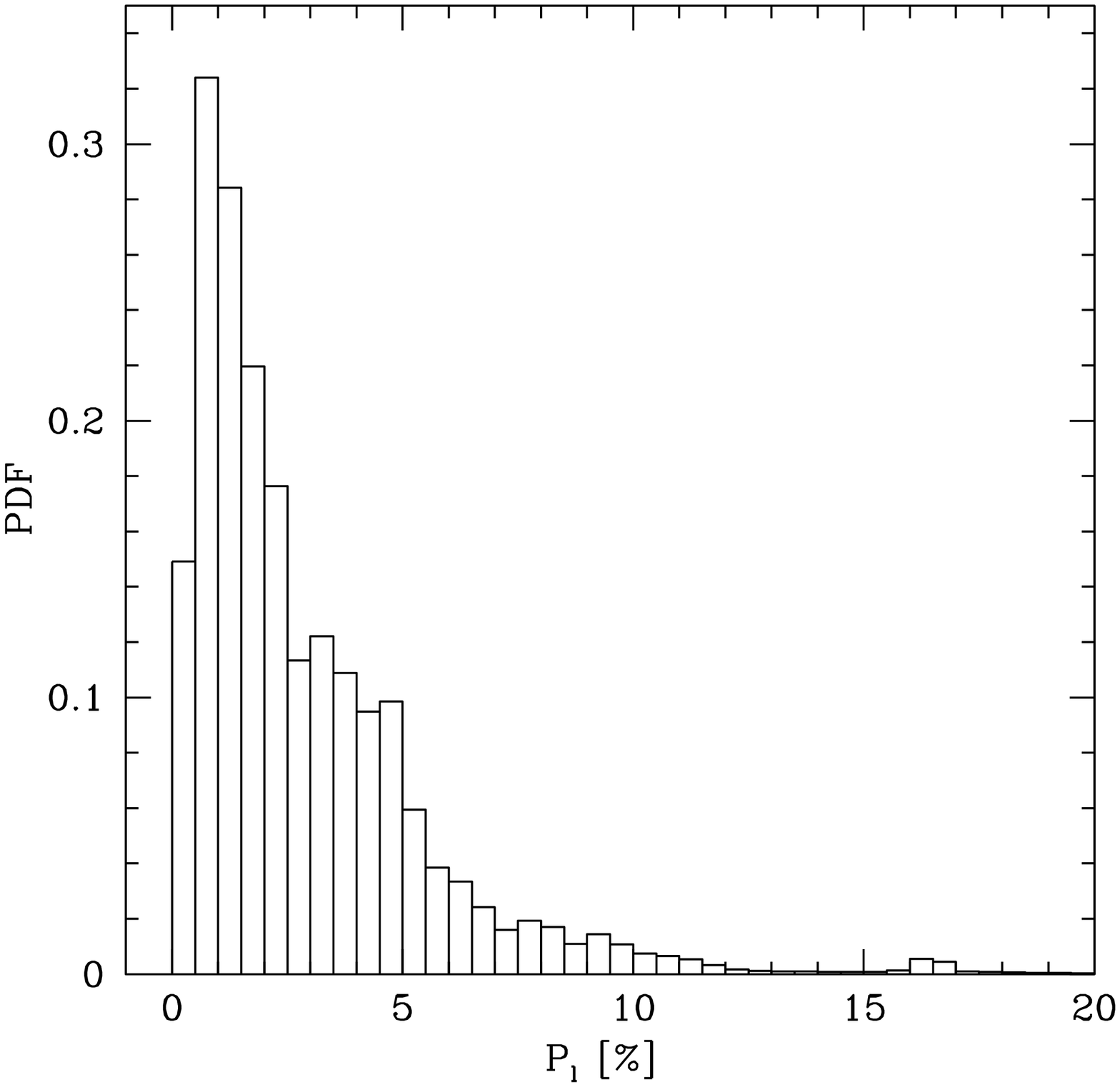}
\includegraphics[width = 6 cm]{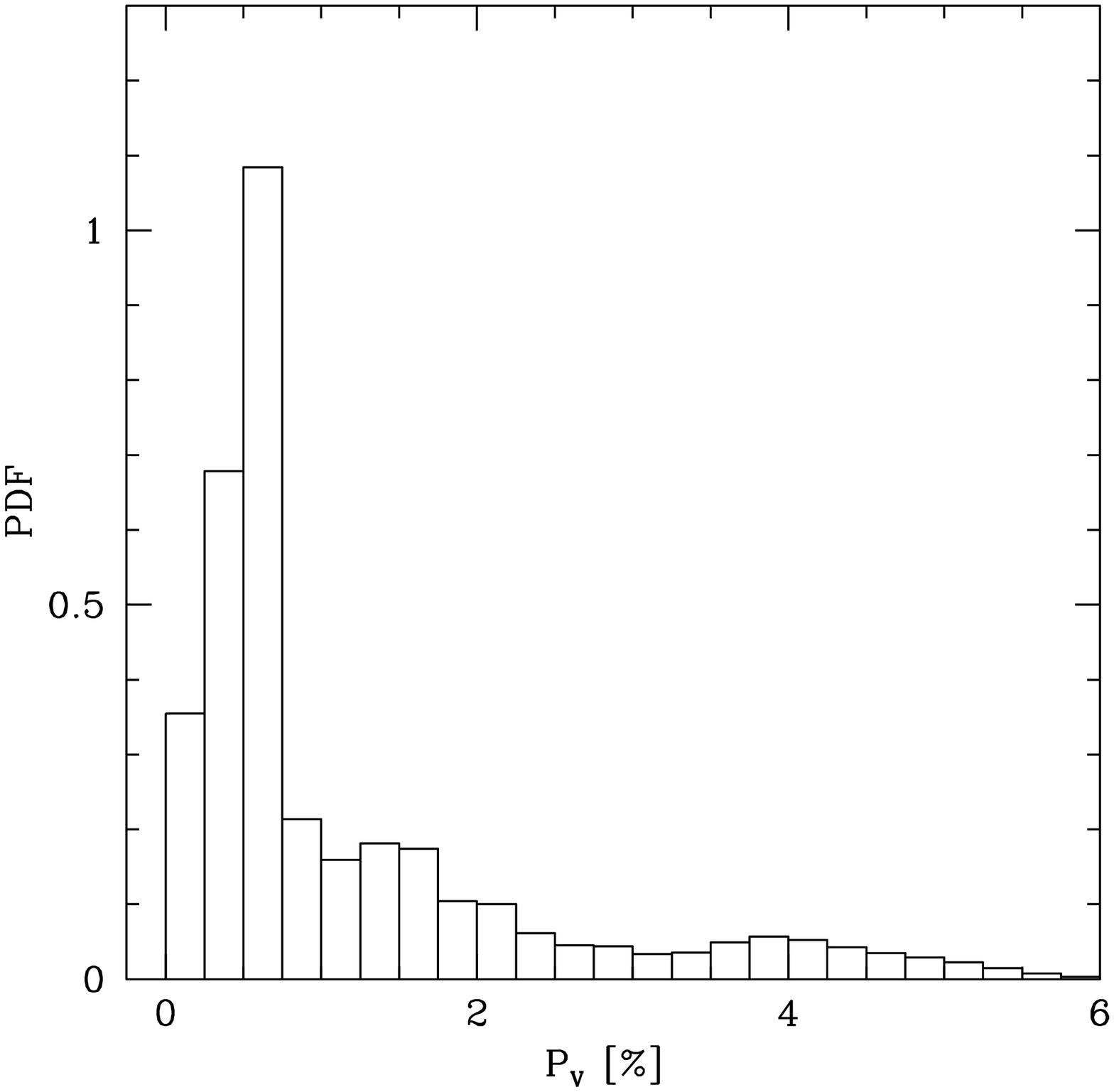}
\includegraphics[width = 6 cm]{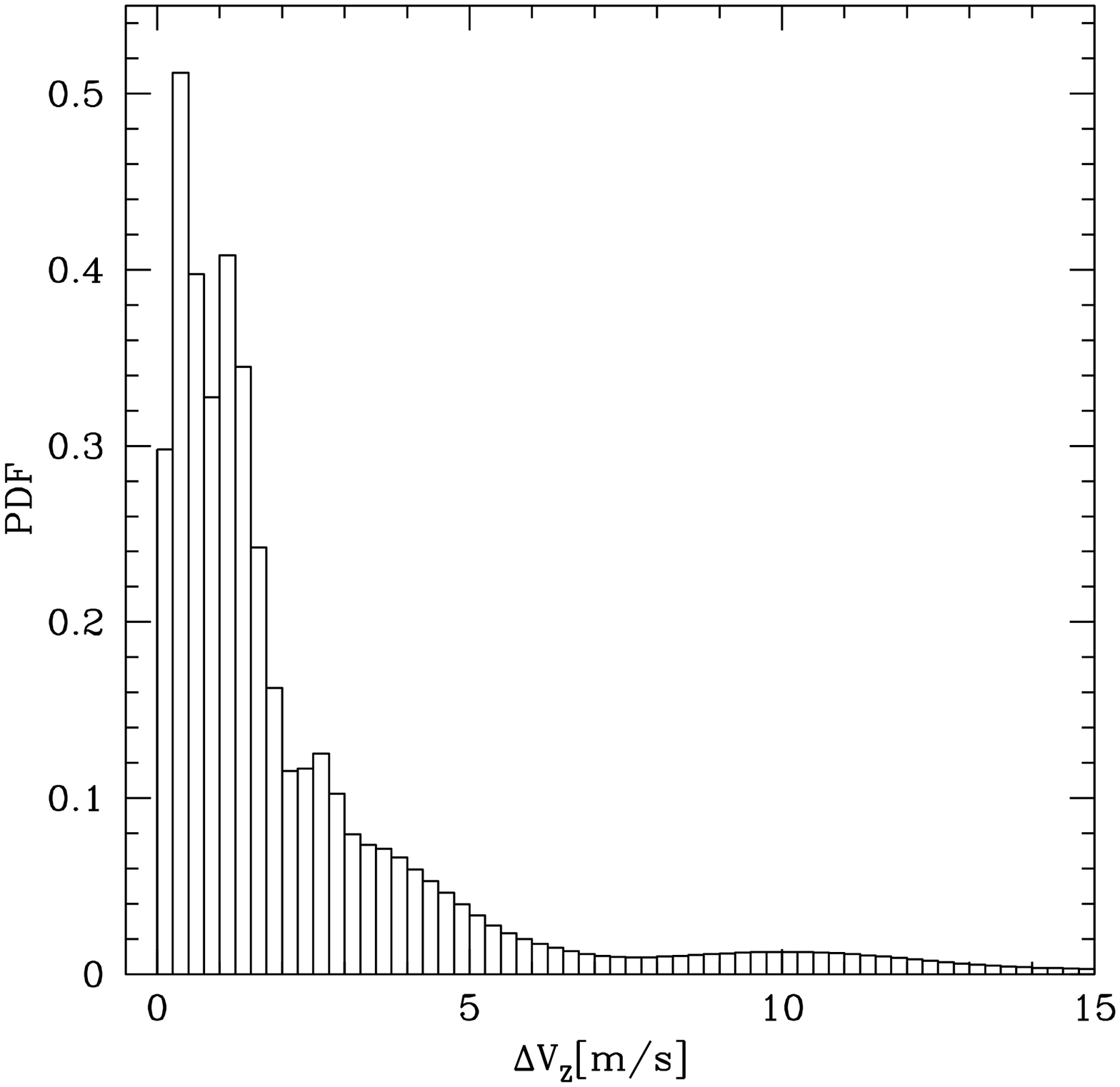}
\caption{Probability density funtion of the linear polarization fraction ($P_{\rm{l}}$, \textit{left panel}), of the circular polarization fraction ($P_{\rm{V}}$, \textit{middle panel}), and of the
Zeeman-splitting (\dvz, \textit{right panel}) of the 6.7~GHz \meth ~maser emission. The interval width of
the histograms is 0.5\% and 0.25\% and 0.25~\ms ~for the $P_{\rm{l}}$,  $P_{\rm{V}}$, and the \dvz ~plots,
respectively. The data are taken from: \cite{sur09,sur112,sur141}, \cite{vle10}, Papers I-IV, and this work.}
\label{pdf_pl_pv_dvz}
\end{figure*}

\subsection{Statistical results}
\label{stats}
The completion of the flux-limited sample, composed of 31 sources, allows us to perform a detailed
statistical analysis. In the previous papers of the series (i.e., Papers II-IV), we compared the 
orientation on the plane of the sky of the error-weighted magnetic field
($\langle\Phi_{\rm{B}}\rangle$) with the orientation on the plane of the sky of the outflows 
($\rm{PA}_{\rm{outflow}}$) and the linear distribution of the detected \meth ~masers 
($\rm{PA}_{\rm{CH_{3}OH}}$) both with the error-weighted value of the linear polarization angles 
($\langle\chi\rangle$) and with the angles $\langle\Phi_{\rm{B}}\rangle$. In addition,
we perform here Monte Carlo simulations and compare the results with our measurements.
Hence, we first performed the nonparametric Kolmogorov-Smirnov (K-S) test on the sample of angles
$|\rm{PA}_{\rm{outflow}}-\langle\Phi_{\rm{B}}\rangle|$,
$|\rm{PA}_{\rm{CH_{3}OH}}-\rm{PA}_{\rm{outflow}}|$,
and $|\rm{PA}_{\rm{CH_{3}OH}}-\langle\chi\rangle|$ to determine whether their distributions are drawn 
from a random probability distribution (see Table~\ref{KS}). Afterward, we studied the 
probability distribution function (PDF) and the cumulative distribution function (CDF) of these 
sample of angles. As done previously in Papers~II-IV, we also added the
nine southern hemisphere sources studied by \cite{dod12} to our
flux-limited sample. Therefore the total number of sources is
40, and the projected angles are listed in Table~\ref{Comp_ang}.\\
\indent The K-S test shows a probability of 2\% that the distribution of the angles 
$|\rm{PA}_{\rm{outflow}}-\langle\Phi_{\rm{B}}\rangle|$ is drawn from a random distribution. This
indicates that the orientations of the magnetic field and of the outflow are likely related
one to the other. To determine whether they are perpendicular or parallel to each other, we compared the 
PDF and CDF of $|\rm{PA}_{\rm{outflow}}-\langle\Phi_{\rm{B}}\rangle|$ with the results of 
Monte Carlo simulations (see Fig.~\ref{CDF_B}). We performed two separated Monte Carlo runs, one 
to determine the distribution of the angles when the outflow and the magnetic field are parallel 
(case 1), and the other when they are perpendicular (case 2). The outflows are usually broader at
the tip than at the origin, that is, they have a certain opening angle that can conservatively be
assumed to be $30$\d. If the magnetic field is associated with the outflow, this might be aligned
either with the outflow axis or with the edges of the outflow, that is, we might expect an offset
of 15\d ~with respect to the outflow axis. If the magnetic field is associated with a disk
structure, we expect that the angle between the magnetic field and the outflow axis is 90\d, but
because of the thickness of the disk, we might expect an offset of 5\d. With this in mind, we draw 
vectors pairs in case 1 that in the 3D space are parallel to each other with a Gaussian sigma of
15\d ~and whose projection on the plane of the sky has an error of 34\d, as observed on

average in our sample. The distribution of these projected angles, called 0 deg distribution, is 
shown as a dotted blue line in Fig.~\ref{CDF_B}. In case 2, we instead draw vectors pairs that in 
the 3D space are perpendicular to each other with a Gaussian sigma of 5\d ~and whose projection on
the plane of the sky has an error of 34\d. The distribution in this case is called 90 deg 
distribution and is shown as a dot-dashed green line in Fig.~\ref{CDF_B}.  Neither the 0 deg 
distribution nor the 90 deg distribution describe the observed projected angles
$|\rm{PA}_{\rm{outflow}}-\langle\Phi_{\rm{B}}\rangle|$, which are represented in Fig.~\ref{CDF_B} 
with a histogram that accounts for the uncertainties of our measurements. However, we found that the observed 
distribution is best fit when we consider a combination of the 0 deg and 90 deg distributions, in
particular, when 50\% of the sources follows the 90 deg distribution and 50\% of the sources follow the
0 deg distribution. This is shown in Fig.~\ref{CDF_B} as a solid red line. Differently from 
the case of low-mass protostars \citep{gal18}, where the bimodal distribution was justified by the 
presence of a single YSO with a small disk or by multiple YSOs with large disks, here the bimodal
distribution found by us might be due to the different evolutionary stages of the YSOs or to where
the 6.7~GHz \meth ~masers arise. The 6.7~GHz \meth ~masers arise in the interface between 
the disk structure and the outflow, hence they might probe the magnetic field either associated 
with the disk structure or with the outflows. This is seen in G23.01-0.41, where the 6.7~GHz \meth
~masers trace the magnetic field at different locations of the interface \citep{san15}. Here, the
masers that arise close to the disk material ($\sim$1000~au from the center of the disk) probe a 
magnetic field along the disk; the  masers that arise at $\sim$1000~au from the disk surface and 
nearer to the jet or outflow probe a magnetic field aligned with the jet or outflow axis
\citep{san15}.\\
\indent The K-S test performed on the angles $|\rm{PA}_{\rm{CH_{3}OH}}-\rm{PA}_{\rm{outflow}}|$ 
indicates that the probability that their distribution is random is 18\%. The left
panel of Fig.~\ref{cdf} shows that the PDF suggests a slightly preferential distribution
of the angles in the range 70\d-90\d ~with a few notable cases at about 25\d. This might 
suggest that the 6.7~GHz \meth ~masers preferentially trace a gas located around and along the disk 
structures rather than outflow material. However, the corresponding CDF does not give any 
clear indication of this. We must note that the correlation coefficient of the linear fit in some
sources of the sample is $|\rho|<0.5$ (Col.~7 of Table~\ref{Comp_ang}). This indicates a very weak linear 
correlation for the maser features. An example is G31.28+0.06 in the present work, for which we 
have $\rho=+0.39$ (see Fig.~\ref{G312_cp}). \\
\indent Although the K-S test for the angles $|\rm{PA}_{\rm{CH_{3}OH}}-\langle\chi\rangle|$ 
provides a probability of 77\% that these angles are drawn from a random distribution, the PDF and
CDF (right panel of Fig.~\ref{cdf}) suggest preferential angles in the range 20\d-50\d ~and
30\d-80\d, respectively. Still, the linear distribution on the plane of the sky of the \meth ~maser
spots is not always perfect, as explained above. 
\begin{figure}[t!]
\centering
\includegraphics[width = 9 cm]{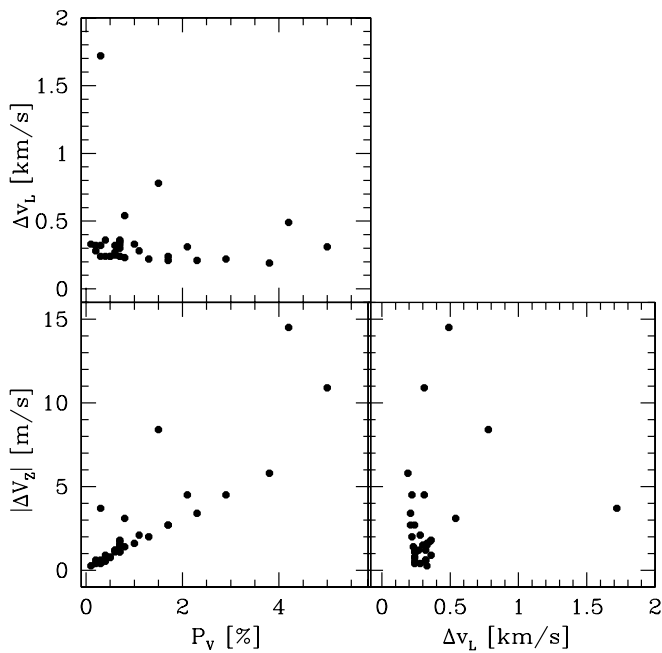}
\caption{Measured Zeeman-splitting ($|$\dvz$|$) of the 6.7~GHz \meth ~maser features as function of
the circular polarization fraction ($P_{\rm{V}}$; \textit{bottom left panel}) and of the FWHM of the
maser lines ($\Delta v_{\rm{L}}$; \textit{bottom right panel}). \textit{Top left panel}: 
$\Delta v_{\rm{L}}$ as function of $P_{\rm{V}}$. The data are taken from Papers IV and this work.
}
\label{pv_dvz_dvl}
\end{figure}
\subsection{Polarimetric characteristics of the 6.7~GHz \meth ~maser.}
\label{maser}
We detected linearly and circularly polarized emissions toward a large number of 6.7~GHz \meth 
~maser features in the flux-limited sample. In total, we measured linear polarization fraction 
($P_{\rm{l}}$) and circular polarization fraction ($P_{\rm{V}}$) towards 233 and 33 maser 
features, respectively (\citealt{sur09,sur111,sur141,vle10}; Papers I-IV; this work). From these 
measurements we can determine a typical value of $P_{\rm{l}}$ and $P_{\rm{V}}$ for the 
6.7~GHz \meth ~maser emission. The PDF of the measured $P_{\rm{l}}$,
which is shown in Fig.~\ref{pdf_pl_pv_dvz} (left panel), indicates that a typical value is in the
range $P_{\rm{l}}=1.0\%-2.5\%$. For $P_{\rm{V}}$, for which we assumed an uncertainty of 10\%, we 
find that a typical value is in the range 0.5\% and 0.75\% (middle panel of 
Fig.~\ref{pdf_pl_pv_dvz}). These ranges agree with the theoretical results of 
\citet{dal20}, who modeled the 6.7~GHz \meth ~maser emission considering the cases where each
hyperfine transition dominates the others and the special case where all the eight hyperfine
transitions contribute equally. However, it is not possible to determine the contribution 
of each hyperfine transition to the observed maser emission on the basis of the measured 
$P_{\rm{l}}$ and $P_{\rm{V}}$.
In addition to measuring $P_{\rm{V}}$, we were able to measure from the
circularly polarized spectra of the \meth ~maser features the Zeeman splitting of the maser lines 
(\dvz). We note from the right panel of Fig.~\ref{pdf_pl_pv_dvz} that \dvz ~is typically in the 
range between 0.5~\ms ~and 2.0~\ms, which would correspond to $9~\rm{mG}<|B_{\rm{||}}|<40~\rm{mG}$ 
if $F=3\rightarrow4$ is the most favored of the eight hyperfine transitions that might
contribute to the maser emission. \\
\indent The circular polarization fraction and the Zeeman splitting are related through the 
following equation:
\begin{equation}
    P_{\rm{V}} = \frac{V_{\rm{max}}-V_{\rm{min}}}{I_{\rm{max}}} = \frac{2 \cdot A_{\rm{FF'}}}{\Delta v_{\rm{L}}} \cdot \frac{\Delta V_{\rm{Z}}}{\alpha_{\rm{Z}}},
    \label{pv_eq}
\end{equation}
\begin{figure}[t!]
\centering
\includegraphics[width = 8 cm]{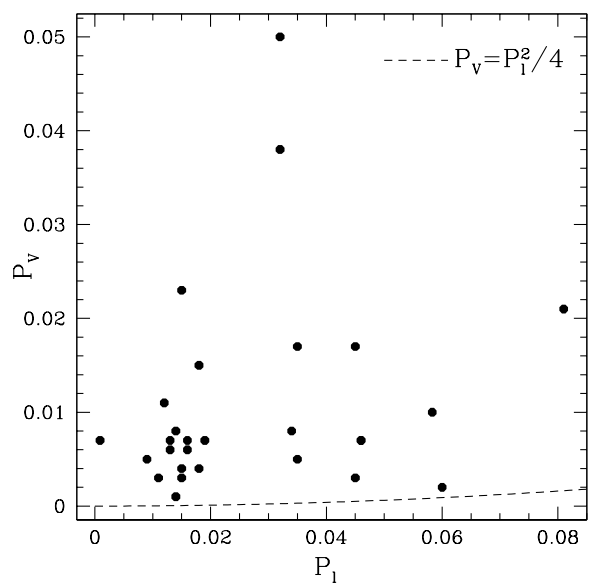}
\caption{Circular polarization fraction ($P_{\rm{V}}$) of the 6.7~GHz \meth ~maser
features as function of the corresponding linear polarization fraction ($P_{\rm{l}}$). 
The dashed line represents the conversion of $P_{\rm{l}}$ into $P_{\rm{V}}$ due to the rotation of 1
rad in the magnetic field along the maser path \citep{dal20}. The data are taken from:
\cite{sur09,sur112,sur141}, \cite{vle10}, Papers I-IV, and this work.
}
\label{plpv}
\end{figure}
where $A_{\rm{FF'}}$ is a coefficient that depends on the masing hyperfine transition 
\citep{vle05} and $\Delta v_{\rm{L}}$ is the full width at half maximum (FWHM) of the maser lines. We
determined the coefficient $A_{\rm{FF'}}$ of the \meth ~maser transition $F=3\rightarrow4$ by 
calculating $P_{\rm{V}}$ from synthetic I and V spectra and for a series of magnetic fields. The 
value that we found is $A_{\rm{3\rightarrow4}}^{\rm{synthetic}}=-0.0364$~\kmsg. 
We measured three quantities from our spectra that are independent of the 
hyperfine transition that we assumed in our analysis. These are $\Delta v_{\rm{L}}$, which is measured
from a Gaussian fit of the $I$ spectrum of the maser features, and $|\Delta V_{\rm{Z}}|$ and 
$P_{\rm{V}}$,
which are measured from the $V$ and $I$ spectra of the maser features. To verify whether these three quantities are related, we plot them in three different panels of
Fig.~\ref{pv_dvz_dvl}. As expected,  $|\Delta V_{\rm{Z}}|$ and $P_{\rm{V}}$ are linearly 
related, whereas $\Delta v_{\rm{L}}$ is independent of  \dvz ~and $P_{\rm{V}}$.\\
\indent The circularly polarized maser emission of nonparamagnetic molecules, such as the \meth
~molecule,
can be affected by several non-Zeeman effects and/or instrumental effects \citep{dal20}. Some of these
effects can be ruled out if the masers are unsaturated \citep{dal20}, for instance, one of them is the 
effect due to the change in quantization axis \citep{ned90}. In our analysis we have always
considered only unsaturated masers (e.g., Papers I-IV), therefore this kind of effect cannot 
influence our Zeeman-splitting measurements. We might look for signs of other non-Zeeman effects 
from our measurements. Unfortunately, we cannot use Fig.~\ref{pv_dvz_dvl} for our purpose. If
any non-Zeeman effect were present, this would have affected both $|\Delta V_{\rm{Z}}|$ and
$P_{\rm{V}}$ because both are measured from the V spectra. Furthermore, 
although $\Delta v_{\rm{L}}$ is independent of  \dvz ~and $P_{\rm{V}}$, any non-Zeeman effect
might be hidden in the two panels of Fig.~\ref{pv_dvz_dvl} where $\Delta v_{\rm{L}}$ is reported. We may expect to see their contribution there only if the magnetic field is constant in our
measurements, in this case, $|\Delta V_{\rm{Z}}|$ would be the same for all the maser features, and
we might verify the relation between $P_{\rm{V}}$ and $\Delta v_{\rm{L}}$ given by Eq.~\ref{pv_eq}, but
unfortunately, this is not the case here.\\
\begin{figure}[t!]
\centering
\includegraphics[width = 8 cm]{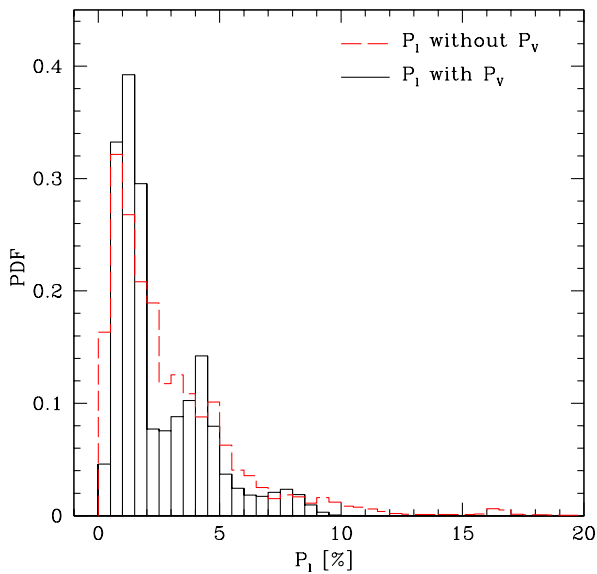}
\caption{Probability distribution function of the linear polarization fraction ($P_{\rm{l}}$) measured toward the 6.7~GHz \meth 
~maser features of the flux-limited sample that do not show (dashed red histogram; distribution
\textit{d1}) and show (solid black histogram; distribution \textit{d2}) circular polarized emission. The
interval width of the histograms is 0.5\%. The data are taken from \cite{sur09,sur112,sur141},
\cite{vle10}, Papers I-IV, and this work. 
}
\label{plwpv}
\end{figure}
\indent A non-Zeeman effect that might be investigated from our measurements is the effect that is due to the 
rotation of the magnetic field along the maser path \citep{wie98}, which converts $P_{\rm{l}}$ into
$P_{\rm{V}}$. For a rotation of 1 rad in the magnetic field, the generated fraction of circular
polarization is given by $P_{\rm{V}}=P_{\rm{l}}^{2}/4$ \citep{wie98}. We plot $P_{\rm{V}}$ as function 
of $P_{\rm{l}}$ in Fig.~\ref{plpv}. Here we considered all the maser features for which we were able to
detect both polarizations. No clear relation between $P_{\rm{V}}$ and $P_{\rm{l}}$ is seen in
Fig.~\ref{plpv}. We also note that the possible contribution of $P_{\rm{l}}$ to $P_{\rm{V}}$ due to the 
1 rad rotation of the magnetic field along the maser path is marginal, as already reported in
previous works \citep{vle11,dal20}. However, we cannot completely rule out a contribution of similar
non-Zeeman effects, that is, a conversion of $P_{\rm{l}}$ into $P_{\rm{V}}$, only on the basis of Fig.~\ref{plpv} because each maser feature might be affected differently depending on the rotation of the magnetic
field in their amplification paths, which could be a fraction of or even more than 1 rad. Another way to
investigate the contribution of this kind of non-Zeeman effect is to compare the distributions of
$P_{\rm{l}}$ when $P_{\rm{V}}$ is not detected (distribution \textit{d1}) and when it is detected
(distribution \textit{d2}).
We plot these two distributions, for which the total number of measurements is different ($n1=205$ for
\textit{d1} and $n2=29$ for \textit{d2}), in Fig.~\ref{plwpv}. The two distributions look quite similar.
In addition to the absence in \textit{d2} of measurements with $P_{\rm{l}}>10\%$, which might be due to the low
number $n2$, there are small differences around $P_{\rm{l}}\approx1.5\%$ and $\approx2.5\%$. These
differences do not allow us to completely rule out the contribution of a non-Zeeman effect of this kind
in all our measurements, but the similarity of the two distributions suggests that the conversion of
$P_{\rm{l}}$ into $P_{\rm{V}}$ is not prevalent in our measurements. This also implies that the
non-Zeeman effect due to the anisotropic resonant scattering \citep{hou13}, which converts $P_{\rm{l}}$ into 
$P_{\rm{V}}$, may not be important in our measurements in the same way as was found to be significant for 
SiO masers \citep{hou13}. 
\section{Summary}
We completed the flux-limited sample by observing the last five SFRs at 6.7~GHz in full 
polarization spectral mode with the EVN to detect the linearly and circularly polarized emission 
of \meth ~masers. We detected linearly polarized emission from all the five sources and the 
circularly polarized emission from three sources: G30.70-0.07, G32.03+0.06, and G69.52-0.97. By 
modeling the linearly polarized emission with the FRTM code, we estimated the orientation of the 
magnetic field in all the sources. However, we were able to compare the orientation of the 
magnetic field with the outflow axis only toward G31.28+0.06 and G69.52-0.97. We found in both 
cases that the magnetic field is almost aligned with the outflow axis. Using the FRTM code to also
model the circular polarization of the \meth ~maser emission, we measured the Zeeman splitting in 
G30.70-0.07 ($\Delta V_{\rm{Z}}^{\rm{G307.E18}}~=+1.5\pm0.2$~\ms), in G32.03+0.06 
($\Delta V_{\rm{Z}}^{\rm{G32.E11}}~=+0.6\pm0.1$~\ms ~and 
$\Delta V_{\rm{Z}}^{\rm{G32.E37}}~=+1.4\pm0.2$~\ms), and in G69.52-0.97 
($\Delta V_{\rm{Z}}^{\rm{G69.E03}}~=+1.2\pm0.2$~\ms ~and 
$\Delta V_{\rm{Z}}^{\rm{G69.E07}}~=+14.5\pm4.4$~\ms). The corresponding estimated magnetic field
strengths along the line of sight are $B_{||}^{\rm{G307.E18}}>30$~mG, $B_{||}^{\rm{G32.E11}}>12$~mG,
$B_{||}^{\rm{G32.E37}}>27$~mG, $B_{||}^{\rm{G69.E03}}>24$~mG, and $B_{||}^{\rm{G69.E07}}>285$~mG,
respectively.\\
\indent Because the flux-limited sample (composed of 31 SFRs) was completed, we performed a
detailed statistical analysis that revealed a bimodal distribution of the 3D angles between the
magnetic field and the outflow axis ($|\rm{PA}_{\rm{outflow}}-\langle\Phi_{\rm{B}}\rangle|$).
We performed two different runs of Monte Carlo simulations of the projection on the plane of the 
sky of two random 3D vectors, one run with 3D parallel vectors (called 0 deg distribution), and the
other run with 3D perpendicular vectors (called 90 deg distribution). The simulation outputs led 
us to the conclusion that the distribution of the observed angles
$|\rm{PA}_{\rm{outflow}}-\langle\Phi_{\rm{B}}\rangle|$ is consistent neither with a pure
90 deg distribution nor with a pure 0 deg distribution, but rather with a combination of the two. 
In particular, 50\% of the angles
drawn from the 0 deg distribution and 50\% from the 90 deg distribution. In addition, by
considering all the maser features that we detected with the EVN toward the sources of 
the flux-limited sample, we were able to determine that typical values of the linear and circular
polarization fraction for 6.7~GHz \meth ~masers are $P_{\rm{l}}=1.0\%-2.5\%$ (based on 233 maser 
features) and $P_{\rm{V}}=0.5\%-0.75\%$ (based on 33 maser features), respectively. From the 
circularly polarized spectra ($V$ spectra) of the \meth ~maser features, we found that a typical 
Zeeman splitting is in the range between 0.5~\ms ~and 2.0~\ms.\\

\noindent \small{\textit{Acknowledgments.} 
We wish to thank the referee Dr. Sandra Etoka for the useful suggestions that have improved the paper. 
G.S. acknowledges James O. Chibueze for his support on the data reduction of G31.28+0.06.
W.H.T.V. acknowledges support from the Swedish Research Council (VR) under grant No. 2014-05713. 
A.B. acknowledges support from the National Science Centre, Poland through grant 2016/21/B/ST9/01455. 
The European VLBI Network is a joint facility of independent European, African, Asian, and North 
American radio astronomy institutes. Scientific results from data presented in this publication are
derived from the following EVN project code(s): ES072.
The research leading to these results has received funding from the European Commission Seventh 
Framework Programme (FP/2007-2013) under grant agreement No. 283393 (RadioNet3).}

\bibliographystyle{aa}
\bibliography{surcis_2021}
%

\begin{appendix}
\normalsize
\section{Tables}
\label{appA}
\begin {table*}[h!]
\caption []{Parameters of the 6.7-GHz \meth ~maser features detected in G30.70-0.07.} 
\begin{center}
\scriptsize
\begin{tabular}{ l c c c c c c c c c c c c c c}
\hline
\hline
\,\,\,\,\,(1)&(2)   & (3)      & (4)            & (5)       & (6)              & (7)         & (8)       & (9)                     & (10)                    & (11)                        & (12)         &(13)                  &(14)         \\
Maser     &  RA\tablefootmark{a}&Dec\tablefootmark{a}& Peak flux & $V_{\rm{lsr}}$& $\Delta v\rm{_{L}}$ &$P_{\rm{l}}\tablefootmark{b}$ &  $\chi\tablefootmark{b}$   & $\Delta V_{\rm{i}}\tablefootmark{c}$ & $T_{\rm{b}}\Delta\Omega\tablefootmark{c}$& $P_{\rm{V}}$ & $\Delta V_{\rm{Z}}$  & $|B_{||}|$  &$\theta\tablefootmark{d}$\\
          & offset &  offset  & Density(I)     &           &                  &             &             &                         &                         &              &                      &      & \\ 
          & (mas)  &  (mas)   & (Jy/beam)      &  (km/s)   &      (km/s)      & (\%)        &   (\d)    & (km/s)                  & (log K sr)              &   ($\%$)     &  (m/s)               & (mG) & (\d)       \\ 
\hline
G307.E01  & -709.951& 427.432   & $0.565\pm0.007$& 87.69     &      $1.03$      & $-$         & $-$       &  $-$                    & $-$                     & $-$     & $-$                  & $-$     &$-$ \\ 
G307.E02  & -707.378& 386.938   & $0.070\pm0.007$& 87.73     &      $0.29$      & $-$         & $-$       &  $-$                    & $-$                     & $-$     & $-$                  & $-$     &$-$ \\ 
G307.E03  & -706.234& 318.524   & $0.075\pm0.005$& 87.30     &      $0.42$      & $-$         & $-$       &  $-$                    & $-$                     & $-$     & $-$                  & $-$     &$-$ \\ 
G307.E04  & -694.283& 547.577   & $0.309\pm0.007$& 86.11     &      $0.41$      & $-$         & $-$       &  $-$                    & $-$                     & $-$     & $-$                  & $-$     &$-$ \\ 
G307.E05  & -687.992& 417.643   & $3.709\pm0.007$& 86.15     &      $0.47$      &  $1.2\pm0.1$ &  $+57\pm5$            & $1.9^{+0.1}_{-0.2}$     & $8.9^{+0.3}_{-0.5}$         &$-$         & $-$                      & $-$     &$90^{+24}_{-24}$ \\ 
G307.E06  & -685.190& 376.825   & $0.588\pm0.007$& 86.11     &      $0.41$      & $-$         & $-$       &  $-$                    & $-$                     & $-$     & $-$                  & $-$     &$-$ \\ 
G307.E07  & -682.674& 434.138   & $0.125\pm0.032$& 86.20     &      $0.35$      & $-$         & $-$       &  $-$                    & $-$                     & $-$     & $-$                  & $-$     &$-$ \\ 
G307.E08  & -273.516&  13.481  & $0.730\pm0.007$& 89.10     &      $0.26$      & $-$         & $-$       &  $-$                    & $-$                     & $-$     & $-$                  & $-$     &$-$ \\ 
G307.E09  & -271.000& -25.930  & $0.078\pm0.007$& 89.10     &      $0.18$      & $-$         & $-$       &  $-$                    & $-$                     & $-$     & $-$                  & $-$     &$-$ \\   
G307.E10  & -42.431 &  95.012  & $0.132\pm0.004$& 89.49     &      $0.45$      & $-$         & $-$       &  $-$                    & $-$                     & $-$     & $-$                  & $-$     &$-$ \\   
G307.E11  & -36.884 & -94.384  & $0.075\pm0.003$& 89.71     &      $0.20$      & $-$         & $-$       &  $-$                    & $-$                     & $-$     & $-$                  & $-$    &$-$ \\  
G307.E12  & -13.210 & -97.353  & $0.066\pm0.006$& 87.38     &      $0.20$      & $-$         & $-$       &  $-$                    & $-$                     & $-$     & $-$                  & $-$    &$-$ \\   
G307.E13  & -12.409 & -93.733  & $0.068\pm0.003$& 90.37     &      $0.28$      & $-$         & $-$       &  $-$                    & $-$                     & $-$     & $-$                  & $-$     &$-$ \\  
G307.E14  & -4.518  & -176.004 & $0.075\pm0.009$& 88.79     &      $1.56$      & $-$         & $-$       &  $-$                    & $-$                     & $-$     & $-$                  & $-$     &$-$ \\   
G307.E15  & -3.431  & -64.735  & $1.104\pm0.011$& 88.44     &      $0.29$      & $-$         & $-$       &  $-$                    & $-$                     & $-$     & $-$                  & $-$     &$-$ \\   
G307.E16  & -2.745  & -62.480  & $0.347\pm0.020$& 88.13     &      $0.30$      & $-$         & $-$       &  $-$                    & $-$                     & $-$     & $-$                  & $-$     &$-$ \\   
G307.E17  & -1.716  &  46.170  & $0.553\pm0.010$& 88.09     &      $0.26$      & $-$         & $-$       &  $-$                    & $-$                     & $-$     & $-$                  & $-$     &$-$ \\   
G307.E18  & -0.114  & -0.262   & $38.158\pm0.018$& 88.31    &      $0.30$      & $0.09\pm0.06$  & $+49\pm18$  &  $1.3^{+0.1}_{-0.5}$    & $7.7^{+0.6}_{-0.5}$        & $0.7$      & $+1.5\pm0.2$    & $>30$   &$\mathbf{45^{+30}_{-28}}$ \\ 
G307.E19  &  0      &  0       & $39.587\pm0.042$& 88.31    &      $0.29$      & $0.07\pm0.02$  & $+87\pm18$  &  $1.3^{+0.4}_{-0.4}$    & $7.6^{+0.5}_{-0.2}$        & $-$        & $-$          & $-$     &$75^{+14}_{-39}$ \\ 
G307.E20  &  1.773  & -43.808  & $0.434\pm0.010$& 88.09     &      $0.13$      & $-$         & $-$       &  $-$                    & $-$                     & $-$     & $-$                  & $-$     &$-$ \\  
G307.E21  &  3.889  &  65.256  & $0.539\pm0.014$& 88.13     &      $0.76$      & $-$         & $-$       &  $-$                    & $-$                     & $-$     & $-$                  & $-$     &$-$ \\   
G307.E22  &  4.003  &  61.158  & $1.056\pm0.029$& 88.57     &      $0.36$      & $-$         & $-$       &  $-$                    & $-$                     & $-$     & $-$                  & $-$     &$-$ \\  
G307.E23  &  5.433  &  60.280  & $0.952\pm0.009$& 88.57     &      $0.36$      & $-$         & $-$       &  $-$                    & $-$                     & $-$     & $-$                  & $-$     &$-$ \\   
G307.E24  &  6.119  & -170.896 & $0.112\pm0.013$& 88.22     &      $0.80$      & $-$         & $-$       &  $-$                    & $-$                     & $-$     & $-$                  & $-$     &$-$ \\   
G307.E25  &  6.748  & -412.904 & $0.101\pm0.018$& 88.57     &      $0.13$      & $-$         & $-$       &  $-$                    & $-$                     & $-$     & $-$                  & $-$     &$-$ \\   
G307.E26  &  6.862  & -413.448 & $0.104\pm0.028$& 88.57     &      $0.14$      & $-$         & $-$       &  $-$                    & $-$                     & $-$     & $-$                  & $-$     &$-$ \\   
G307.E27  &  11.609 &  333.790 & $0.103\pm0.007$& 88.31     &      $0.72$      & $-$         & $-$       &  $-$                    & $-$                     & $-$     & $-$                  & $-$     &$-$ \\   
G307.E28  &  19.672 &  91.844  & $0.070\pm0.006$& 88.31     &      $0.20$      & $-$         & $-$       &  $-$                    & $-$                     & $-$     & $-$                  & $-$     &$-$ \\   
G307.E29  &  127.866&  342.210 & $0.056\pm0.003$& 89.32     &      $0.23$      & $-$         & $-$       &  $-$                    & $-$                     & $-$     & $-$                  & $-$     &$-$ \\   
G307.E30  &  127.923&  398.027 & $0.085\pm0.005$& 88.96     &      $0.24$      & $-$         & $-$       &  $-$                    & $-$                     & $-$     & $-$                  & $-$     &$-$ \\   
G307.E31  &  266.024& -42.141  & $0.052\pm0.004$& 90.72     &      $0.23$      & $-$         & $-$       &  $-$                    & $-$                     & $-$     & $-$                  & $-$     &$-$ \\   
\hline
\end{tabular} \end{center}
\tablefoot{
\tablefoottext{a}{The reference position is $\alpha_{2000}=18^{\rm{h}}47^{\rm{m}}36^{\rm{s}}\!.900$ and 
$\delta_{2000}=-02^{\circ}01'05''\!\!.025$ (see Sec.~\ref{obssect}).}
\tablefoottext{b}{$P_{\rm{l}}$ and $\chi$ are the mean values of the linear polarization fraction and the linear polarization angle measured across the spectrum, respectively.}
\tablefoottext{c}{The best-fitting results obtained using a model based on the radiative transfer theory of methanol masers 
for $\Gamma+\Gamma_{\nu}=1~\rm{s^{-1}}$ \citep{vle10,sur111}. The errors were determined 
by analyzing the full PDF.}
\tablefoottext{d}{The angle between the magnetic field and the maser propagation direction is determined using the observed $P_{\rm{l}}$ 
and the fitted emerging brightness temperature. The errors were determined by analyzing the full PDF. The boldface indicates that $|\theta^{\rm{+}}-55$\d$|<|\theta^{\rm{-}}-55$\d$|$, i.e., the magnetic field is parallel to the linear polarization vector (see Sect.~\ref{res}).}
}
\label{G307_tab}
\end{table*}
\begin {table*}[t!]
\caption []{Parameters of the 6.7-GHz \meth ~maser features detected in G30.76-0.05.} 
\begin{center}
\scriptsize
\begin{tabular}{ l c c c c c c c c c c c c c}
\hline
\hline
\,\,\,\,\,(1)&(2)   & (3)      & (4)            & (5)       & (6)              & (7)         & (8)       & (9)                     & (10)                    & (11)                        & (12)         &(13)                  &(14)         \\
Maser     & RA\tablefootmark{a}&Dec\tablefootmark{a}& Peak flux & $V_{\rm{lsr}}$& $\Delta v\rm{_{L}}$ &$P_{\rm{l}}\tablefootmark{b}$ &  $\chi\tablefootmark{b}$   & $\Delta V_{\rm{i}}\tablefootmark{c}$ & $T_{\rm{b}}\Delta\Omega\tablefootmark{c}$& $P_{\rm{V}}$ & $\Delta V_{\rm{Z}}$  & $|B_{||}|$  &$\theta\tablefootmark{d}$\\
          &  offset &  offset  & Density(I)     &           &                  &             &            &                         &                         &              &                      &      & \\ 
          &  (mas)  &  (mas)   & (Jy/beam)      &  (km/s)   &      (km/s)      & (\%)        &   (\d)    & (km/s)                  & (log K sr)              &   ($\%$)     &  (m/s)               & (mG) & (\d)       \\ 
\hline
G308.E01  & -168.187& -21.082  & $0.215\pm0.005$& 89.28     &      $0.50$      & $-$         & $-$       &  $-$                    & $-$                     & $-$     & $-$                  & $-$     &$-$ \\ 
G308.E02  & -163.384& -23.605  & $0.274\pm0.012$& 88.22     &      $0.29$      & $-$         & $-$       &  $-$                    & $-$                     & $-$     & $-$                  & $-$     &$-$ \\ 
G308.E03  & -81.835 &  116.776 & $0.113\pm0.004$& 92.61     &      $0.27$      & $-$         & $-$       &  $-$                    & $-$                     & $-$     & $-$                  & $-$     &$-$ \\ 
G308.E04  & -72.856 &  103.867 & $0.159\pm0.006$& 89.54     &      $0.44$      & $-$         & $-$       &  $-$                    & $-$                     & $-$     & $-$                  & $-$     &$-$ \\ 
G308.E05  & -41.918 & -19.604  & $0.578\pm0.007$& 89.89     &      $0.45$      & $-$         & $-$       &  $-$                    & $-$                     & $-$     & $-$                  & $-$     &$-$ \\ 
G308.E06  & -31.682 & -239.437 & $0.239\pm0.029$& 91.08     &      $0.99$      & $-$         & $-$       &  $-$                    & $-$                     & $-$     & $-$                  & $-$     &$-$ \\ 
G308.E07  & -30.824 &  61.823  & $0.436\pm0.048$& 91.21     &      $0.33$      & $-$         & $-$       &  $-$                    & $-$                     & $-$     & $-$                  & $-$     &$-$ \\ 
G308.E08  & -30.709 & -0.284   & $13.831\pm0.067$& 91.38    &      $0.41$      & $0.4\pm0.1$ & $-81\pm19$&  $1.9^{+0.1}_{-0.2}$       & $8.3^{+0.5}_{-1.2}$         & $-$       & $-$                  & $-$     &$\mathbf{62^{+6}_{-44}}$ \\
G308.E09  & -29.165 &  62.798  & $0.222\pm0.018$& 90.90     &      $0.57$      & $-$         & $-$       &  $-$                    & $-$                     & $-$     & $-$                  & $-$     &$-$ \\ 
G308.E10  & -27.850 &  111.902 & $0.287\pm0.019$& 90.94     &      $0.15$      & $-$         & $-$       &  $-$                    & $-$                     & $-$     & $-$                  & $-$     &$-$ \\ 
G308.E11  & -19.386 &  10.185  & $0.077\pm0.008$& 91.78     &      $1.87$      & $-$         & $-$       &  $-$                    & $-$                     & $-$     & $-$                  & $-$     &$-$ \\ 
G308.E12  & -17.213 & -3.788   & $0.183\pm0.006$& 90.24     &      $0.33$      & $-$         & $-$       &  $-$                    & $-$                     & $-$     & $-$                  & $-$     &$-$ \\ 
G308.E13  & -16.813 &  2.300   & $0.076\pm0.008$& 90.59     &      $2.74$      & $-$         & $-$       &  $-$                    & $-$                     & $-$     & $-$                  & $-$     &$-$ \\ 
G308.E14  & -14.526 & -66.776  & $0.081\pm0.006$& 90.46     &      $0.97$      & $-$         & $-$       &  $-$                    & $-$                     & $-$     & $-$                  & $-$     &$-$ \\ 
G308.E15  & -11.437 & -16.159  & $1.070\pm0.029$& 91.12     &      $0.41$      & $-$         & $-$       &  $-$                    & $-$                     & $-$     & $-$                  & $-$     &$-$ \\ 
G308.E16  & -10.694 &  19.527  & $0.829\pm0.006$& 93.71     &      $0.24$      & $-$         & $-$       &  $-$                    & $-$                     & $-$     & $-$                  & $-$     &$-$ \\
G308.E17  & -2.173  & -175.463 & $0.093\pm0.018$& 92.17     &      $0.87$      & $-$         & $-$       &  $-$                    & $-$                     & $-$     & $-$                  & $-$     &$-$ \\ 
G308.E18  & -1.601  &  13.784  & $1.531\pm0.009$& 93.32     &      $0.34$      & $-$         & $-$       &  $-$                    & $-$                     & $-$     & $-$                  & $-$     &$-$ \\ 
G308.E19  & -0.515  &  64.768  & $0.303\pm0.040$& 91.82     &      $0.48$      & $-$         & $-$       &  $-$                    & $-$                     & $-$     & $-$                  & $-$     &$-$ \\ 
G308.E20  &  0      &  0       & $14.074\pm0.085$& 91.82    &      $0.38$      & $1.9\pm0.4$ & $8\pm3$   &  $1.5^{+0.1}_{-0.5}$       & $9.2^{+1.1}_{-0.5}$         & $-$       & $-$         & $-$     &$\mathbf{70^{+9}_{-41}}$ \\ 
G308.E21  &  5.947  &  18.745  & $0.088\pm0.005$& 92.53     &      $0.27$      & $-$         & $-$       &  $-$                    & $-$                     & $-$     & $-$                  & $-$     &$-$ \\ 
G308.E22  &  11.495 & -400.833 & $0.519\pm0.074$&  91.87    &      $0.38$      & $-$         & $-$       &  $-$                    & $-$                     & $-$     & $-$                  & $-$     &$-$ \\ 
G308.E23  &  31.396 & -26.608  & $0.166\pm0.023$& 92.13     &      $0.19$      & $-$         & $-$       &  $-$                    & $-$                     & $-$     & $-$                  & $-$     &$-$ \\ 
G308.E24  &  45.807 & -77.103  & $0.098\pm0.007$& 93.93     &      $0.23$      & $-$         & $-$       &  $-$                    & $-$                     & $-$     & $-$                  & $-$     &$-$ \\  
\hline
\end{tabular} \end{center}
\tablefoot{
\tablefoottext{a}{The reference position is $\alpha_{2000}=18^{\rm{h}}47^{\rm{m}}39^{\rm{s}}\!.732$ and 
$\delta_{2000}=-01^{\circ}57'21''\!\!.975$ (see Sec.~\ref{obssect}).}
\tablefoottext{b}{$P_{\rm{l}}$ and $\chi$ are the mean values of the linear polarization fraction and the linear polarization angle measured across the spectrum, respectively.}
\tablefoottext{c}{The best-fitting results obtained using a model based on the radiative transfer theory of methanol masers 
for $\Gamma+\Gamma_{\nu}=1~\rm{s^{-1}}$ \citep{vle10,sur111}. The errors were determined 
by analyzing the full PDFprobability distribution function.}
\tablefoottext{d}{The angle between the magnetic field and the maser propagation direction is determined using the observed $P_{\rm{l}}$ 
and the fitted emerging brightness temperature. The errors were determined by analyzing the full PDF. The boldface indicates that $|\theta^{\rm{+}}-55$\d$|<|\theta^{\rm{-}}-55$\d$|$, i.e., the magnetic field is parallel to the linear polarization vector (see Sect.~\ref{res}).}
}
\label{G308_tab}
\end{table*}

\clearpage
\onecolumn
\scriptsize
\setlength{\tabcolsep}{6pt}
\begin{longtable}{@ {} l c c c c c c c c c c c c c}
\caption[]{Parameters of the 6.7-GHz \meth ~maser features detected in G31.28+0.06.} \\
\setlength{\tabcolsep}{3pt}
\cr\hline\hline
\scriptsize
\,\,\,\,\,(1)&(2)   & (3)      & (4)            & (5)       & (6)              & (7)         & (8)       & (9)                     & (10)                    & (11)                        & (12)         &(13)                  &(14)         \\
Maser     & RA\tablefootmark{a}&Dec\tablefootmark{a}& Peak flux & $V_{\rm{lsr}}$& $\Delta v\rm{_{L}}$ &$P_{\rm{l}}\tablefootmark{b}$ &  $\chi\tablefootmark{b}$   & $\Delta V_{\rm{i}}\tablefootmark{c}$ & $T_{\rm{b}}\Delta\Omega\tablefootmark{c}$& $P_{\rm{V}}$ & $\Delta V_{\rm{Z}}$  & $|B_{||}|$  &$\theta\tablefootmark{d}$\\
          &  offset &  offset  & Density(I)     &           &                  &             &            &                         &                         &              &                      &      & \\ 
          &  (mas)  &  (mas)   & (Jy/beam)      &  (km/s)   &      (km/s)      & (\%)        &   (\d)    & (km/s)                  & (log K sr)              &   ($\%$)     &  (m/s)               & (mG) & (\d)       \\ \hline
\endfirsthead
\caption{(continued.)}\\
\hline
\hline
\\
\,\,\,\,\,(1)&(2)   & (3)      & (4)            & (5)       & (6)              & (7)         & (8)       & (9)                     & (10)                    & (11)                        & (12)         &(13)                  &(14)         \\
Maser     & RA\tablefootmark{a}&Dec\tablefootmark{a}& Peak flux & $V_{\rm{lsr}}$& $\Delta v\rm{_{L}}$ &$P_{\rm{l}}\tablefootmark{b}$ &  $\chi\tablefootmark{b}$   & $\Delta V_{\rm{i}}\tablefootmark{c}$ & $T_{\rm{b}}\Delta\Omega\tablefootmark{c}$& $P_{\rm{V}}$ & $\Delta V_{\rm{Z}}$  & $|B_{||}|$  &$\theta\tablefootmark{d}$\\
          &  offset &  offset  & Density(I)     &           &                  &             &            &                         &                         &              &                      &      & \\ 
          &  (mas)  &  (mas)   & (Jy/beam)      &  (km/s)   &      (km/s)      & (\%)        &   (\d)    & (km/s)                  & (log K sr)              &   ($\%$)     &  (m/s)               & (mG) & (\d)       \\ 
\hline
\endhead
\hline
\endfoot
\hline
\noalign{\noindent\small{\textbf{Notes.}$^{(a)}$ The reference position is $\alpha_{2000}=18^{\rm{h}}48^{\rm{m}}12^{\rm{s}}\!.390$ and $\delta_{2000}=-01^{\circ}26'22''\!\!.629$ (see Sec.~\ref{obssect}).
$^{(b)}$ $P_{\rm{l}}$ and $\chi$ are the mean values of the linear polarization fraction and the linear polarization angle measured across the spectrum, respectively.
$^{(c)}$ The best-fitting results obtained using a model based on the radiative transfer theory of methanol masers 
for $\Gamma+\Gamma_{\nu}=1~\rm{s^{-1}}$ \citep{vle10,sur111}. The errors were determined 
by analyzing the full PDF.
$^{(d)}$ The angle between the magnetic field and the maser propagation direction is determined using the observed $P_{\rm{l}}$ 
and the fitted emerging brightness temperature. The errors were determined by analyzing the full PDF. The boldface indicates that $|\theta^{\rm{+}}-55$\d$|<|\theta^{\rm{-}}-55$\d$|$, i.e., the magnetic field is parallel to the linear polarization vector (see Sect.~\ref{res}).}
}
\endlastfoot
G312.E01  & -274.572& -273.005 & $0.422\pm0.015$& 109.52     &      $0.28$      & $-$         & $-$       &  $-$                    & $-$                     & $-$     & $-$                  & $-$     &$-$ \\ 
G312.E02  & -242.638& -289.921 & $3.006\pm0.035$& 110.00     &      $0.38$      & $-$         & $-$       &  $-$                    & $-$                     & $-$     & $-$                  & $-$     &$-$ \\   
G312.E03  & -242.338& -193.659 & $1.014\pm0.049$& 111.23     &      $0.25$      & $-$         & $-$       &  $-$                    & $-$                     & $-$     & $-$                  & $-$     &$-$ \\   
G312.E04  & -241.723& -178.473 & $2.557\pm0.029$& 111.67     &      $0.19$      & $-$         & $-$       &  $-$                    & $-$                     & $-$     & $-$                  & $-$    &$-$ \\  
G312.E05  & -237.590& -502.018 & $0.298\pm0.026$& 112.37     &      $0.31$      & $-$         & $-$       &  $-$                    & $-$                     & $-$     & $-$                  & $-$     &$-$ \\   
G312.E06  & -234.272& -277.576 & $0.859\pm0.008$& 107.89     &      $0.20$      & $-$         & $-$       &  $-$                    & $-$                     & $-$     & $-$                  & $-$     &$-$ \\  
G312.E07  & -233.529& -181.396 & $0.102\pm0.006$& 109.08     &      $0.33$      & $-$         & $-$       &  $-$                    & $-$                     & $-$     & $-$                  & $-$     &$-$ \\   
G312.E08  & -232.070& -183.102 & $2.656\pm0.051$& 111.01     &      $0.65$      & $2.6\pm0.3$ & $+43\pm4$ &  $1.4^{+0.1}_{-0.6}$      & $9.2^{+0.2}_{-1.0}$         & $-$       & $-$                  & $-$     &$90^{+13}_{-13}$ \\
G312.E09  & -231.741& -160.955 & $3.196\pm0.037$& 112.37     &      $0.29$      & $-$         & $-$       &  $-$                    & $-$                     & $-$     & $-$                  & $-$     &$-$ \\   
G312.E10  & -231.269&  18.898  & $0.066\pm0.006$& 107.98     &      $0.39$      & $-$         & $-$       &  $-$                    & $-$                     & $-$     & $-$                  & $-$     &$-$ \\   
G312.E11  & -229.267& -139.851 & $0.298\pm0.028$& 111.58     &      $1.35$      & $-$         & $-$       &  $-$                    & $-$                     & $-$     & $-$                  & $-$     &$-$ \\ 
G312.E12  & -228.223& -168.879 & $4.423\pm0.070$& 110.17     &      $0.30$      & $-$         & $-$       &  $-$                    & $-$                     & $-$     & $-$                  & $-$     &$-$ \\ 
G312.E13  & -227.565& -297.728 & $0.369\pm0.024$& 109.65     &      $0.83$      & $-$         & $-$       &  $-$                    & $-$                     & $-$     & $-$                  & $-$     &$-$ \\  
G312.E14  & -226.493& -179.544 & $1.242\pm0.025$& 111.49     &      $0.28$      & $-$         & $-$       &  $-$                    & $-$                     & $-$     & $-$                  & $-$     &$-$ \\   
G312.E15  & -225.277& -279.278 & $0.095\pm0.005$& 108.07     &      $0.34$      & $-$         & $-$       &  $-$                    & $-$                     & $-$     & $-$                  & $-$     &$-$ \\  
G312.E16  & -219.857& -163.776 & $3.403\pm0.025$& 111.80     &      $0.22$      & $2.5\pm0.5$ & $+38\pm1$ &  $0.9^{+0.1}_{-0.1}$      & $9.2^{+0.4}_{-0.1}$         & $-$       & $-$                  & $-$     &$90^{+23}_{-23}$ \\
G312.E17  & -208.903& -282.822 & $2.836\pm0.017$& 108.68     &      $0.27$      & $-$         & $-$       &  $-$                    & $-$                     & $-$     & $-$                  & $-$     &$-$ \\   
G312.E18  & -200.580& -290.514 & $0.117\pm0.007$& 108.98     &      $0.22$      & $-$         & $-$       &  $-$                    & $-$                     & $-$     & $-$                  & $-$     &$-$ \\   
G312.E19  & -186.737& -144.226 & $0.368\pm0.019$& 112.11     &      $0.14$      & $-$         & $-$       &  $-$                    & $-$                     & $-$     & $-$                  & $-$     &$-$ \\   
G312.E20  & -184.050& -137.276 & $0.290\pm0.020$& 111.89     &      $0.18$      & $-$         & $-$       &  $-$                    & $-$                     & $-$     & $-$                  & $-$     &$-$ \\   
G312.E21  & -183.534&  16.922  & $0.068\pm0.005$& 110.74     &      $3.94$      & $-$         & $-$       &  $-$                    & $-$                     & $-$     & $-$                  & $-$     &$-$ \\   
G312.E22  & -176.569& -128.181 & $1.078\pm0.019$& 112.11     &      $0.20$      & $-$         & $-$       &  $-$                    & $-$                     & $-$     & $-$                  & $-$     &$-$ \\   
G312.E23  & -160.438&  25.583  & $0.413\pm0.035$& 112.28     &      $0.21$      & $-$         & $-$       &  $-$                    & $-$                     & $-$     & $-$                  & $-$     &$-$ \\   
G312.E24  & -159.180& -3.201   & $0.069\pm0.006$& 112.72     &      $0.22$      & $-$         & $-$       &  $-$                    & $-$                     & $-$     & $-$                  & $-$     &$-$ \\
G312.E25  & -156.520& -102.699 & $0.908\pm0.014$& 112.24     &      $0.40$      & $-$         & $-$       &  $-$                    & $-$                     & $-$     & $-$                  & $-$     &$-$ \\  
G312.E26  & -152.559& -99.792  & $3.846\pm0.038$& 112.33     &      $0.39$      & $4.1\pm0.5$ & $+37\pm8$ &  $1.4^{+0.1}_{-0.5}$       & $9.5^{+0.4}_{-0.1}$         & $-$       & $-$                  & $-$     &$86^{+4}_{-24}$ \\
G312.E27  & -135.684&  49.168  & $0.548\pm0.025$& 109.69     &      $0.26$      & $-$         & $-$       &  $-$                    & $-$                     & $-$     & $-$                  & $-$     &$-$ \\  
G312.E28  & -128.276& -76.635  & $0.746\pm0.005$& 112.85     &      $0.32$      & $5.4\pm0.1$ & $+39\pm2$ &  $0.8^{+0.1}_{-0.3}$      & $9.7^{+0.2}_{-0.1}$         & $-$       & $-$                  & $-$     &$90^{+14}_{-14}$ \\
G312.E29  & -18.691 &  129.513 & $0.370\pm0.018$& 109.56     &      $0.23$      & $-$         & $-$       &  $-$                    & $-$                     & $-$     & $-$                  & $-$     &$-$ \\   
G312.E30  &  0      &  0       & $32.441\pm0.137$& 110.35    &      $0.25$      & $8.1\pm1.3$ & $+40\pm1$ &  $0.7^{+0.5}_{-0.4}$       & $10.0^{+0.3}_{-0.6}$        & $-$        & $-$                  & $-$     &$88^{+1}_{-1}$ \\
G312.E31  &  0.057  & -42.929  & $1.763\pm0.017$& 111.93     &      $0.29$      & $-$         & $-$       &  $-$                    & $-$                     & $-$     & $-$                  & $-$     &$-$ \\   
G312.E32  &  3.690  & -17.494  & $1.485\pm0.010$& 108.37     &      $0.39$      & $16.7\pm0.3$& $+63\pm1$ &  $0.5^{+0.1}_{-0.1}$       & $11.4^{+0.1}_{-0.1}$        & $-$        & $-$                  & $-$     &$90^{+1}_{-1}$ \\
G312.E33  &  14.172 &  17.973  & $0.136\pm0.007$& 109.25     &      $0.19$      & $-$         & $-$       &  $-$                    & $-$                     & $-$     & $-$                  & $-$     &$-$ \\   
G312.E34  &  29.874 & -237.848 & $0.333\pm0.010$& 108.42     &      $0.26$      & $-$         & $-$       &  $-$                    & $-$                     & $-$     & $-$                  & $-$     &$-$ \\   
G312.E35  &  36.738 &  124.414 & $0.086\pm0.008$& 109.34     &      $0.37$      & $-$         & $-$       &  $-$                    & $-$                     & $-$     & $-$                  & $-$     &$-$ \\   
G312.E36  &  42.802 &  263.100 & $0.143\pm0.011$& 106.95     &      $0.29$      & $-$         & $-$       &  $-$                    & $-$                     & $-$     & $-$                  & $-$     &$-$ \\ 
G312.E37  &  53.870 &  114.351 & $0.767\pm0.010$& 108.86     &      $0.37$      & $-$         & $-$       &  $-$                    & $-$                     & $-$     & $-$                  & $-$     &$-$ \\  
G312.E38  &  59.176 &  107.553 & $0.707\pm0.009$& 109.38     &      $0.50$      & $-$         & $-$       &  $-$                    & $-$                     & $-$     & $-$                  & $-$     &$-$ \\   
G312.E39  &  72.218 &  117.935 & $0.645\pm0.006$& 103.02     &      $0.30$      & $-$         & $-$       &  $-$                    & $-$                     & $-$     & $-$                  & $-$     &$-$ \\  
G312.E40  &  76.351 &  113.398 & $0.270\pm0.004$& 102.45     &      $0.38$      & $-$         & $-$       &  $-$                    & $-$                     & $-$     & $-$                  & $-$     &$-$ \\   
G312.E41  &  79.211 & -2.188   & $0.049\pm0.007$& 103.28     &      $0.35$      & $-$         & $-$       &  $-$                    & $-$                     & $-$     & $-$                  & $-$     &$-$ \\   
G312.E42  &  81.242 &  98.152  & $0.044\pm0.004$& 102.87     &      $0.29$      & $-$         & $-$       &  $-$                    & $-$                     & $-$     & $-$                  & $-$     &$-$ \\   
G312.E43  &  91.066 &  102.911 & $1.824\pm0.012$& 104.42     &      $0.32$      & $-$         & $-$       &  $-$                    & $-$                     & $-$     & $-$                  & $-$     &$-$ \\   
G312.E44  &  93.068 &  110.018 & $0.900\pm0.012$& 104.47     &      $0.28$      & $-$         & $-$       &  $-$                    & $-$                     & $-$     & $-$                  & $-$     &$-$ \\   
G312.E45  &  97.158 & -11.244  & $0.037\pm0.004$& 104.73     &      $0.32$      & $-$         & $-$       &  $-$                    & $-$                     & $-$     & $-$                  & $-$     &$-$ \\   
G312.E46  &  98.245 & -288.097 & $0.067\pm0.007$& 104.25     &      $0.67$      & $-$         & $-$       &  $-$                    & $-$                     & $-$     & $-$                  & $-$     &$-$ \\   
G312.E47  &  115.005&  81.486  & $0.061\pm0.007$& 104.60     &      $0.32$      & $-$         & $-$       &  $-$                    & $-$                     & $-$     & $-$                  & $-$     &$-$ \\   
G312.E48  &  117.036& -261.021 & $0.525\pm0.022$& 105.30     &      $0.26$      & $-$         & $-$       &  $-$                    & $-$                     & $-$     & $-$                  & $-$     &$-$ \\ 
G312.E49  &  120.368& -250.776 & $1.742\pm0.021$& 105.26     &      $0.26$      & $2.3\pm0.5$ & $+8\pm2$  &  $1.0^{+0.1}_{-0.2}$       & $9.2^{+0.2}_{-1.1}$          & $-$      & $-$                  & $-$     &$90^{+20}_{-20}$ \\
G312.E50  &  121.784& -255.949 & $0.819\pm0.022$& 105.61     &      $0.21$      & $-$         & $-$       &  $-$                    & $-$                     & $-$     & $-$                  & $-$     &$-$ \\   
G312.E51  &  122.613&  51.416  & $0.245\pm0.004$& 103.46     &      $0.31$      & $-$         & $-$       &  $-$                    & $-$                     & $-$     & $-$                  & $-$     &$-$ \\  
G312.E52  &  132.867&  43.779  & $0.094\pm0.004$& 102.71     &      $0.22$      & $-$         & $-$       &  $-$                    & $-$                     & $-$     & $-$                  & $-$     &$-$ \\   
G312.E53  &  136.814& -106.281 & $2.610\pm0.060$& 110.88     &      $0.42$      & $-$         & $-$       &  $-$                    & $-$                     & $-$     & $-$                  & $-$     &$-$ \\   
G312.E54  &  139.088& -62.532  & $0.075\pm0.006$& 108.02     &      $0.20$      & $-$         & $-$       &  $-$                    & $-$                     & $-$     & $-$                  & $-$     &$-$ \\   
G312.E55  &  139.888& -32.051  & $0.297\pm0.010$& 106.09     &      $0.56$      & $-$         & $-$       &  $-$                    & $-$                     & $-$     & $-$                  & $-$     &$-$ \\   
G312.E56  &  140.689& -76.920  & $0.162\pm0.010$& 108.86     &      $0.33$      & $-$         & $-$       &  $-$                    & $-$                     & $-$     & $-$                  & $-$     &$-$ \\   
G312.E57  &  140.918&  10.685  & $0.050\pm0.004$& 102.53     &      $0.47$      & $-$         & $-$       &  $-$                    & $-$                     & $-$     & $-$                  & $-$     &$-$ \\   
G312.E58  &  141.333&  10.635  & $0.052\pm0.004$& 102.53     &      $0.47$      & $-$         & $-$       &  $-$                    & $-$                     & $-$     & $-$                  & $-$     &$-$ \\   
G312.E59  &  143.220& -236.525 & $0.164\pm0.010$& 106.84     &      $0.31$      & $-$         & $-$       &  $-$                    & $-$                     & $-$     & $-$                  & $-$     &$-$ \\   
G312.E60  &  144.822& -145.670 & $0.518\pm0.020$& 111.89     &      $0.27$      & $-$         & $-$       &  $-$                    & $-$                     & $-$     & $-$                  & $-$     &$-$ \\
G312.E61  &  145.880& -30.191  & $3.785\pm0.023$& 105.52     &      $0.53$      & $1.3\pm1.0$ & $-43\pm14$&  $1.7^{+0.1}_{-0.4}$       & $9.7^{+0.2}_{-0.1}$         & $-$       & $-$                  & $-$     &$\mathbf{59^{+6}_{-50}}$ \\
G312.E62  &  146.181& -221.058 & $7.026\pm0.041$& 105.87     &      $0.31$      & $0.5\pm0.1$ & $+23\pm1$ &  $1.3^{+0.1}_{-0.4}$       & $8.5^{+0.1}_{-0.1}$         & $-$       & $-$                  & $-$     &$90^{+23}_{-23}$ \\
G312.E63  &  148.197& -116.352 & $5.537\pm0.075$& 110.48     &      $0.26$      & $-$         & $-$       &  $-$                    & $-$                     & $-$     & $-$                  & $-$     &$-$ \\  
G312.E64  &  148.326& -129.145 & $8.260\pm0.055$& 111.14     &      $0.32$      & $0.9\pm0.1$ & $+74\pm10$&  $1.4^{+0.1}_{-0.5}$       & $8.7^{+0.4}_{-0.1}$         & $-$       & $-$                  & $-$     &$87^{+2}_{-47}$ \\
G312.E65  &  150.085& -121.626 & $0.181\pm0.009$& 109.38     &      $0.38$      & $-$         & $-$       &  $-$                    & $-$                     & $-$     & $-$                  & $-$     &$-$ \\   
G312.E66  &  151.043& -36.299  & $0.227\pm0.007$& 104.25     &      $0.29$      & $-$         & $-$       &  $-$                    & $-$                     & $-$     & $-$                  & $-$     &$-$ \\   
G312.E67  &  151.972& -47.279  & $0.392\pm0.004$& 103.54     &      $0.32$      & $-$         & $-$       &  $-$                    & $-$                     & $-$     & $-$                  & $-$     &$-$ \\   
G312.E68  &  152.416& -82.766  & $0.433\pm0.042$& 110.88     &      $0.39$      & $-$         & $-$       &  $-$                    & $-$                     & $-$     & $-$                  & $-$     &$-$ \\   
G312.E69  &  154.146& -146.338 & $3.087\pm0.043$& 110.88     &      $0.26$      & $-$         & $-$       &  $-$                    & $-$                     & $-$     & $-$                  & $-$     &$-$ \\ 
G312.E70  &  158.765& -153.168 & $8.425\pm0.060$& 110.66     &      $0.35$      & $-$         & $-$       &  $-$                    & $-$                     & $-$     & $-$                  & $-$     &$-$ \\   
G312.E71  &  159.409& -172.253 & $0.398\pm0.007$& 107.80     &      $0.24$      & $-$         & $-$       &  $-$                    & $-$                     & $-$     & $-$                  & $-$     &$-$ \\   
G312.E72  &  160.710& -162.045 & $0.145\pm0.008$& 107.89     &      $0.23$      & $-$         & $-$       &  $-$                    & $-$                     & $-$     & $-$                  & $-$     &$-$ \\ 
G312.E73  &  161.926& -177.849 & $0.160\pm0.007$& 108.20     &      $0.15$      & $-$         & $-$       &  $-$                    & $-$                     & $-$     & $-$                  & $-$     &$-$ \\  
G312.E74  &  163.356& -174.149 & $0.090\pm0.007$& 109.25     &      $0.26$      & $-$         & $-$       &  $-$                    & $-$                     & $-$     & $-$                  & $-$     &$-$ \\   
G312.E75  &  166.731& -162.006 & $0.121\pm0.012$& 107.01     &      $0.35$      & $-$         & $-$       &  $-$                    & $-$                     & $-$     & $-$                  & $-$     &$-$ \\  
G312.E76  &  167.975& -242.701 & $0.154\pm0.020$& 107.54     &      $0.54$      & $-$         & $-$       &  $-$                    & $-$                     & $-$     & $-$                  & $-$     &$-$ \\   
G312.E77  &  168.718& -132.042 & $4.021\pm0.022$& 107.54     &      $0.31$      & $1.8\pm0.2$ & $-78\pm3$ &  $1.3^{+0.1}_{-0.4}$       & $9.0^{+0.3}_{-0.1}$         & $-$       & $-$                  & $-$     &$90^{+25}_{-25}$\\
G312.E78  &  170.492& -137.651 & $1.447\pm0.009$& 106.79     &      $0.50$      & $-$         & $-$       &  $-$                    & $-$                     & $-$     & $-$                  & $-$     &$-$ \\   
G312.E79  &  173.409& -159.010 & $0.697\pm0.010$& 108.86     &      $0.31$      & $-$         & $-$       &  $-$                    & $-$                     & $-$     & $-$                  & $-$     &$-$ \\   
G312.E80  &  177.814& -172.522 & $2.119\pm0.019$& 106.09     &      $0.25$      & $-$         & $-$       &  $-$                    & $-$                     & $-$     & $-$                  & $-$     &$-$ \\   
G312.E81  &  319.475&  486.170 & $0.178\pm0.012$& 107.27     &      $0.47$      & $-$         & $-$       &  $-$                    & $-$                     & $-$     & $-$                  & $-$     &$-$ \\   
G312.E82  &  320.548&  189.465 & $1.461\pm0.012$& 107.14     &      $0.26$      & $-$         & $-$       &  $-$                    & $-$                     & $-$     & $-$                  & $-$     &$-$ \\   
G312.E83  &  326.740&  184.299 & $7.008\pm0.045$& 107.23     &      $0.28$      & $-$         & $-$       &  $-$                    & $-$                     & $-$     & $-$                  & $-$     &$-$ \\   
G312.E84  &  329.185& -112.333 & $0.357\pm0.027$& 107.19     &      $0.30$      & $-$         & $-$       &  $-$                    & $-$                     & $-$     & $-$                  & $-$     &$-$ \\   
G312.E85  &  449.997&  98.328  & $0.454\pm0.023$& 105.65     &      $0.17$      & $-$         & $-$       &  $-$                    & $-$                     & $-$     & $-$                  & $-$     &$-$ \\   
G312.E86  &  484.090& -22.982  & $0.324\pm0.012$& 108.81     &      $0.25$      & $-$         & $-$       &  $-$                    & $-$                     & $-$     & $-$                  & $-$     &$-$ \\   
G312.E87  &  489.767& -9.336   & $1.914\pm0.026$& 109.78     &      $0.29$      & $4.1\pm0.4$ & $+88\pm2$ &  $1.0^{+0.3}_{-0.1}$       & $9.5^{+0.4}_{-0.2}$         & $-$       & $-$                  & $-$     &$90^{+17}_{-17}$ \\
G312.E88  &  504.868&  18.246  & $5.967\pm0.075$& 110.17     &      $0.21$      & $2.1\pm0.6$ & $-52\pm2$ &  $0.9^{+0.1}_{-0.2}$      & $9.3^{+0.6}_{-0.4}$         & $-$       & $-$                  & $-$     &$71^{+15}_{-37}$ 
\label{G312_tab}
\end{longtable}
\clearpage
\twocolumn
\normalsize
\begin {table*}[]
\caption []{Parameters of the 6.7-GHz \meth ~maser features detected in G32.03+0.06.} 
\begin{center}
\scriptsize
\begin{tabular}{ l c c c c c c c c c c c c c}
\hline
\hline
\,\,\,\,\,(1)&(2)   & (3)      & (4)            & (5)       & (6)              & (7)         & (8)       & (9)                     & (10)                    & (11)                        & (12)         &(13)                  &(14)         \\
Maser     & RA\tablefootmark{a}&Dec\tablefootmark{a}& Peak flux & $V_{\rm{lsr}}$& $\Delta v\rm{_{L}}$ &$P_{\rm{l}}\tablefootmark{b}$ &  $\chi\tablefootmark{b}$   & $\Delta V_{\rm{i}}\tablefootmark{c}$ & $T_{\rm{b}}\Delta\Omega\tablefootmark{c}$& $P_{\rm{V}}$ & $\Delta V_{\rm{Z}}$  & $|B_{||}|$  &$\theta\tablefootmark{d}$\\
          &  offset &  offset  & Density(I)     &           &                  &             &            &                         &                         &              &                      &      & \\ 
          &  (mas)  &  (mas)   & (Jy/beam)      &  (km/s)   &      (km/s)      & (\%)        &   (\d)    & (km/s)                  & (log K sr)              &   ($\%$)     &  (m/s)               & (mG) & (\d)       \\ 
\hline
G32.E01   & -649.967& 218.353 & $0.357\pm0.015$& 99.68      &      $0.23$      & $-$         & $-$       &  $-$                    & $-$                     & $-$     & $-$                  & $-$     &$-$ \\ 
G32.E02   & -69.631 & 732.437 & $0.243\pm0.009$& 95.42      &      $0.35$      & $-$         & $-$       &  $-$                    & $-$                     & $-$     & $-$                  & $-$     &$-$ \\ 
G32.E03   & -67.343 & 209.625 & $0.288\pm0.010$& 95.46      &      $0.26$      & $-$         & $-$       &  $-$                    & $-$                     & $-$     & $-$                  & $-$     &$-$ \\ 
G32.E04   & -61.335 & -291.874 & $2.065\pm0.010$& 95.55      &      $0.34$      & $-$         & $-$       &  $-$                    & $-$                     & $-$     & $-$                  & $-$     &$-$ \\ 
G32.E05   & -59.618 & -346.397 & $0.865\pm0.015$& 94.67      &      $0.22$      & $-$         & $-$       &  $-$                    & $-$                     & $-$     & $-$                  & $-$     &$-$ \\ 
G32.E06   & -58.589 &   -1.057 & $0.544\pm0.005$& 93.62      &      $0.31$      & $-$         & $-$       &  $-$                    & $-$                     & $-$     & $-$                  & $-$     &$-$ \\ 
G32.E07   & -58.074 & 171.963 & $0.536\pm0.021$& 93.09      &      $0.32$      & $-$         & $-$       &  $-$                    & $-$                     & $-$     & $-$                  & $-$     &$-$ \\ 
G32.E08   & -54.469 &   -11.765 & $5.265\pm0.016$& 93.22      &      $0.31$      & $1.5\pm0.2$ & $+21\pm4$ &  $1.3^{+0.1}_{-0.2}$    & $8.9^{+0.6}_{-0.6}$     & $-$           & $-$                  & $-$     &$90^{+27}_{-27}$ \\ 
G32.E09   & -54.412 &   7.839 & $0.054\pm0.004$& 93.71      &      $0.27$      & $-$         & $-$       &  $-$                    & $-$                     & $-$     & $-$                  & $-$     &$-$ \\ 
G32.E10   & -50.178 &  62.977 & $1.178\pm0.097$& 92.61      &      $0.20$      & $-$         & $-$       &  $-$                    & $-$                     & $-$     & $-$                  & $-$     &$-$ \\ 
G32.E11   & -49.148 &  -22.091 & $53.392\pm0.151$& 92.83     &      $0.32$      & $-$         & $-$       &  $-$                    & $-$                     & $0.2$           & $0.6\pm0.1\tablefootmark{e}$   & $>12\tablefootmark{e}$    &$-$ \\ 
G32.E12   & -43.426 &  -17.902 & $8.656\pm0.033$& 92.39      &      $0.30$      & $0.4\pm0.1$ & $+11\pm3$ &  $1.3^{+0.1}_{-0.2}$    & $8.4^{+1.2}_{-0.9}$     & $-$           & $-$                  & $-$     &$74^{+12}_{-40}$ \\ 
G32.E13   & -43.369 &  -67.810 & $0.036\pm0.003$& 91.42      &      $0.30$      & $-$         & $-$       &  $-$                    & $-$                     & $-$     & $-$                  & $-$     &$-$ \\ 
G32.E14   & -41.310 & 198.444 & $1.468\pm0.096$& 92.61      &      $0.23$      & $-$         & $-$       &  $-$                    & $-$                     & $-$     & $-$                  & $-$     &$-$ \\ 
G32.E15   & -40.623 &   -8.327 & $8.203\pm0.145$& 92.70      &      $0.28$      & $-$         & $-$       &  $-$                    & $-$                     & $-$     & $-$                  & $-$     &$-$ \\ 
G32.E16   & -31.755 &  -385.883 & $0.149\pm0.003$& 96.52      &      $0.19$      & $-$         & $-$       &  $-$                    & $-$                     & $-$     & $-$                  & $-$     &$-$ \\ 
G32.E17   & -10.814 &  37.399 & $0.050\pm0.003$& 94.14      &      $0.23$      & $-$         & $-$       &  $-$                    & $-$                     & $-$     & $-$                  & $-$     &$-$ \\ 
G32.E18   & -5.550  &   -4.475 & $0.201\pm0.003$& 94.36      &      $0.36$      & $-$         & $-$       &  $-$                    & $-$                     & $-$     & $-$                  & $-$     &$-$ \\ 
G32.E19   &   0     &    0     & $18.832\pm0.044$& 94.89     &      $0.29$      & $1.2\pm0.1$ & $+50\pm2$ &  $1.2^{+0.1}_{-0.1}$    & $8.8^{+0.8}_{-0.5}$     & $-$           & $-$                  & $-$     &$83^{+6}_{-41}$ \\ 
G32.E20   &   1.716 &   4.063 & $0.550\pm0.003$& 93.93      &      $0.33$      & $-$         & $-$       &  $-$                    & $-$                     & $-$     & $-$                  & $-$     &$-$ \\ 
G32.E21   &  19.396 &   4.753 & $0.414\pm0.006$& 95.73      &      $0.33$      & $-$         & $-$       &  $-$                    & $-$                     & $-$     & $-$                  & $-$     &$-$ \\ 
G32.E22   &  21.112 &  -403.988 & $0.177\pm0.005$& 95.29      &      $0.19$      & $-$         & $-$       &  $-$                    & $-$                     & $-$     & $-$                  & $-$     &$-$ \\ 
G32.E23   &  34.558 &  17.071 & $0.064\pm0.004$& 95.81      &      $0.30$      & $-$         & $-$       &  $-$                    & $-$                     & $-$     & $-$                  & $-$     &$-$ \\ 
G32.E24   &  34.615 &    -8.762 & $0.081\pm0.006$& 98.14      &      $0.28$      & $-$         & $-$       &  $-$                    & $-$                     & $-$     & $-$                  & $-$     &$-$ \\ 
G32.E25   &  36.160 &   9.945 & $0.436\pm0.009$& 95.64      &      $0.41$      & $-$         & $-$       &  $-$                    & $-$                     & $-$     & $-$                  & $-$     &$-$ \\ 
G32.E26   &  36.446 &    -3.906 & $0.030\pm0.003$& 102.35     &      $1.00$      & $-$         & $-$       &  $-$                    & $-$                     & $-$     & $-$                  & $-$     &$-$ \\ 
G32.E27   &  37.648 &    -0.950 & $0.338\pm0.003$& 96.87      &      $0.32$      & $-$         & $-$       &  $-$                    & $-$                     & $-$     & $-$                  & $-$     &$-$ \\ 
G32.E28   &  41.824 &    -0.465 & $0.086\pm0.011$& 100.64     &      $0.79$      & $-$         & $-$       &  $-$                    & $-$                     & $-$     & $-$                  & $-$     &$-$ \\ 
G32.E29   &  43.884 &   1.823 & $0.132\pm0.003$& 98.49      &      $0.97$      & $-$         & $-$       &  $-$                    & $-$                     & $-$     & $-$                  & $-$     &$-$ \\ 
G32.E30   &  46.116 &   4.936 & $0.134\pm0.003$& 96.56      &      $0.22$      & $-$         & $-$       &  $-$                    & $-$                     & $-$     & $-$                  & $-$     &$-$ \\ 
G32.E31   &  51.265 &  10.368 & $0.172\pm0.003$& 96.78      &      $0.34$      & $-$         & $-$       &  $-$                    & $-$                     & $-$     & $-$                  & $-$     &$-$ \\ 
G32.E32   &  77.870 &    -9.899 & $0.204\pm0.003$& 98.40      &      $0.42$      & $-$         & $-$       &  $-$                    & $-$                     & $-$     & $-$                  & $-$     &$-$ \\ 
G32.E33   & 186.408 & 409.798 & $0.279\pm0.003$& 100.29     &      $0.29$      & $-$         & $-$       &  $-$                    & $-$                     & $-$     & $-$                  & $-$     &$-$ \\ 
G32.E34   & 193.045 & 449.825 & $0.193\pm0.008$& 100.82     &      $0.49$      & $-$         & $-$       &  $-$                    & $-$                     & $-$     & $-$                  & $-$     &$-$ \\ 
G32.E35   & 193.216 & 399.990 & $5.390\pm0.045$& 101.21     &      $0.68$      & $2.1\pm0.6$ & $+7\pm4$  &  $1.2^{+0.1}_{-0.2}$    & $6.1^{+2.7}_{-0.3}$     & $-$           & $-$                  & $-$     &$90^{+25}_{-25}$ \\ 
G32.E36   & 193.903 & 403.492 & $9.027\pm0.026$& 99.54      &      $2.58$      & $0.7\pm0.2$ & $+40\pm10$&  $1.2^{+0.1}_{-0.1}$    & $8.6^{+1.0}_{-1.2}$     & $-$           & $-$                  & $-$     &$79^{+10}_{-37}$ \\ 
G32.E37   & 196.478 & 401.653 & $18.740\pm0.046$& 101.21    &      $0.23$      & $3.4\pm1.4$ & $+6\pm3$  &  $0.8^{+0.1}_{-0.2}$    & $9.4^{+0.1}_{-3.1}$     & $0.8$         & $1.4\pm0.2$          & $>27$   &$90^{+53}_{-53}$ \\ 
G32.E38   & 197.107 & 216.026 & $0.342\pm0.005$& 100.51     &      $0.17$      & $-$         & $-$       &  $-$                    & $-$                     & $-$     & $-$                  & $-$     &$-$ \\ 
G32.E39   & 197.336 & 405.983 & $18.92\pm0.059$& 100.91     &      $0.35$      & $-$         & $-$       &  $-$                    & $-$                     & $-$     & $-$                  & $-$     &$-$ \\ 
G32.E40   & 199.968 & 449.001 & $1.765\pm0.046$& 101.08     &      $0.17$      & $3.8\pm1.3$ & $-1\pm3$  &  $0.6^{+0.1}_{-0.1}$    & $9.5^{+0.1}_{-3.2}$     & $-$           & $-$                  & $-$     &$90^{+27}_{-27}$ \\ 
G32.E41   & 200.940 & 278.004 & $1.344\pm0.044$& 101.30     &      $0.20$      & $3.1\pm0.4$ & $+1\pm88$ &  $0.7^{+0.1}_{-0.1}$    & $9.3^{+0.6}_{-1.2}$     & $-$           & $-$                  & $-$     &$90^{+14}_{-14}$ \\ 
G32.E42   & 201.341 & 386.169 & $6.562\pm0.059$& 101.17     &      $0.39$      & $1.9\pm0.4$ & $-7\pm4$  &  $1.3^{+0.2}_{-0.5}$    & $9.7^{+1.5}_{-0.4}$     & $-$           & $-$                  & $-$     &$\mathbf{67^{+7}_{-44}}$ \\ 
G32.E43   & 202.485 & 424.084 & $0.848\pm0.018$& 100.69     &      $0.22$      & $-$         & $-$       &  $-$                    & $-$                     & $-$     & $-$                  & $-$     &$-$ \\ 
G32.E44   & 208.150 & 208.946 & $0.040\pm0.003$& 101.35     &      $0.42$      & $-$         & $-$       &  $-$                    & $-$                     & $-$     & $-$                  & $-$     &$-$ \\ 
G32.E45   & 211.411 & 393.990 & $0.614\pm0.057$& 98.89      &      $0.34$      & $-$         & $-$       &  $-$                    & $-$                     & $-$     & $-$                  & $-$     &$-$ \\ 
G32.E46   & 212.612 & 394.341 & $0.147\pm0.005$& 101.61     &      $0.19$      & $-$         & $-$       &  $-$                    & $-$                     & $-$     & $-$                  & $-$     &$-$ \\ 
G32.E47   & 222.053 & -858.109 & $0.219\pm0.022$& 98.49      &      $1.54$      & $-$         & $-$       &  $-$                    & $-$                     & $-$     & $-$                  & $-$     &$-$ \\ 
G32.E48   & 226.344 & 391.407 & $0.803\pm0.011$& 99.37      &      $1.55$      & $9.6\pm0.3$ & $-1\pm1$  &  $0.5^{+0.1}_{-0.1}$    & $10.2^{+0.1}_{-0.3}$    & $-$            & $-$                  & $-$     &$90^{+5}_{-5}$ \\ 
G32.E49   & 231.665 & 388.920 & $0.409\pm0.005$& 101.61     &      $0.33$      & $-$         & $-$       &  $-$                    & $-$                     & $-$     & $-$                  & $-$     &$-$ \\ 
G32.E50   & 235.212 & 393.272 & $52.457\pm0.142$& 98.48     &      $0.46$      & $10.0\pm0.8$& $+80\pm2$ &  $1.1^{+0.4}_{-0.1}$    & $10.2^{+0.6}_{-0.3}$    & $-$            & $-$                  & $-$     &$76^{+11}_{-1}$ \\ 
G32.E51   & 236.586 & 392.563 & $0.172\pm0.003$& 97.61      &      $0.31$      & $-$         & $-$       &  $-$                    & $-$                     & $-$     & $-$                  & $-$     &$-$ \\ 
G32.E52   & 241.220 & 398.182 & $0.095\pm0.003$& 97.44      &      $1.26$      & $-$         & $-$       &  $-$                    & $-$                     & $-$     & $-$                  & $-$     &$-$ \\ 
G32.E53   & 294.945 & 227.856 & $0.230\pm0.003$& 98.49      &      $0.46$      & $-$         & $-$       &  $-$                    & $-$                     & $-$     & $-$                  & $-$     &$-$ \\ 
\hline
\end{tabular} \end{center}
\tablefoot{
\tablefoottext{a}{The reference position is $\alpha_{2000}=18^{\rm{h}}49^{\rm{m}}36^{\rm{s}}\!.580$ and 
$\delta_{2000}=-00^{\circ}45'46''\!\!.891$ (see Sec.~\ref{obssect}).}
\tablefoottext{b}{$P_{\rm{l}}$ and $\chi$ are the mean values of the linear polarization fraction and the linear polarization angle measured across the spectrum, respectively.}
\tablefoottext{c}{The best-fitting results obtained using a model based on the radiative transfer theory of methanol masers 
for $\Gamma+\Gamma_{\nu}=1~\rm{s^{-1}}$ \citep{vle10,sur111}. The errors were determined 
by analyzing the full PDF.}
\tablefoottext{d}{The angle between the magnetic field and the maser propagation direction is determined using the observed $P_{\rm{l}}$ 
and the fitted emerging brightness temperature. The errors were determined by analyzing the full PDF. The boldface indicates that $|\theta^{\rm{+}}-55$\d$|<|\theta^{\rm{-}}-55$\d$|$, i.e., the magnetic field is parallel to the linear polarization vector (see Sect.~\ref{res}).}
\tablefoottext{e} {To model the circularly polarized emission, we considered the values of \tbo$=2.0\cdot10^8~\rm{K sr}$ and
\dvi$=1.5$~\kms ~that best fit the total intensity emission.}
}
\label{G32_tab}
\end{table*}
\begin {table*}[t!]
\caption []{Parameters of the 6.7-GHz \meth ~maser features detected in G69.52-0.97.} 
\begin{center}
\scriptsize
\begin{tabular}{ l c c c c c c c c c c c c c c}
\hline
\hline
\,\,\,\,\,(1)&(2)   & (3)      & (4)            & (5)       & (6)              & (7)         & (8)       & (9)                     & (10)                    & (11)                        & (12)         &(13)                  &(14) & (15)        \\
Maser     & cluster & RA\tablefootmark{a}&Dec\tablefootmark{a}& Peak flux & $V_{\rm{lsr}}$& $\Delta v\rm{_{L}}$ &$P_{\rm{l}}\tablefootmark{b}$ &  $\chi\tablefootmark{b}$   & $\Delta V_{\rm{i}}\tablefootmark{c}$ & $T_{\rm{b}}\Delta\Omega\tablefootmark{c}$& $P_{\rm{V}}$ & $\Delta V_{\rm{Z}}$  & $|B_{||}|$  &$\theta\tablefootmark{d}$\\
          &         & offset &  offset  & Density(I)     &           &                  &             &           &                         &                         &              &                      &      & \\ 
          &         & (mas)  &  (mas)   & (Jy/beam)      &  (km/s)   &      (km/s)      & (\%)        &   (\d)    & (km/s)                  & (log K sr)              &   ($\%$)     &  (m/s)               & (mG) & (\d)       \\ 
\hline
 G69.E01  & II & -45.092 &  126.293 & $0.065\pm0.002$& 15.66     &      $0.19$      & $-$         & $-$       &  $-$                    & $-$                     & $-$     & $-$                  & $-$     &$-$ \\ 
 G69.E02  & II & -0.890  &  6.123   & $0.817\pm0.024$& 14.83     &      $0.16$      & $-$         & $-$       &  $-$                    & $-$                     & $-$     & $-$                  & $-$     &$-$ \\ 
 G69.E03  & II & 0       &  0       & $11.524\pm0.122$& 14.65    &      $0.27$      & $1.3\pm0.3$ & $-34\pm9$ &  $1.1^{+0.1}_{-0.1}$       & $9.2^{+1.4}_{-0.6}$         & $0.6$     & $1.2\pm0.2$          & $>24$     &$69^{+18}_{-34}$ \\ 
 G69.E04  & II & 16.303  &  7.305   & $0.286\pm0.010$& 14.35     &      $0.18$      & $-$         & $-$       &  $-$                    & $-$                     & $-$     & $-$                  & $-$     &$-$ \\ 
 G69.E05  & I & 338.887 &  883.820 & $3.238\pm0.065$& -0.01     &      $0.48$      & $1.5\pm0.3$ & $-27\pm3$ &  $1.9^{+0.1}_{-0.6}$       & $9.2^{+0.5}_{-0.4}$           & $-$     & $-$                  & $-$     &$\mathbf{62^{+14}_{-45}}$ \\ 
 G69.E06  & I & 345.581 &  928.917 & $0.527\pm0.012$& 1.44      &      $0.20$      & $-$         & $-$       &  $-$                    & $-$                     & $-$     & $-$                  & $-$     &$-$ \\ 
 G69.E07  & I & 345.679 &  896.439 & $0.525\pm0.065$& 0.03      &      $0.49$      & $-$         & $-$       &  $-$                    & $-$                     & $4.2$           & $14.5\pm4.4\tablefootmark{e}$        & $>285\tablefootmark{e}$     &$-$ \\ 
 G69.E08  & I & 346.557 &  896.175 & $0.515\pm0.051$& 0.03      &      $0.48$      & $-$         & $-$       &  $-$                    & $-$                     & $-$     & $-$                  & $-$     &$-$ \\
 G69.E09  & I & 350.520 &  928.394 & $0.058\pm0.005$& 1.13      &      $2.30$      & $-$         & $-$       &  $-$                    & $-$                     & $-$     & $-$                  & $-$     &$-$ \\ 
 G69.E10  & I & 378.187 &  943.081 & $0.033\pm0.002$& 2.32      &      $0.29$      & $-$         & $-$       &  $-$                    & $-$                     & $-$     & $-$                  & $-$     &$-$ \\ 
 G69.E11  & I & 389.637 &  950.752 & $0.032\pm0.002$& 2.58      &      $0.18$      & $-$         & $-$       &  $-$                    & $-$                     & $-$     & $-$                  & $-$     &$-$ \\ 
 G69.E12  & III & 401.380 & -134.197 & $1.394\pm0.047$& 14.52     &      $1.64$      & $-$         & $-$       &  $-$                    & $-$                     & $-$      & $-$                  & $-$     &$-$ \\ 
 G69.E13  & III & 417.475 & -126.911 & $0.194\pm0.017$& 14.39     &      $0.15$      & $-$         & $-$       &  $-$                    & $-$                     & $-$      & $-$                  & $-$     &$-$ \\ 
\hline
\end{tabular} \end{center}
\tablefoot{
\tablefoottext{a}{The reference position is $\alpha_{2000}=20^{\rm{h}}10^{\rm{m}}09^{\rm{s}}\!.0699$ and 
$\delta_{2000}=+31^{\circ}31'34''\!\!.399$ (see Sec.~\ref{obssect}).}
\tablefoottext{b}{$P_{\rm{l}}$ and $\chi$ are the mean values of the linear polarization fraction and the linear polarization angle measured across the spectrum, respectively.}
\tablefoottext{c}{The best-fitting results obtained using a model based on the radiative transfer theory of methanol masers 
for $\Gamma+\Gamma_{\nu}=1~\rm{s^{-1}}$ \citep{vle10,sur111}. The errors were determined 
by analyzing the full PDF.}
\tablefoottext{d}{The angle between the magnetic field and the maser propagation direction is determined using the observed $P_{\rm{l}}$ 
and the fitted emerging brightness temperature. The errors were determined by analyzing the full PDF. The boldface indicates that $|\theta^{\rm{+}}-55$\d$|<|\theta^{\rm{-}}-55$\d$|$, i.e., the magnetic field is parallel to the linear polarization vector (see Sect.~\ref{res}).}
\tablefoottext{e}{To model the circularly polarized emission, we assumed \tbo$=6.3\cdot10^9$~K~sr and \dvi$=1.0$~\kms ~that best fit the total intensity emission.}
}
\label{G69_tab}
\end{table*}

\end{appendix}

\end{document}